\documentclass{article} 
\usepackage{graphicx}
\usepackage{color}
\usepackage{subfig}
\usepackage{amssymb}
\usepackage{amsthm}
\usepackage{amsmath} 
\usepackage{mathrsfs}

\begin{document}

\title{Constrained quantum mechanics: chaos in  non-planar billiards}
\author{R. Salazar and G.~T\'ellez\\
Departamento de F\'{\i}sica, Universidad de los Andes\\
A.A. 4976, Bogot\'a Colombia
}

\date{}


\maketitle

\begin{abstract}
We illustrate some of the techniques to identify chaos signatures at
the quantum level using as a guiding examples some systems where a
particle is constrained to move on a radial symmetric, but non planar,
surface. In particular, two systems are studied: the case of a cone
with an arbitrary contour or \textit{dunce hat billiard} and the
rectangular \textit{billiard with an inner Gaussian
  surface}. 
\end{abstract}


\section{Introduction}
Hard wall billiards are among the simplest and most studied systems in
the field of classical and quantum chaos
\cite{MethodForBilliards}-\cite{annularBilliardQSM}. A two dimensional
quantum billiard is analogous to a vibrating membrane because of the
mathematical equivalence of the stationary Schr\"odinger and Helmholtz
equation with the same Dirichlet boundary conditions. This similarity
between billiards and classical waves has been used in experiments
e.g.~the quantum microwave
cavities~\cite{stadiumBilliardExperimental}. When the classical
counterpart of these systems are chaotic, some signatures of this chaos
appear at the quantum level. One of them is the statistical properties
of the spectrum, which follow the \textit{Bohigas, Giannoni and
  Schmit} conjecture \cite{sinaiBilliardAndBGSConjecture}. This
conjecture states that the nearest neighbour energy level statistics of
a time reversal invariant chaotic system follows the same statistical
as the ensemble of random orthogonal matrices with Gaussian
distributed elements, i.e. the \textit{Gaussian Orthogonal Ensemble}
(GOE)~\cite{Mehta}. For $2\times2$ matrices this distribution, also
known as the Wigner distribution, is given by
\begin{equation}
P_{GOE}(s) = \frac{\pi}{2} s \exp\left(\frac{-\pi s^2}{4}\right).
\label{eq:GOE}
\end{equation}
To be more precise, the energy levels should be first classified
according to the partial symmetries that the system has, then the statistical
analysis should performed on each symmetry class of states, and for
each of those the nearest neighbour spacing distribution of the energy
level, properly normalized, follows the GOE
distribution~(\ref{eq:GOE}).  On the other hand, for systems with
enough symmetries to be integrable, the nearest neighbour energy level
statistics follows a Poisson distribution.

Additionally, quantum chaotic billiards exhibit the phenomenon
of scarring of their wavefunction specially when the corresponding
state is deep in the semiclassical limit \cite{HellerPaper}. Although,
the scars in the wavefunction were first observed in the seventies
\cite{McDonaldPaper} they have become a phenomenon of great interest
because they represent a connection between the quantum chaotic
billiard with its periodic orbits. 

The chaotic features of two dimensional billiards are often studied
confining them on a plane with a contour $\partial\mathfrak{D}$ where
chaos emerges when an ``irregular" contour is used. However, the
system present interesting features if we set a non-planar geometry in
the billiard internal region $\mathfrak{D}$, even when the contour is
regular as a circle, a rectangle or an ellipse.

\begin{figure}[h]
  \centering   
  \subfloat{\includegraphics[width=0.5\textwidth]{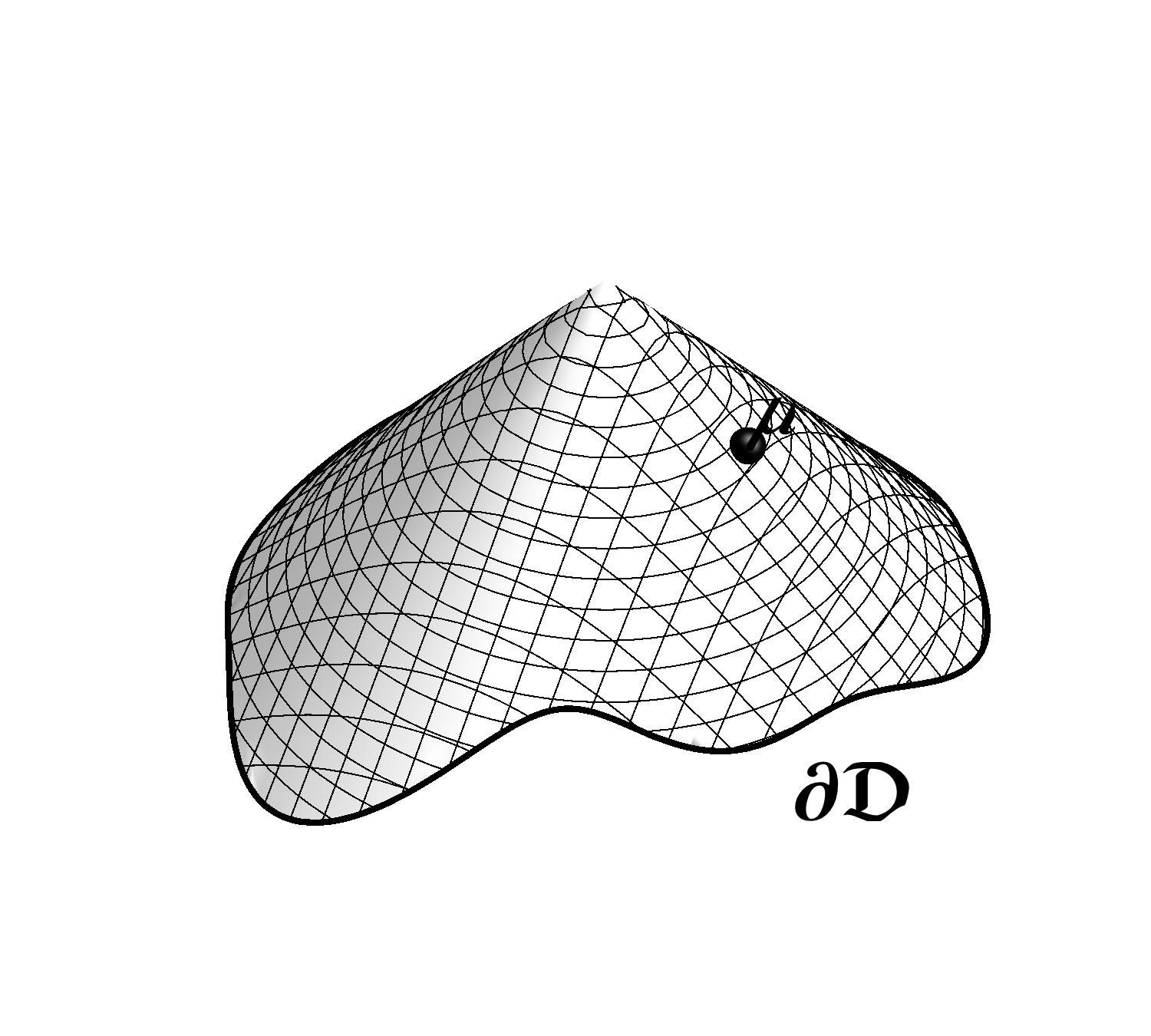}}     
  \subfloat{\includegraphics[width=0.5\textwidth]{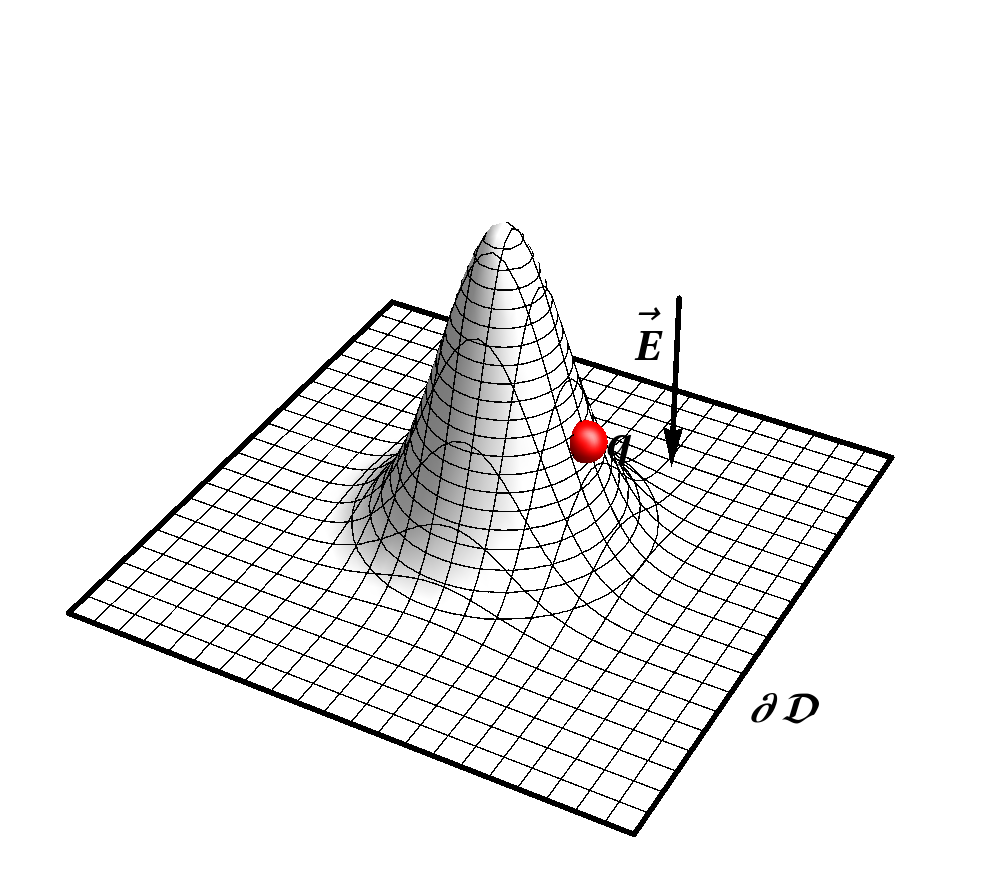}}
  \caption[The systems]
 {\textbf{The systems:} \textit{(left)} the
  dunce hat billiard with an arbitrary contour, \textit{(right)} the
  rectangular billiard with a Gaussian surface immersed in a uniform electric field.}
  \label{boundaryFig}
\end{figure}

The aim of this work is to illustrate some of the characteristic
signatures of chaos at the quantum level in non-planar billiards. As
such, this paper is an introduction for graduate and advanced
undergraduate students to the topic of chaos in quantum systems,
illustrated in systems different than the usual planar billiards.  It
can serve as a guiding example to the techniques used to analyze and
identify chaos signatures in quantum systems. Two systems are
studied: the \textit{dunce hat billiard} (or conical billiard) and the
rectangular \textit{billiard with a Gaussian surface}. These billiards
are shown in figure~\ref{boundaryFig}.

We start in section~\ref{sec:geom} with a geometrical background
needed to study non-planar systems. Although most of this work is
about quantum systems, in section~\ref{sec:classic}, we study the
classical dynamics on these billiards and present some of the chaos
characteristic signatures at the classical level. Then we proceed, in
section~\ref{sec:qh}, to present how the quantum Hamiltonian is
constructed on non-planar surfaces, and in sections~\ref{sec:dh}
and~\ref{sec:bigs} both types of billiards are studied at the quantum level.

\section{Geometrical background}
\label{sec:geom}

Consider a surface $\Sigma$ embedded in the three-dimensional Euclidean
space $E_3$ defined by
\begin{equation}
\Sigma: y^i=y^i(u^1,u^2), \hspace{0.5cm} (i=1,2,3)
\end{equation}
\\where $\left\{y^1,y^1,y^3\right\}$ are the Cartesian coordinates in $E_3$ and $\left\{u^1,u^2\right\}$ are the Gaussian curvilinear coordinates defined on the surface. A particle restricted to move on a two-dimensional surface has only two degrees of freedom, and its position $\vec{r}\in\mathbb{R}^3$ may be expressed as follows
\begin{equation}
\vec{r}=\vec{r}\left(u^1,u^2\right) = y^1(u^1,u^2)\hat{\mathbf{i}} + y^2(u^1,u^2) \hat{\mathbf{j}} + y^3(u^1,u^2) \hat{\mathbf{k}} \hspace{0.1cm}.
\end{equation}
The motion of a particle will depend of the surface geometry which is specified by the first and second fundamental forms. Several of the intrinsic properties of $\Sigma$ are determined by its metric $a_{\alpha\beta}$ included in the \textit{first fundamental form} $\Phi_I$ given by 

\begin{equation}
\Phi_I := d\vec{r}\cdot d\vec{r} = a_{\alpha\beta}du^{\alpha}du^{\beta} \hspace{0.5cm}\mbox{with}\hspace{0.5cm} a_{\alpha\beta}=\frac{\partial y^i}{\partial u^\alpha}\frac{\partial y^i}{\partial u^\beta}, \hspace{0.5cm} (i=1,2,3) (\alpha,\beta=1,2)\hspace{0.1cm}. 
\end{equation}
\\We are interested in billiards with a radial symmetric surface defined by

\begin{equation}
y^1=u^1\cos\left(u^2\right) \hspace{0.5cm} y^2=u^1\sin\left(u^2\right) \hspace{0.5cm} y^3 = f(u^1)\hspace{0.1cm}. 
\end{equation}
\\For this particular case the metric is a diagonal tensor

\begin{equation}
\left(a_{\alpha\beta}\right)= 
\left( \begin{array}{lcr}
            1+\left(\partial_{u^1}f\right)^2      & 0    \\
            0      & (u^1)^2                 
           \end{array}
    \right)\hspace{0.1cm}.
\label{theMetricUsed}
\end{equation}
\\Therefore, the system is orthogonal and the \textit{Gaussian curvature} $K$ may be computed with  

\begin{equation}
K = -\frac{1}{2\sqrt{a}}\left[\frac{\partial}{\partial u^1}\left(\frac{1}{\sqrt{a}}\frac{\partial a_{22}}{\partial u^1}\right)+\frac{\partial}{\partial u^2}\left(\frac{1}{\sqrt{a}}\frac{\partial a_{11}}{\partial u^2}\right)\right] 
\end{equation}
\\where $a:=\det(a_{\alpha\beta})$. Replacing the metric given by the equation (\ref{theMetricUsed}) we find
\begin{equation}
K = \frac{\left(\partial_{u^1}f\right)\left(\partial^2_{u^1}f\right)}{u^1\left[1+\left(\partial_{u^1}f\right)^2\right]}\hspace{0.1cm}.
\label{GaussianCurvatureEq}
\end{equation}
Classically, the motion of a particle which lives on $\Sigma$ is
affected by the surface curvature because the surface metric is
included explicitly in the kinetic energy, which is proportional to
$a_{\alpha\beta}\dot{u}^\beta \dot{u}^\beta$. At the quantum level,
this kinetic energy is associated with the Laplace-Beltrami operator
on the surface. Other extrinsic properties of $\Sigma$ are specified
in its second fundamental form $\Phi_{II}$ given by
\begin{equation}
\Phi_{II} := b_{\alpha\beta}du^{\alpha}du^{\beta} \hspace{0.5cm}\mbox{with}\hspace{0.5cm} b_{\alpha\beta}=\frac{1}{\sqrt{a}}\left(\frac{\partial\vec{r}}{\partial u^1}\times\frac{\partial\vec{r}}{\partial u^2}\right)\cdot\frac{\partial^2\vec{r}}{\partial u^{\alpha} \partial u^{\beta}}\hspace{0.1cm},
\end{equation}
and the matrix $(b_{\alpha\beta})$ for a radial symmetric surface is
\begin{equation}
\left(b_{\alpha\beta}\right)=\frac{1}{u^1\sqrt{1+\left(\partial_{u^1} f\right)^2}} 
\left( \begin{array}{lcr}
            \partial^2_{u^1}f      & 0    \\
            0      & u^1\partial_{u^1}f                
           \end{array}
    \right)\hspace{0.1cm}.
\end{equation}
This enables to compute the \textit{principal curvatures} $k_1$ and $k_2$. If the principal directions coincide with coordinate curves, then  
\begin{equation}
k_1 = \frac{b_{11}}{a_{11}}\hspace{0.5cm}\mbox{and}\hspace{0.5cm}k_2 = \frac{b_{22}}{a_{22}}\hspace{0.1cm}.
\end{equation}
Finally, if the surface is radially symmetric, then
\begin{equation}
k_1 = \frac{\partial^2_{u^1} f}{\left[1+\left(\partial_{u_1} f\right)^2\right]^\frac{3}{2}}\hspace{0.5cm}\mbox{and}\hspace{0.5cm} k_2 = \frac{\partial_{u^1} f}{u^1\sqrt{1+\left(\partial_{u_1} f\right)^2}}\hspace{0.1cm}.
\label{principalDirectionsEq}
\end{equation}

\section{Classical motion}
\label{sec:classic}

The Lagrangian in cylindrical coordinates of a particle of mass $\mu$
with a time independent potential $V$ is
\begin{equation}
\mathscr{L} =
\frac{\mu}{2}\left(\dot{r}^2+r^2\dot{\phi}^2+\dot{z}^2\right) 
- V(\vec{r}) \hspace{0.1cm}. 
\end{equation}
The position of particle constrained to move on a radial symmetric surface defined by $z := f(r)$  is $\vec{r}=\vec{r}\left(u_1,u_2\right)$. Then, the Lagrangian takes the form
\begin{equation}
\mathscr{L} = \frac{\mu}{2}a_{\alpha\beta}\dot{u}^\alpha\dot{u}^\beta 
- V\left(\vec{r}\right)  \hspace{1.0cm} (\alpha,\beta = 1,2)\hspace{0.1cm}
\end{equation}
with
\begin{equation}
a_{\alpha\beta} = \delta_{\alpha\beta}\left(h_\beta\right)^2\hspace{0.5cm}\mbox{with}\hspace{0.5cm}
\left(h_1,h_2\right)=\left(\sqrt{1+\left(\partial_r f\right)^2}, r\right)\hspace{0.1cm},
\label{metricEq}
\end{equation}
so the canonical momentum is 
\begin{equation}
p_\alpha = \partial_{\dot{u}^\alpha}\mathscr{L} = \mu a_{\alpha\beta} \dot{u}^\beta \hspace{0.1cm}.
\end{equation}
Therefore, the Hamiltonian is
\begin{equation}
\mathscr{H} = p_\alpha \dot{u}^\beta(\vec{u},\vec{p}) - \mathscr{L} = \frac{1}{2\mu}a^{\alpha\beta}p_\alpha p_\beta + V\left(\vec{r}\right) \hspace{0.5cm}\mbox{with}\hspace{0.5cm}
a^{\alpha\beta} = \delta^{\alpha\beta}\left(h_\beta\right)^{-2}\hspace{0.1cm}. 
\end{equation}
If the particle has a charge $q<0$, and the system is placed in a
uniform electric field $\vec{E}=E_o\hat{k}$ (with $E_0>0$), then
$V(\vec{r})=-q E_o f(r)$. Finally, the Hamiltonian is
\begin{equation}
\mathscr{H}(r,p_r) = \frac{1}{2\mu}\frac{p_r^2}{1+\left(\partial_r f\right)^2} + \frac{L_z^2}{2\mu r^2} - q E_o f(r) \hspace{0.1cm},
\label{classicalHamiltonianEq}
\end{equation}
and the Hamilton equations of motion for the $r$-coordinate are
\begin{eqnarray}
\dot{r} &=& \partial_{p_r}\mathscr{H} =
\frac{p_r}{\mu\left[1+\left(\partial_r f\right)^2\right]} \\
 \dot{p}_r
&=& -\partial_{r}\mathscr{H} =
\left[\left(\frac{p_r}{1+\left(\partial_r
    f\right)^2}\right)^2\frac{\partial^2_r
  f}{\mu}+qE_o\right]\partial_r f + \frac{L_z^2}{\mu
  r^3}\,.
\end{eqnarray}
The $z$-component of the angular momentum remains constant in between
two successive collisions of the particle with the billiard walls. As
a result, for the $\phi$-coordinate
\begin{equation}
\dot{\phi} = \partial_{p_\phi}\mathscr{H} = \frac{L_z}{\mu r^2}\hspace{0.5cm}\mbox{with}\hspace{0.5cm} L_z = L_{z_o} \hspace{0.5cm} \mbox{if} \hspace{0.5cm} r(t) < r_c(\phi)\,,
\end{equation}
where the billiard contour is $\partial\mathfrak{D} =
\left\{\left(r_c(\phi),\phi\right): 0<\phi\leq 2\pi\right\}$ and
$L_{z_o}$ is a constant. The value of $L_{z_o}$ depends of the
billiard contour. For instance, $L_{z_o}$ has the same value for all
the collisions in a circular contour but in general it must be updated
after each collision when the contour is not a circular one.

\subsection{The dunce hat billiard map}

\begin{figure}[h]
  \centering
  \includegraphics[scale=0.4]{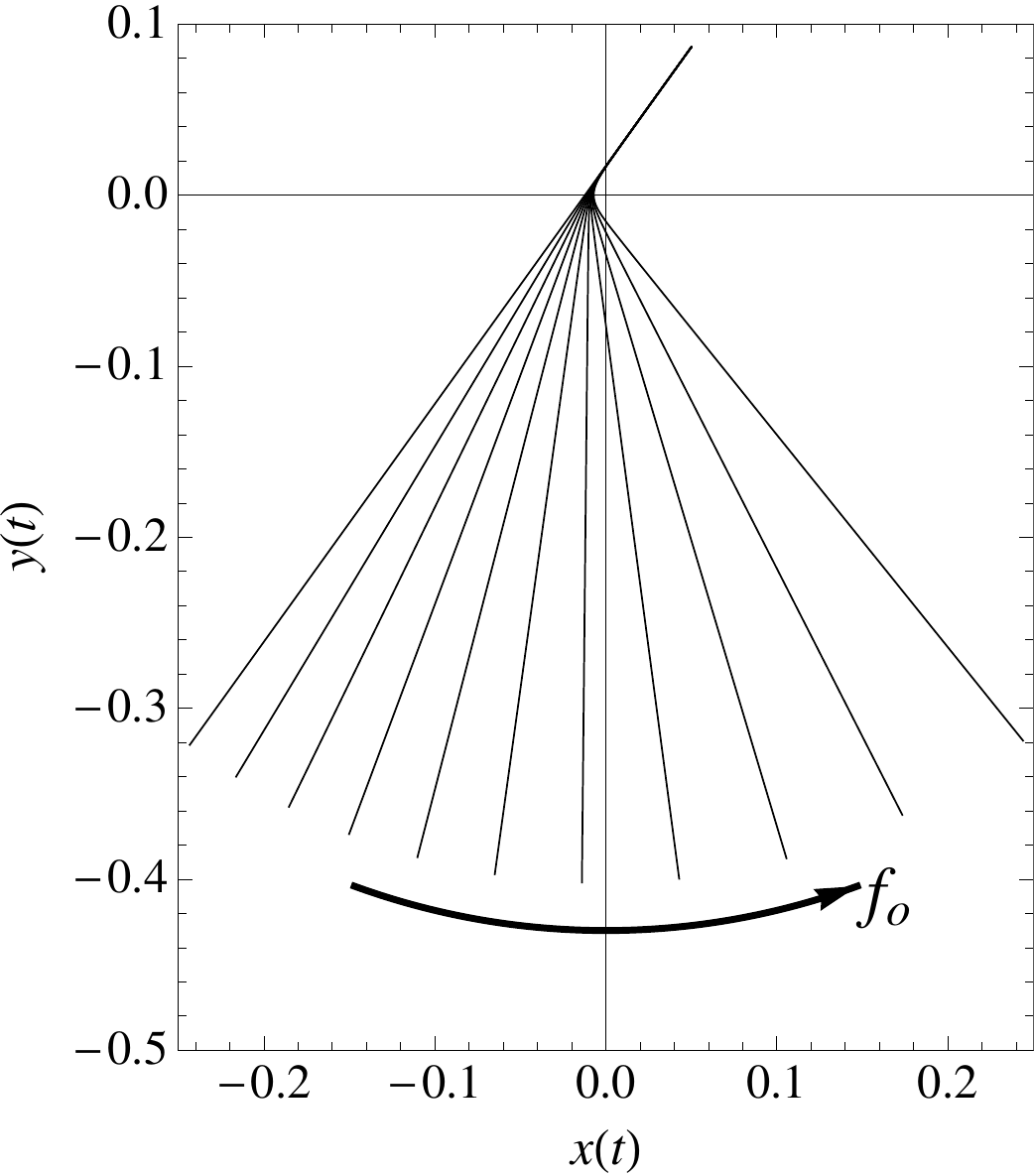} 
  \caption[Trajectory deflection of the free particle living on the cone.]%
  {\textbf{Trajectory deflection of the free particle living on the cone.} This plot is a top view of the trajectories followed by a particle when the cone height varies from $f_o=0$ to $f_o=R$. The typical rectilinear trajectory of the free particle on the plane $f_o = 0$ is deflected in a counter-clockwise sense as the cone emerges by increasing $f_o$ specially near the origin where the cone is located.  } 
  \label{deflectionFig} 
\end{figure}

After the plane surface, one of the simplest case of study with zero Gaussian curvature is the cone. Let us define the \textit{dunce hat billiard} as a cone $f(r):=f_o(1-\frac{r}{R})$ with a circular contour $\partial\mathfrak{D} = \left\{\left(R,\phi\right): 0<\phi\leq 2\pi\right\}$. The Hamiltonian of this billiard in absence of the electric field according the equation (\ref{classicalHamiltonianEq}) is
\begin{equation}
\mathscr{H}(r,p_r) = \frac{1}{2\mu}\frac{p_r^2}{1+\left(\frac{f_o}{R}\right)^2} + \frac{L_z^2}{2\mu r^2} \hspace{0.1cm}.
\end{equation}
This Hamiltonian differs from the free particle on the plane surface Hamiltonian for the constant term $\left(\frac{f_o}{R}\right)^2$, and avoiding the circular contour the free particle solution on the cone is found by direct integration of the equations of motion. The position is given by
\begin{equation}
r(t)=\sqrt{\left(r_o+v_{r_o}t\right)^2+\frac{\left(r_o\omega_o t\right)^2}{1+\left(\frac{f_o}{R}\right)^2}}
\end{equation}
and
\begin{eqnarray}
\phi(t)&=&\phi_o+\sqrt{1+\left(\frac{f_o}{R}\right)^2}
\left\{\arctan\left[\sqrt{1+\left(\frac{f_o}{R}\right)^2}\frac{r_o
    v_{r_o}
    +\left[v_{r_o}^2+
      \frac{\left(r_o\omega_o\right)^2}{1+\left(\frac{f_o}{R}\right)^2}\right] 
    t}{r_o^2\omega}\right]
\right.
\nonumber\\
&& \left.
 -
 \arctan\left[\sqrt{1+\left(\frac{f_o}{R}\right)^2}\frac{v_{r_o}}{r_o\omega_o}\right]
 \right\} \hspace{0.1cm} 
\end{eqnarray}
where $(r_o, \phi_o)$ is the initial position of the particle in polar
coordinates. The radial and angular velocity at $t=0$ are $v_{r_o}$
and $\omega_o$ respectively. Although, the structure of $r(t)$ and
$\phi(t)$ on the cone is similar to the one obtained for the free
particle on the plane surface, the cone may deflect the particle
trajectory as it is shown in Figure \ref{deflectionFig}.

The solution of the equations of motion on the cone, including the
circular contour, may be found defining the following variable
rescaling
\begin{equation}
T: \tilde{r}(t) = \frac{r}{\sqrt{\xi_o}} \hspace{0.5cm} \mbox{and} \hspace{0.5cm} \tilde{\phi}(t) = \sqrt{\xi_o}\phi(t) \hspace{0.5cm} \mbox{with} \hspace{0.5cm} \xi_o := \frac{1}{1+\left(\frac{f_o}{R}\right)^2}.
\label{transformationEq}
\end{equation}
The Lagrangian with the new variables takes the form
\begin{equation}
\mathscr{L} = \frac{\mu}{2}\left[\dot{\tilde{r}}^2+\left(\tilde{r}\dot{\tilde{\phi}}\right)^2\right] = \frac{\mu}{2}\left(\dot{\tilde{x}}^2+\dot{\tilde{y}}^2\right) 
\end{equation}
with $\tilde{x} = \tilde{r}\cos\tilde{\phi}$ and $\tilde{y} =
\tilde{r}\sin\tilde{\phi}$. The new canonical momenta are:
\begin{equation}
\tilde{p}_r = \sqrt{\xi_o} p_r \hspace{0.5cm} \mbox{and} \hspace{0.5cm} \tilde{p}_\phi = \frac{p_\phi}{\sqrt{\xi_o}}.  
\label{newMomentaEqu}
\end{equation}
By computing the Poisson brackets of the new coordinates it can be
checked that the transformation is canonical.

\begin{figure}[h]
  \centering   
  \subfloat{\includegraphics[width=0.25\textwidth]{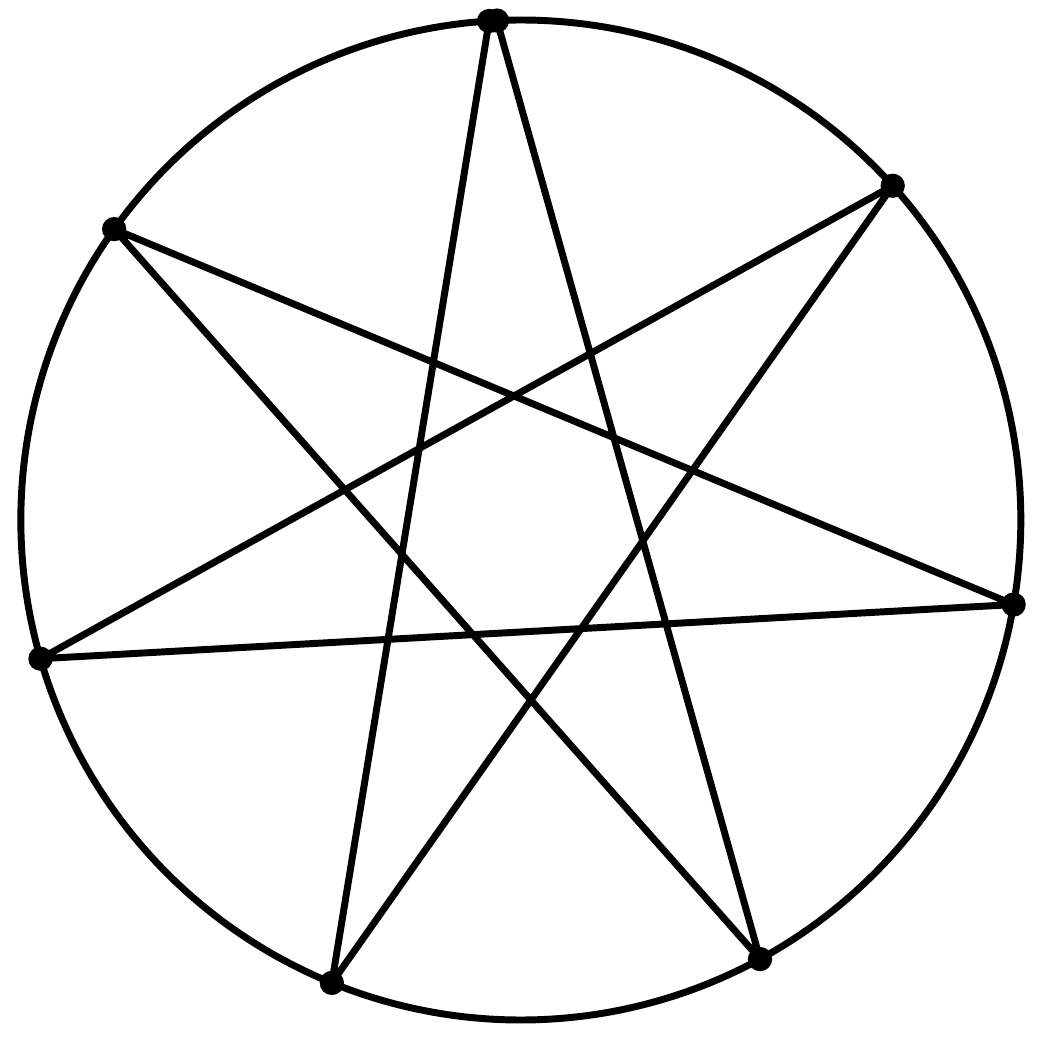}} \hspace{0.1cm}
  \subfloat{\includegraphics[width=0.25\textwidth]{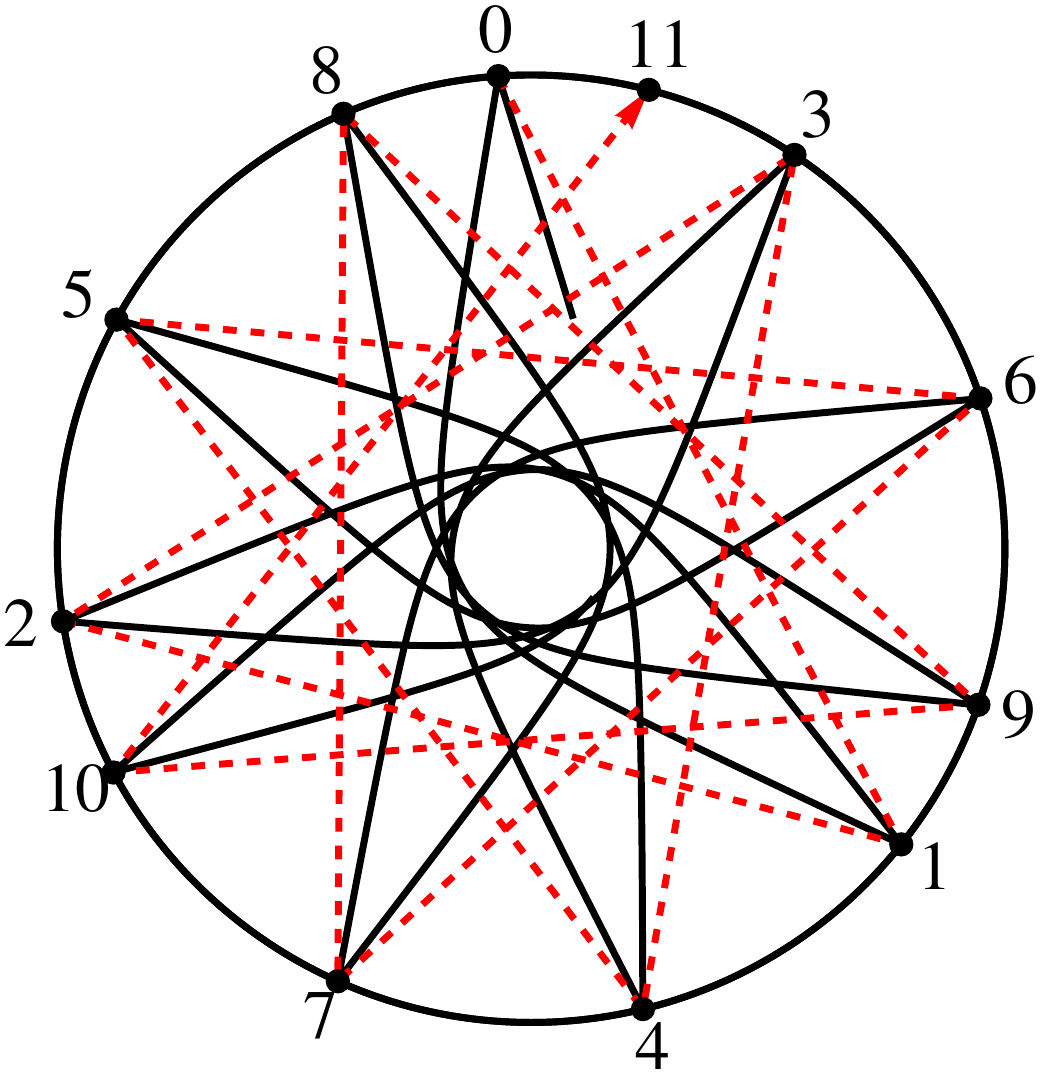}}
  \subfloat{\includegraphics[width=0.30\textwidth]{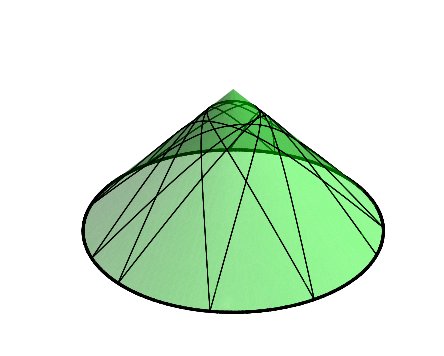}}\\
  \subfloat{\includegraphics[width=0.25\textwidth]{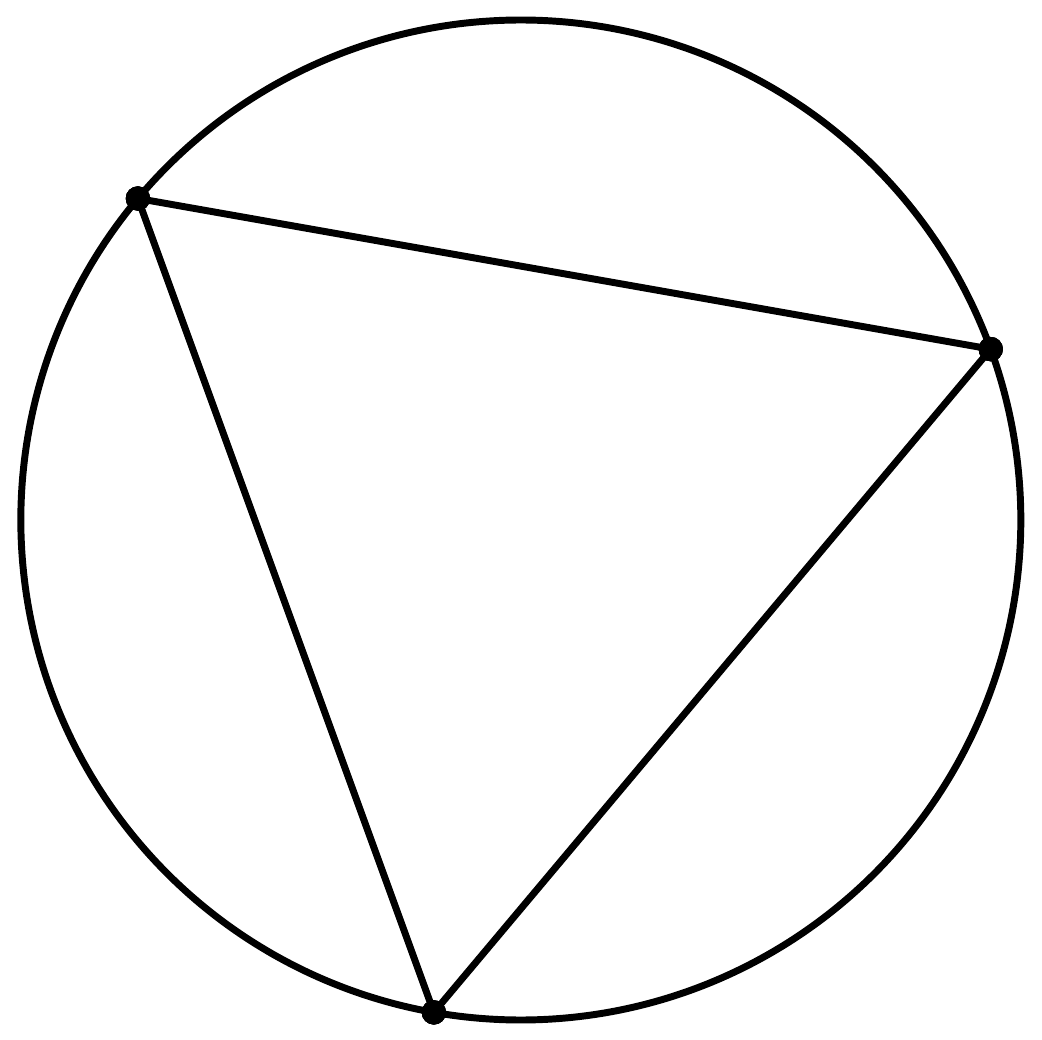}}\hspace{0.1cm}
  \subfloat{\includegraphics[width=0.25\textwidth]{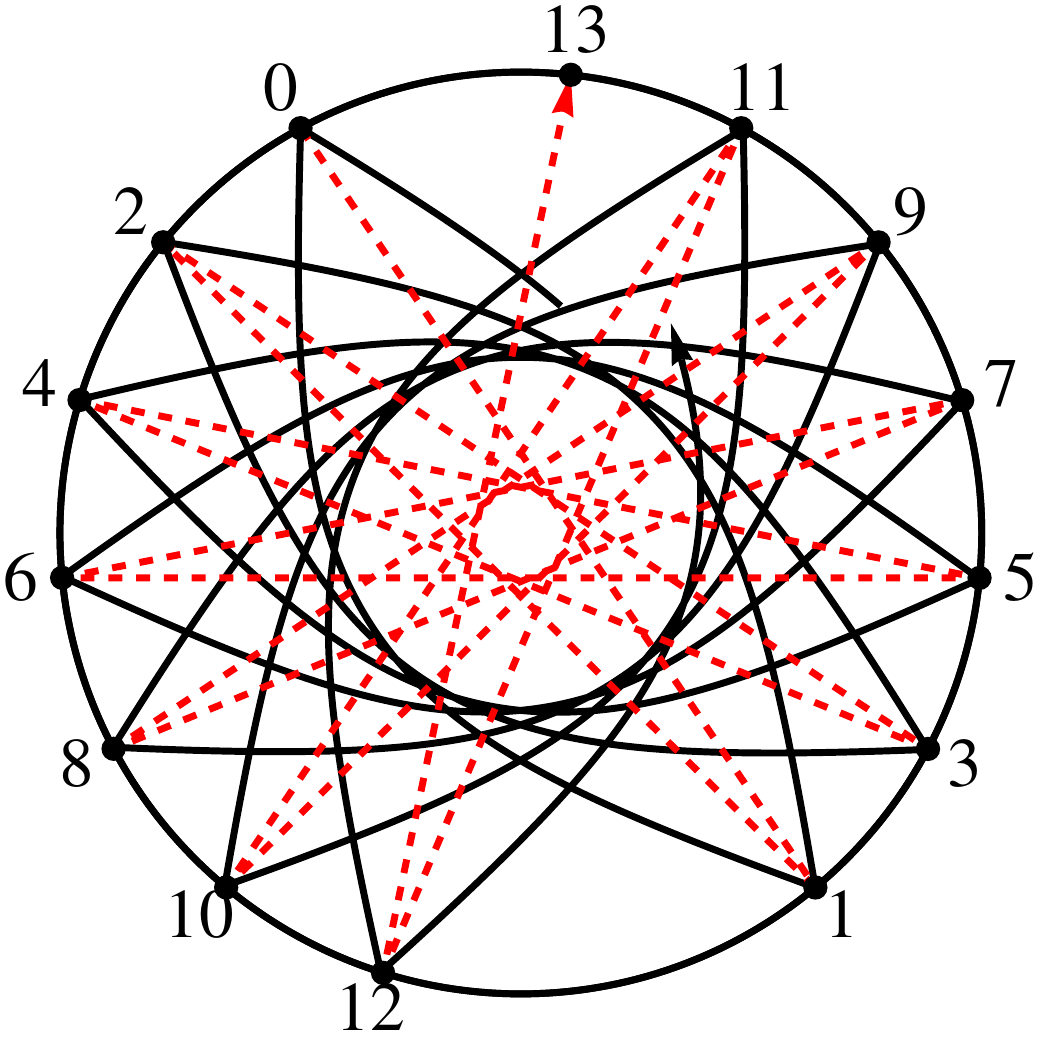}}   
  \subfloat{\includegraphics[width=0.30\textwidth]{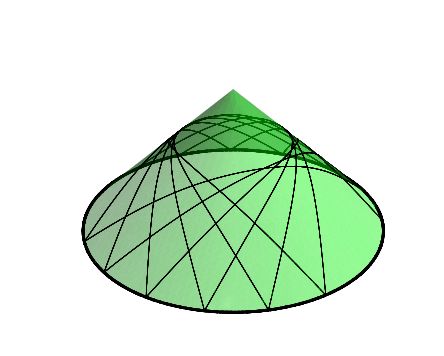}}\\ 
  \subfloat{\includegraphics[width=0.25\textwidth]{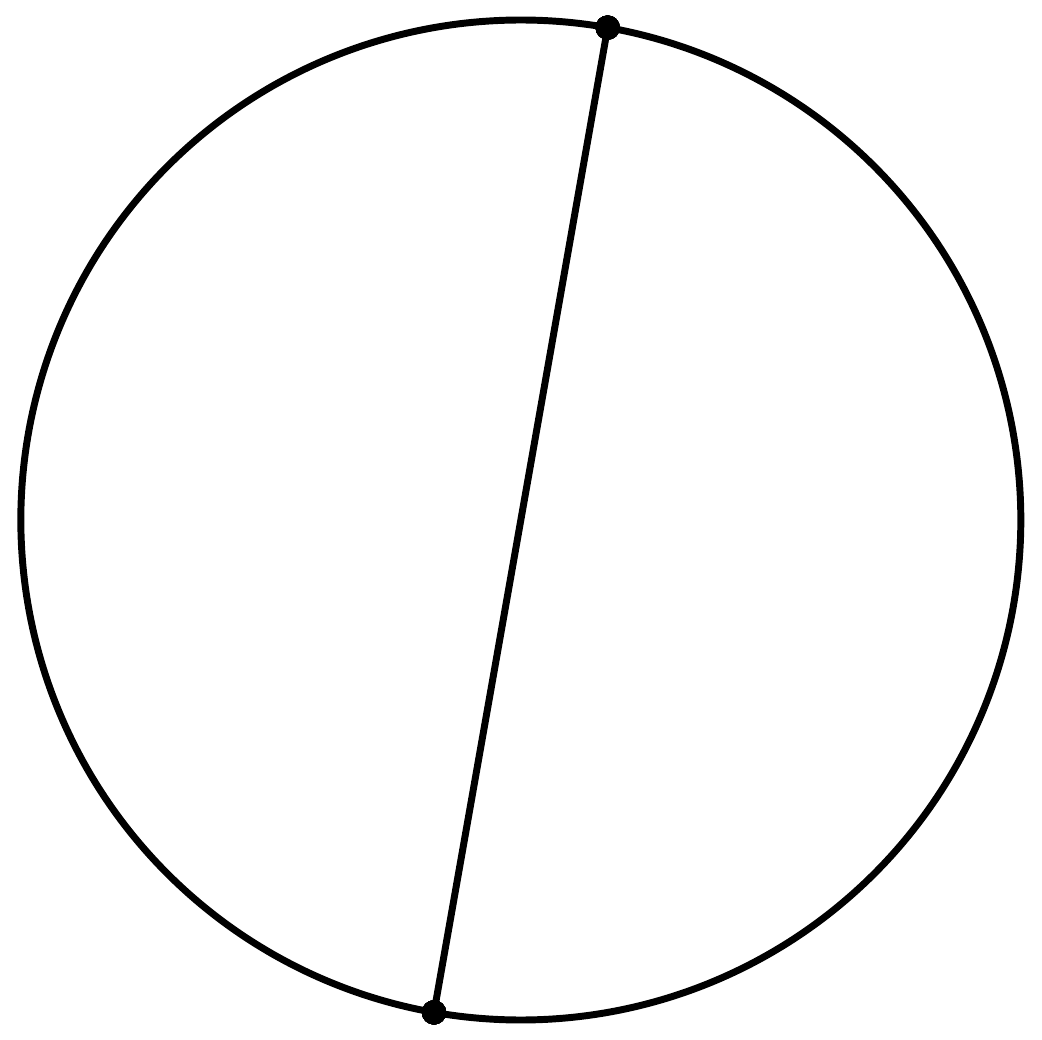}}\hspace{0.1cm}
  \subfloat{\includegraphics[width=0.25\textwidth]{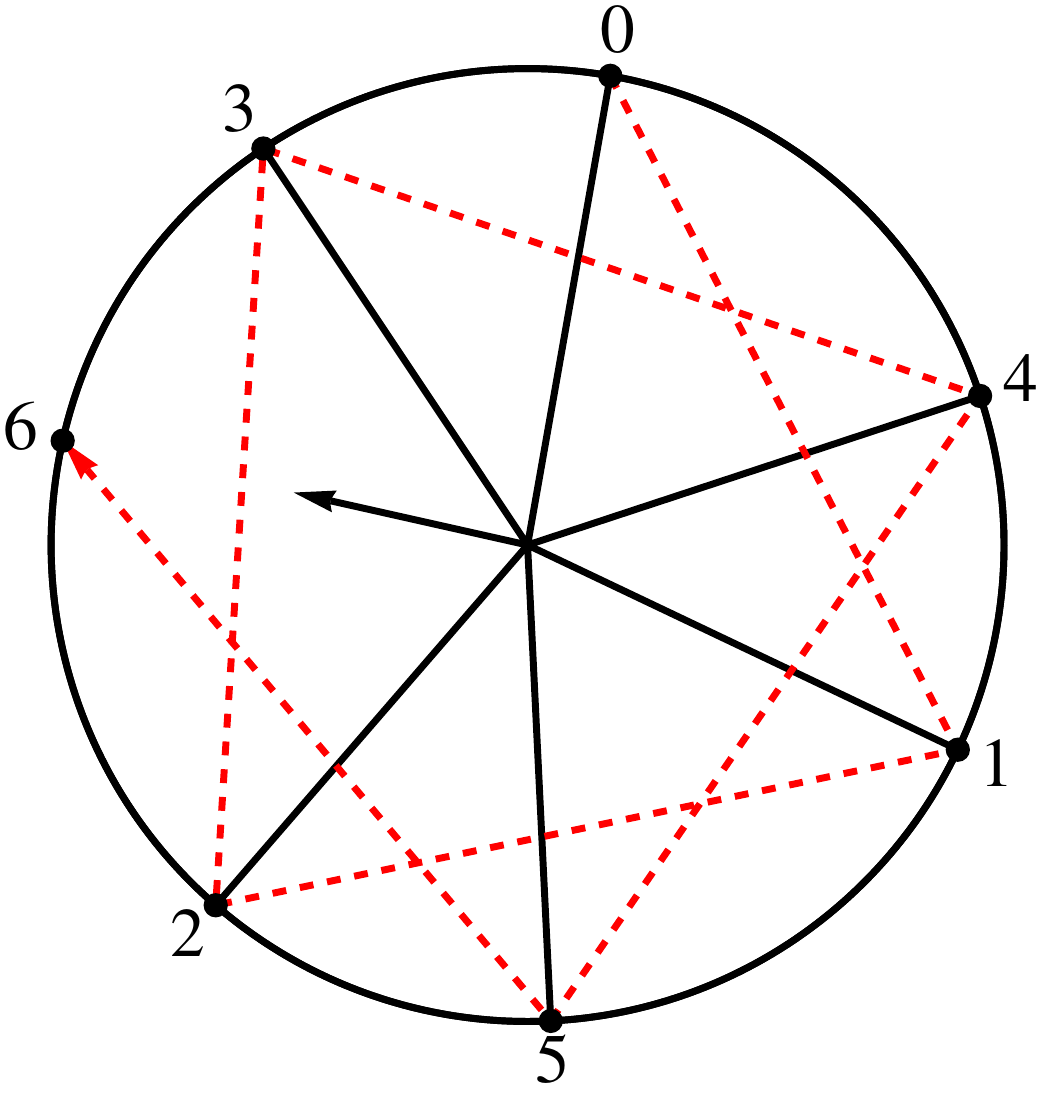}}   
  \subfloat{\includegraphics[width=0.30\textwidth]{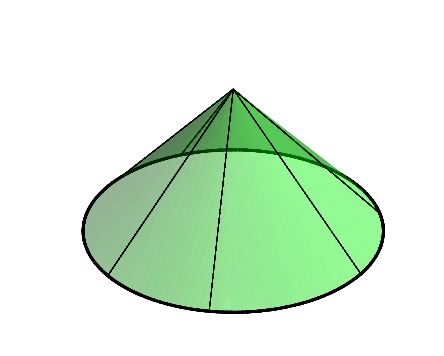}}   
  \caption[Classical trajectories of the circular dunce hat billiard]
  {\textbf{Classical trajectories of the circular dunce hat billiard.} (left) Periodic orbit on the $\tilde{X}\tilde{Y}$ frame. (center) Mapping from the $\tilde{X}\tilde{Y}$ to the $XY$ frame. (right) Trajectory of the particle on the surface. We have set $f_o=\frac{R}{2}$ and $f_o=\frac{3R}{2}$ for the upper and lower panel respectively. The trajectories in the circular billiard are periodic orbits. Note that cone may generate orbits which are not periodic.}
\label{DHBilliardTrajectoriesFig}
\end{figure}
The region $\mathfrak{D}$ remains as a disk:
$x^2+y^2=\tilde{x}^2+\tilde{y}^2<R^2$. However, the contour is
$\partial\tilde{\mathfrak{D}} = \left\{\left(R,\tilde{\phi}\right):
0<\tilde{\phi}\leq
\frac{2\pi}{\sqrt{1+\left(\frac{f_o}{R}\right)^2}}\right\}$. The
upper limit of $\tilde{\phi}$ may be written as
$\tilde{\phi}\leq\sqrt{\xi_o}2\pi = 2\pi - \tilde{\alpha}$ where
$\tilde{\alpha}:=\left(1-\sqrt{\xi_o}\right)2\pi$. If we take a circle
of radius $R':=\sqrt{f_o^2+R^2}$, then we may build a cone of radius
$R$ and height $f_o$ by removing a circular section with angle
$2\pi\left[1-\frac{1}{\sqrt{1+\left(\frac{f_o}{R}\right)^2}}\right]$. As
a result, $\tilde{\alpha}$ is just the angle related with the circular
section removed from the circle in order to construct the cone. In the
$\tilde{X}\tilde{Y}$-plane we have a circular planar
billiard. Therefore, if $\tilde{\phi}_n$ locates the n-collision and
$\tilde{\beta}_n$ is the angle of the incident velocity of the
n-collision, then the next collision is connected with the billiard
map $M_n:\left(\tilde{\phi}_n,\tilde{\beta}_n\right)\rightarrow
\left(\tilde{\phi}_{n+1},\tilde{\beta}_{n+1}\right)$. This map for a
circular billiard is given by \cite{annularMapAndOrbits}

\begin{equation}
\left( \begin{array}{lcr}
            \tilde{\phi}_{n+1} \\
            \tilde{\beta}_{n+1}                
           \end{array}
    \right)=
\left( \begin{array}{lcr}
            1      & 2    \\
            0      & 1                
           \end{array}
    \right)    
    \left( \begin{array}{lcr}
            \tilde{\phi}_{n} \\
            \tilde{\beta}_{n}                
           \end{array}
    \right)+
    \left( \begin{array}{lcr}
            \epsilon\pi \\
            0                
           \end{array}
    \right)
    \label{circularMapEq}
\end{equation}
\\with $(\epsilon = +1 \mbox{ if the motion is clockwise, -1 otherwise})$. Using the equation (\ref{circularMapEq}) the particle position is
\begin{equation}
\tilde{x}(t) = R\left[\left(\cos\tilde{\phi}_{n(t)+1}-\cos\tilde{\phi}_{n(t)}\right)\left(\frac{t}{\tau}-n(t)\right) + \cos\tilde{\phi}_{n(t)} \right] \mbox{ and}
\end{equation}
\begin{equation}
\tilde{y}(t) = R\left[\left(\sin\tilde{\phi}_{n(t)+1}-\sin\tilde{\phi}_{n(t)}\right)\left(\frac{t}{\tau}-n(t)\right) + \sin\tilde{\phi}_{n(t)} \right] \hspace{0.1cm}.
\end{equation}
Where $\tau$ is the time between successive collisions, and
$n:=\mbox{\textrm{Int}}\left(t/\tau\right)$ is the integer
part of ${t}/{\tau}$. According to the circular billiard map
$\tilde{\phi}_n = \tilde{\phi}_o +
n(2\tilde{\beta}_o+\epsilon\pi)$. Therefore, at the $n$-th collision the
particle is located at
\begin{equation}
\phi_n = \phi_o + n\left[2\beta_o+\epsilon\pi\sqrt{1+\left(\frac{f_o}{R}\right)^2}\right]\hspace{0.5cm} \mbox{with} \hspace{0.5cm} \beta_n=\beta_o\hspace{0.1cm}.
\label{dhBilliardMap}
\end{equation}
The last expression gives us the position of each collision. However,
this map does not specify how successive collisions are connected. As
we show in Figure \ref{deflectionFig} the cone deflects the
particle. Therefore, two successive collisions predicted
by~(\ref{dhBilliardMap}) must be connected by a curve (see Figure
\ref{DHBilliardTrajectoriesFig}). This is an expected difference with
the plane billiards where collisions are connected by rectilinear
trajectories.

\subsection{Classical rectangular billiard with an inner Gaussian surface}
This billiard is a rectangular box with an inner Gaussian surface
defined as
$z=f(r)=\frac{V_o}{2\pi\sigma^2}\exp\left(-\frac{r^2}{2\sigma^2}\right)$
where $\sigma$ is the standard deviation and $V_o$ is a constant which
we will set equal to one.
\begin{figure}[h]
  \centering   
  \subfloat[$E_o=0, \sigma=10$]{\includegraphics[width=0.25\textwidth]{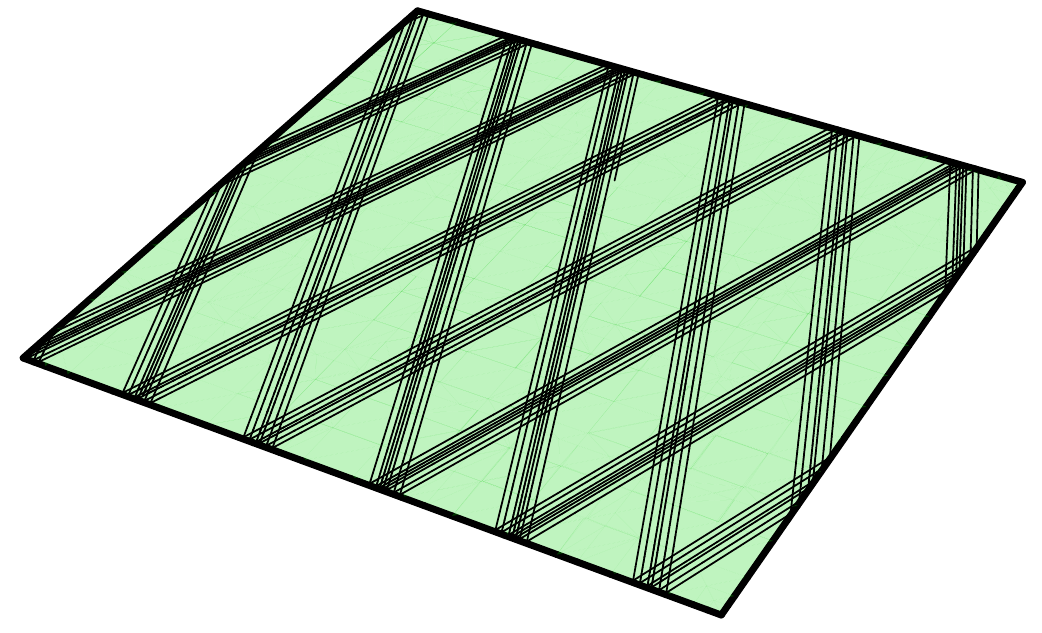}}
  \subfloat[$E_o=0, \sigma=0.8$]{\includegraphics[width=0.25\textwidth]{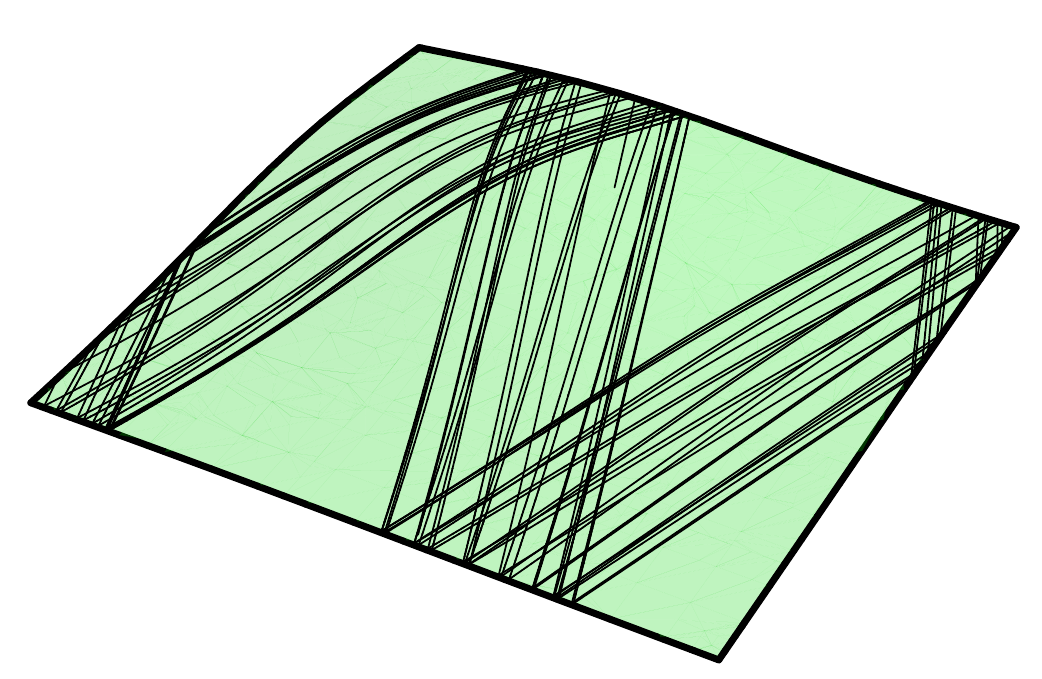}}
  \subfloat[$E_o=0, \sigma=0.6$]{\includegraphics[width=0.25\textwidth]{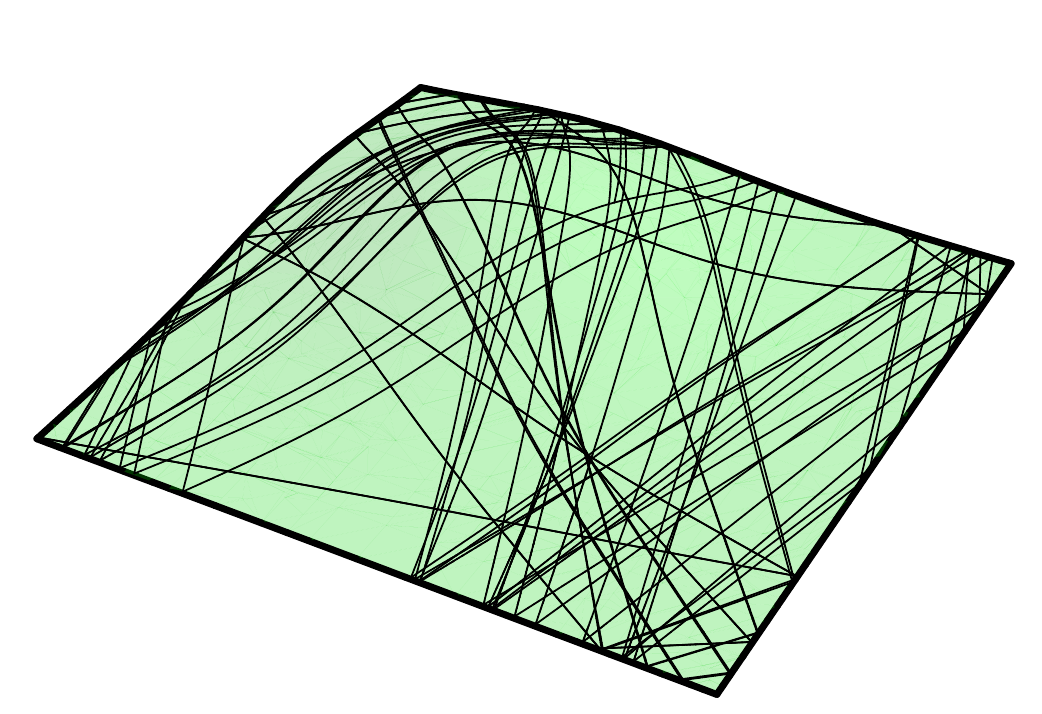}}
  \subfloat[$E_o=0, \sigma=0.4$]{\includegraphics[width=0.25\textwidth]{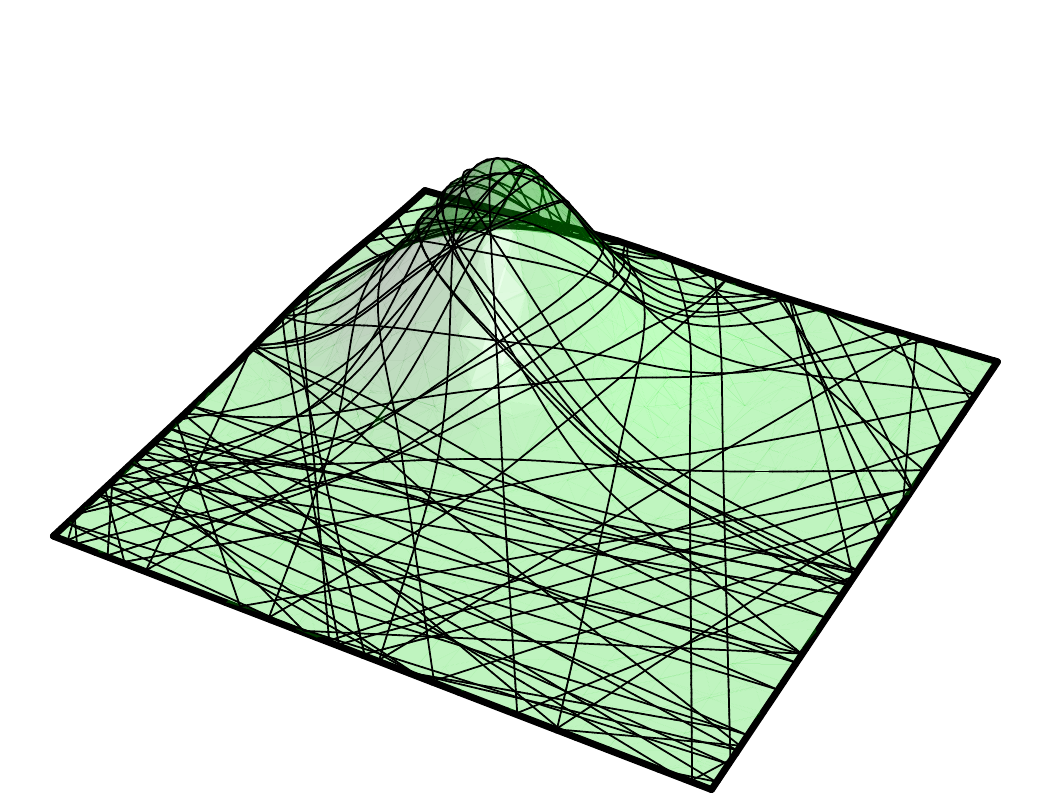}}\\
  \subfloat[$E_o=0, \sigma=0.3$]{\includegraphics[width=0.25\textwidth]{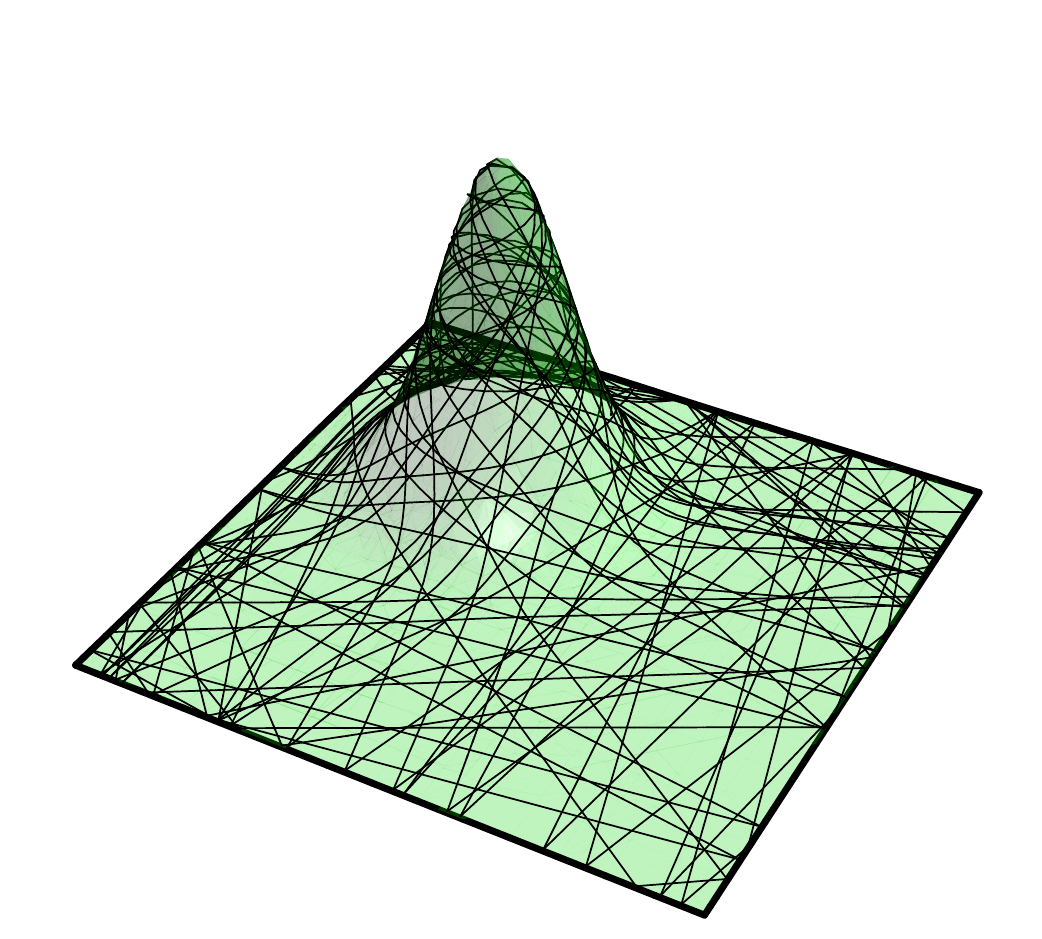}}   
  \subfloat[$E_o=0, \sigma=0.25$]{\includegraphics[width=0.25\textwidth]{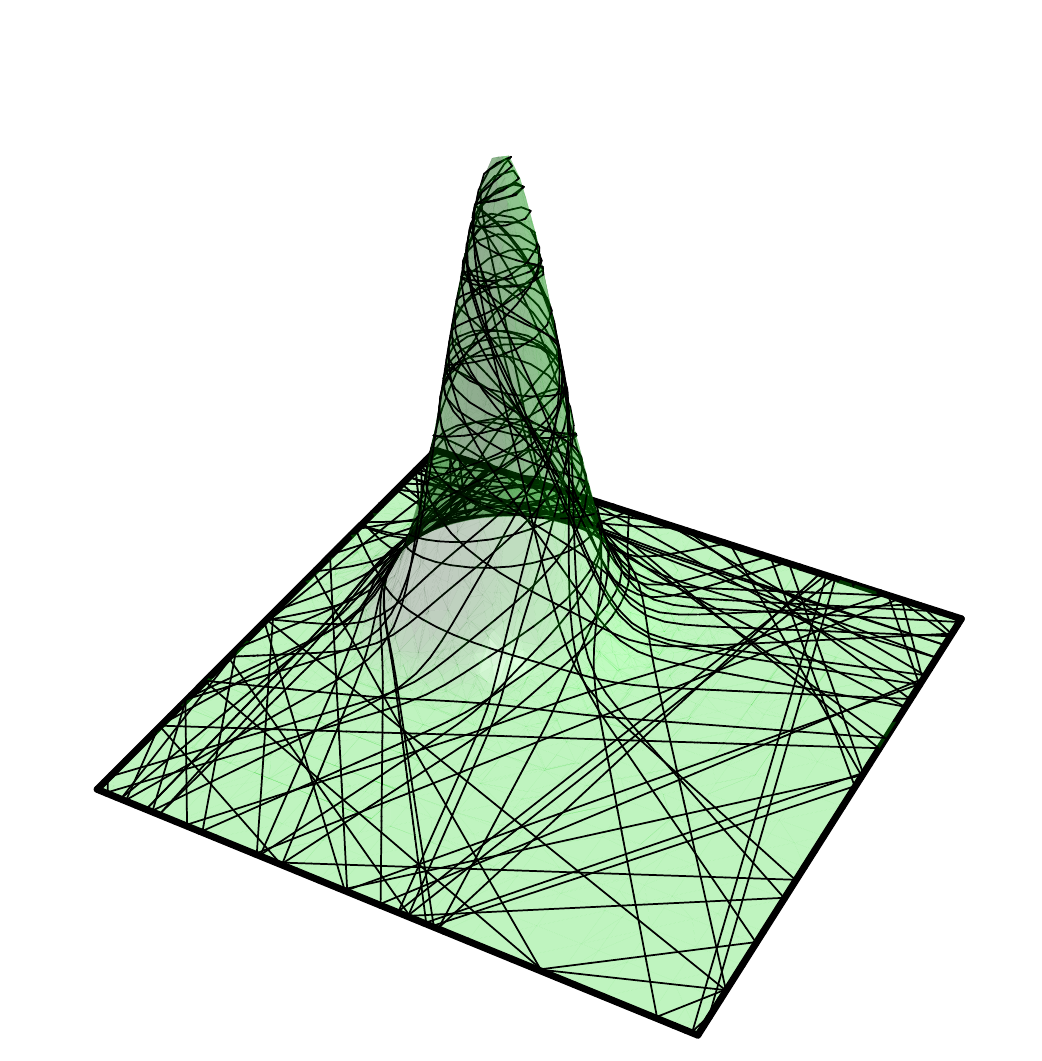}} 
  \subfloat[$E_o=10, \sigma=0.25$]{\includegraphics[width=0.25\textwidth]{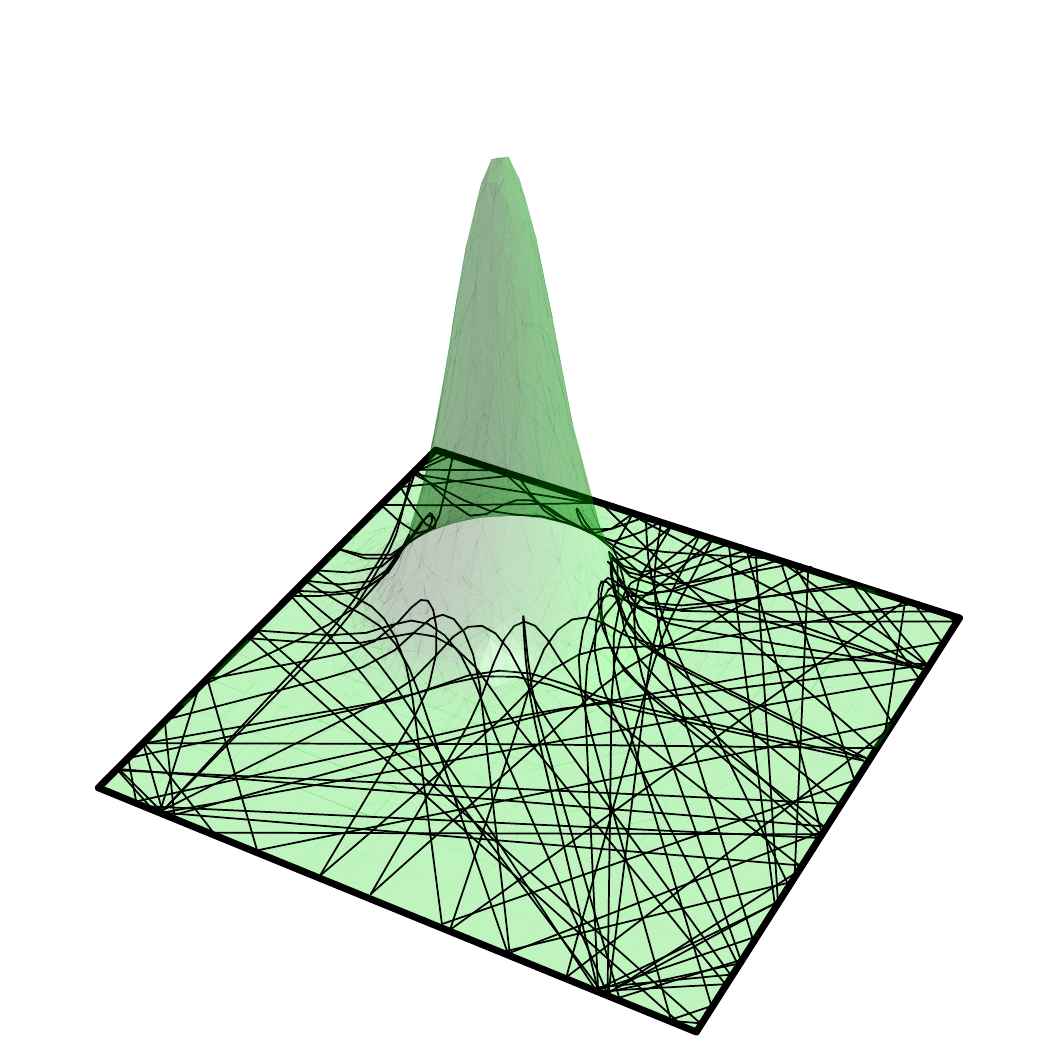}}
  \subfloat[$E_o=10, \sigma=0.8$]{\includegraphics[width=0.2\textwidth]{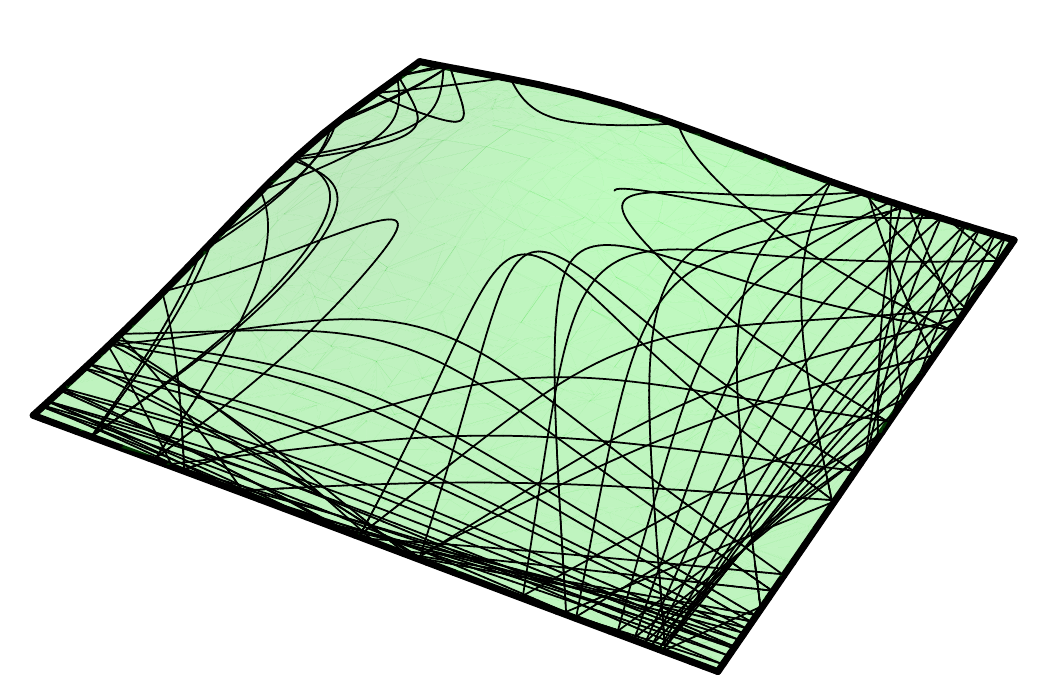}} 
  \caption[Classical rectangular billiard with an inner Gaussian
    surface] {\textbf{Classical rectangular billiard with an inner
  Gaussian surface} (a)-(f) The billiard in absence of an electric
    field. (g) and (h) The billiard is placed in a constant electric
    field. The particle has a negative charge, and it could not climb
    the surface depending if the initial kinetic energy is lower than
    energy required to move against the electric field. This case may
    be understood as a Sinai billiard with an inner soft disk.}
\label{classicalTrayectoriesFig}
\end{figure}
The classical trajectories become irregular as the Gaussian surface
emerges in the interior of the rectangular billiard (Figure
\ref{classicalTrayectoriesFig}). The trajectories were found solving
the equations of motion using the fourth order Runge-Kutta method
between successive collisions. When the particle collided with the
frontier, then its new momentum was found and the new trajectory
before the next collision was computed numerically.

\subsection{The phase space}

The billiards described in this document have two degrees of
freedom because the particle moves on a two-dimensional
surface. Therefore, the phase space is four-dimensional. Taking into
account that system is Hamiltonian then the motion takes place on a
three dimensional hypersurface of the phase space and we need only
three variables in order to describe it.

\begin{figure}[ht]
  \centering
  \subfloat[$\frac{f_o}{R}=0$]{\includegraphics[width=0.33\textwidth]{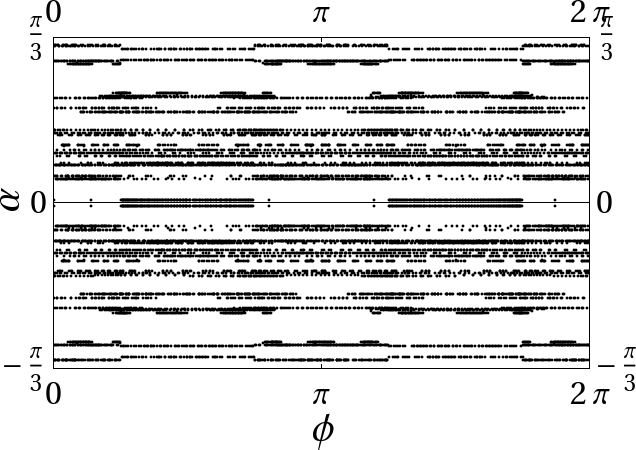}}\hspace{1.0cm}
  \subfloat[$\frac{f_o}{R}=0.1$]{\includegraphics[width=0.33\textwidth]{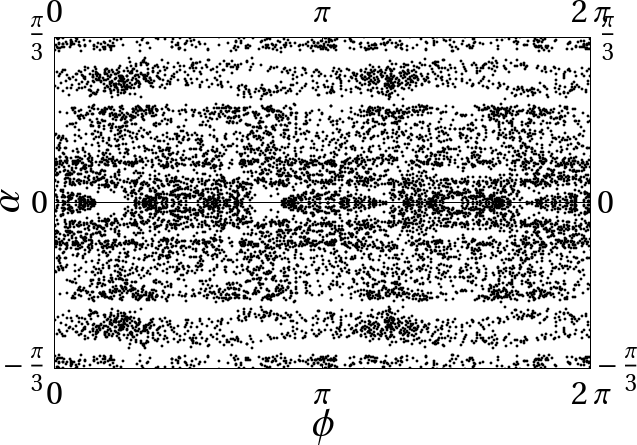}}\\ \subfloat[$\frac{f_o}{R}=0.45$]{\includegraphics[width=0.33\textwidth]{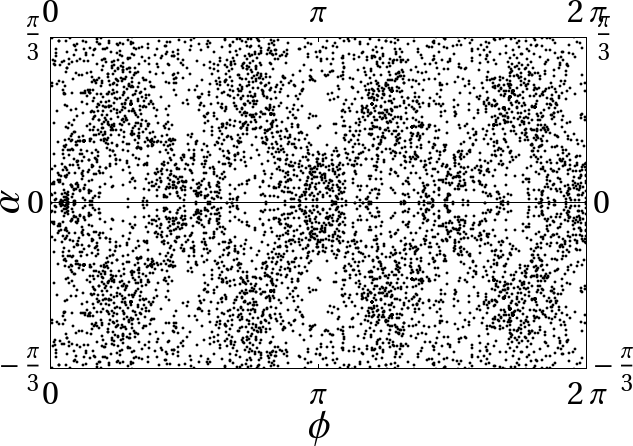}}\hspace{1.0cm}
  \subfloat[Classical trajectory on the rectangular dunce hat
    billiard.]{\includegraphics[width=0.33\textwidth]{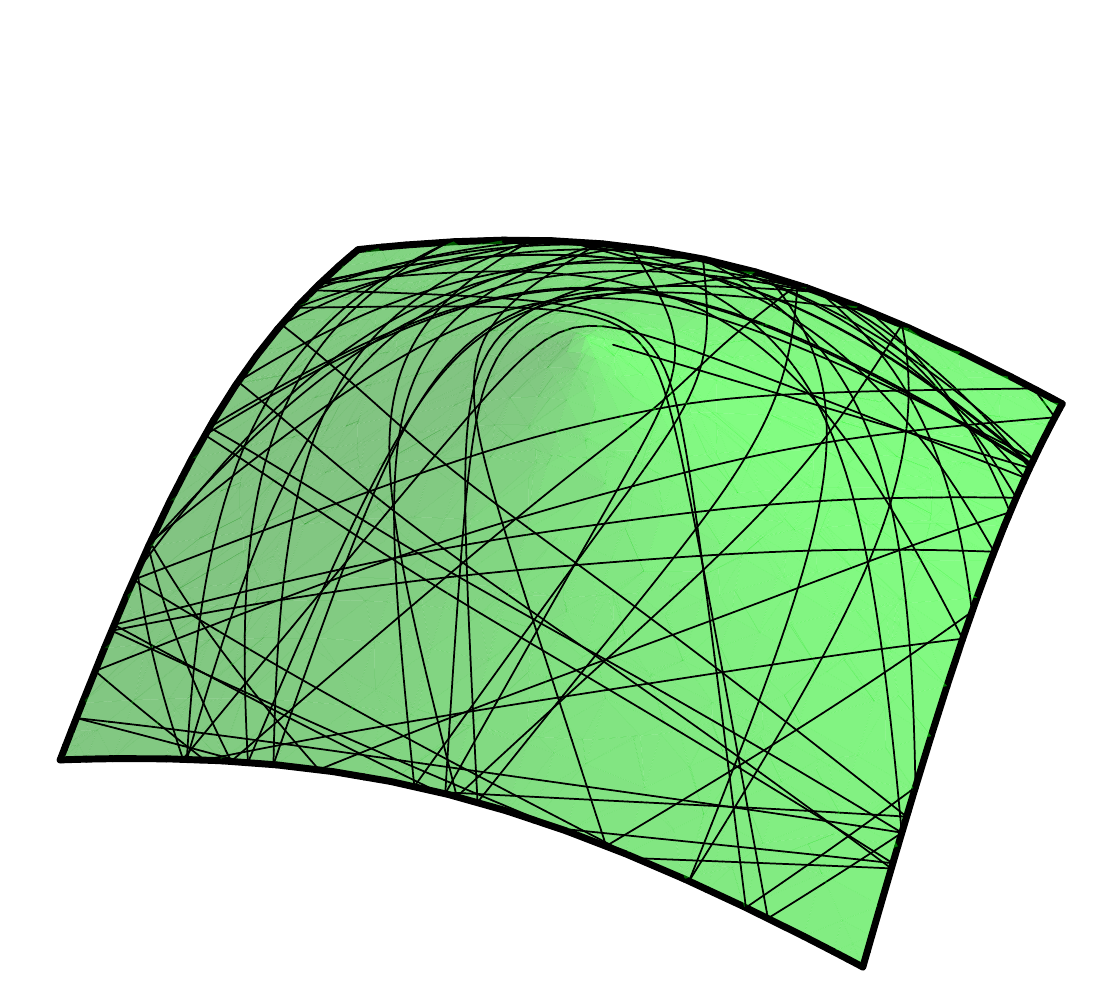}}\\ 
\subfloat[$\sigma/l_x=\sigma/l_y=0.3$.]{\includegraphics[width=0.33\textwidth]{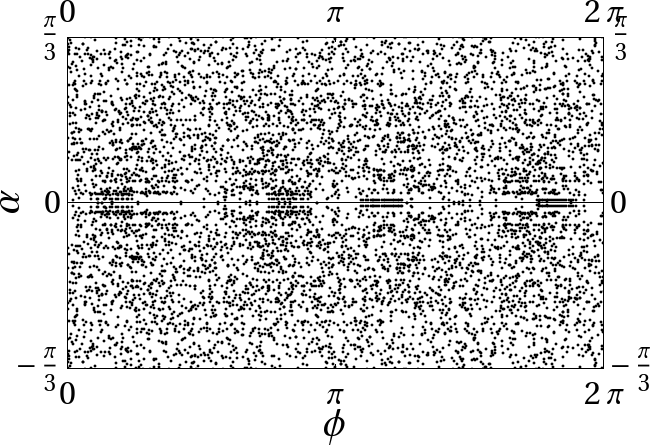}} \hspace{1.0cm}
  \subfloat[Classical trajectory on the rectangle billiard with
    Gaussian
    surface]{\includegraphics[width=0.33\textwidth]{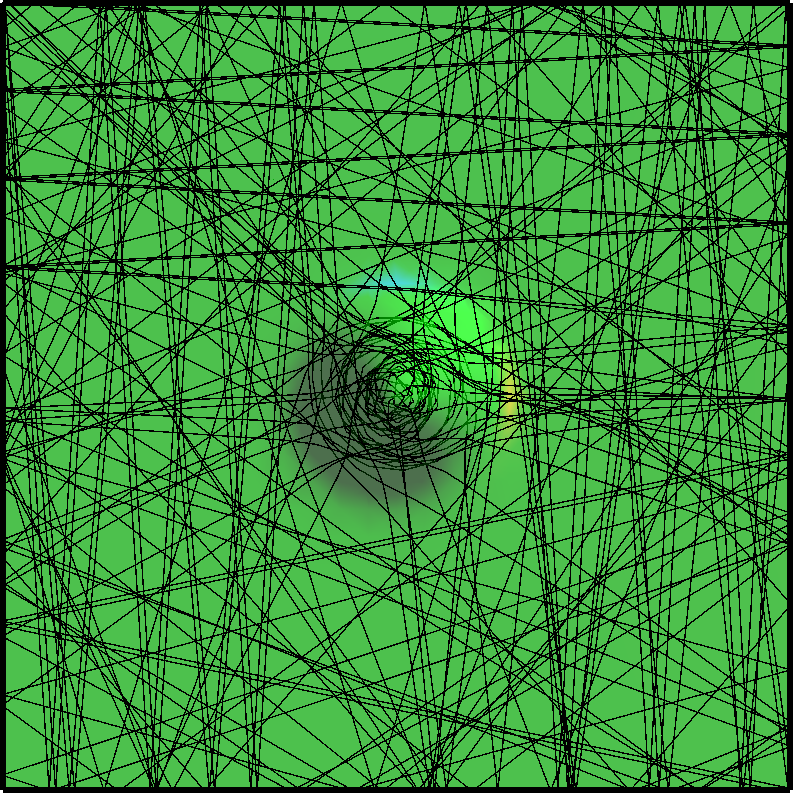}}
  \caption[The reduced phase space] {\textbf{The reduced phase space.} Each
    reduced phase space was built using 30 random initial conditions for the
    first 200 collisions with the boundary. The phase space of the
    rectangular dunce hat billiard is shown in (a)-(c) for different
    values of the ratio $\frac{f_o}{R}$. (d) Classical trajectory in
    the rectangular dunce hat billiard setting
    $\frac{f_o}{R}=0.45$. The phase space and a typical classical
    trajectory of the rectangular billiard with Gaussian surface
    setting $\vec{E}=0$ are shown in (e) and (f) respectively.}
\label{phaseSpaceFig}
\end{figure}

The Poincar\'e section is commonly used in order to map the system and
to obtain its dynamical information. This is equivalent to consider
two variables which define where and how the collision occurs. These
variables are the collision location $\phi\in\left[0,\pi\right]$ and
the angle $\alpha\in\left[-\frac{\pi}{2},\frac{\pi}{2}\right]$ between
the incident vector velocity projected in the plane with the normal
vector of $\partial\mathfrak{D}$. The two variables of the mapping
then define a reduced phase space. In the dunce hat billiard with
circular contour the trajectories are \textbf{\textit{periodic
    orbits}} if
$2\beta_o+\epsilon\pi\sqrt{1+\left(\frac{f_o}{R}\right)^2}:=C$ is a
rational multiple of $\pi$. These type of trajectories are presented
in the reduced phase space by a set of $n'-1$ points located in a
horizontal line $\phi$ = constant where $n'$ is the closed orbit
period. On the other hand, if $C$ is a non-rational multiple of $\pi$,
the billiard will have a \textbf{\textit{quasi periodic orbit}}. This
trajectory is tangent to an inner circle in the surface and it fills
it generating a caustic. A quasi-periodic orbit is represented as a
straight horizontal and densely filled line in the reduced phase
space. The phase space of the dunce hat billiard with rectangular
contour and the rectangular billiard with Gaussian surface were built
for a sample of trajectories starting from different random initial
conditions (see Figure \ref{phaseSpaceFig}). When $f_o=0$ or
$\sigma\to\infty$, both billiards become the rectangular planar
billiard. The phase space of this billiard has straight horizontal
lines typical of a non chaotic billiard. On the other hand, the phase
space is filled by a chaotic sea when a cone or a Gaussian surface
emerge in the rectangular billiard.

\section{Quantum Hamiltonian}
\label{sec:qh}

At the quantum level, the Hamiltonian of the non-planar billiards
considered here has the form
\begin{equation}
\hat{H} = -\frac{\hbar^2}{2\mu}\vec{\nabla}^2_\Sigma + \hat{U}_\Sigma + \hat{V}\hspace{0.1cm}.
\label{generalBilliardHamiltonianEq}
\end{equation}
The terms involved in the Hamiltonian are: 

\noindent (i) \textit{The kinetic energy on the surface}.  The
Laplace-Beltrami operator is  
\begin{equation}
\vec{\nabla}^2_\Sigma = \frac{1}{1+(\partial_r f)^2}\vec{\nabla}^2_{plane} - \frac{\partial_rf\partial_r^2f}{\left[1+(\partial_r f)^2\right]^2}\partial_r+\left[1-\frac{1}{1+(\partial_r f)^2}\right]\frac{1}{r^2}\partial_\phi^2
\end{equation}
with
$\vec{\nabla}^2_{plane}=\frac{1}{r}\partial_r+\partial^2_r+\frac{1}{r^2}\partial_\phi^2$. Then,
the kinetic energy may written as 
\begin{equation}
-\frac{\hbar^2}{2\mu}\vec{\nabla}^2_\Sigma = \xi(r)\hat{H}_o + \frac{\hbar^2}{2\mu}\kappa(r)\partial_r+\zeta(r)\frac{\hat{L}_{z}^2}{2\mu r^2}\hspace{0.1cm},
\label{surfaceKineticEnergyEq}
\end{equation}
where $\hat{H}_o$ is the free particle Hamiltonian on the plane, and
$\kappa(r)$ is a function with dimensions of wave vector defined by
\begin{equation}
\kappa(r) := \xi(r)^2\frac{df}{dr}\frac{d^2f}{dr^2} \hspace{0.5cm}\mbox{with}\hspace{0.5cm}\xi(r):=\frac{1}{1+(\partial_r f)^2}\hspace{0.1cm}.
\end{equation} 
Using the equation (\ref{GaussianCurvatureEq}) the radial kinetic
energy on the surface $\frac{\hbar^2}{2\mu}\kappa(r)\partial_r$ may be
also expressed in terms of the normal Gaussian curvature as
$\frac{\hbar^2}{2\mu}K\vec{x}\cdot\partial_{\vec{x}}$. Therefore, this
term is null for some radially symmetric surfaces e.g.~the plane and the
cone. Finally, the non-dimensional function $\zeta(r)$ is defined by
$\zeta(r) :=
\left(\frac{df}{dr}\right)^2\xi(r)=1-\xi(r)$.

\noindent (ii) \textit{The surface confining potential}. The constrain of classical
mechanics which ties the particle to the surface, in quantum mechanics
may be introduced by a confining potential which enforces the particle
to stay on the surface. This known as \textit{the confining approach},
and it predicts the following contribution to the quantum
Hamiltonian \cite{daCosta, QuantizationOnSurfaces}
\begin{equation}
U_\Sigma = -\frac{\hbar^2}{8\mu}\left({k}_1-{k}_2\right)^2 
\end{equation}
where ${k}_1$ and ${k}_2$ are the principal curvatures of the surface. For a radial symmetric two dimensional surface the principal directions are along $\hat{r}$ and $\hat{\phi}$, then (see equation (\ref{principalDirectionsEq}))
\begin{equation}
k_r = \frac{1}{\left[1+\left(\frac{df}{dr}\right)^2\right]^{\frac{3}{2}}}\frac{d^2f}{dr^2} = \frac{d^2f}{dr^2}\xi(r)^{\frac{3}{2}}
\end{equation}  
and
\begin{equation}
k_\phi = \frac{1}{r\left[1+\left(\frac{df}{dr}\right)^2\right]^{\frac{1}{2}}}\frac{df}{dr} = \frac{1}{r}\frac{df}{dr}\xi(r)^{\frac{1}{2}}.
\end{equation}  

\noindent (ii) \textit{The external potential}. This term includes the
potential of the external electric field $\vec{E}_o$, and the
corresponding potential due to the boundary of the billiard
\begin{equation}
\hat{V} = qE_of(r) + V_{box}\left(r\right)
\end{equation}
where  
\begin{equation}
V_{box}(r) := 
\begin{cases}
0 &\text{if } r<r_c(\phi)\\
\infty &\text{otherwise,}
\end{cases}
\end{equation}
and $r_c(\phi)$ defines the contour $\partial\mathfrak{D}$ of the
billiard. 

\section{The quantum dunce hat billiard}
\label{sec:dh}

\subsection{The quantum dunce hat billiard with a circular contour}

\begin{figure}[h]
   \centering   
   \subfloat[Top view of a caustic]{\includegraphics[width=0.25\textwidth]{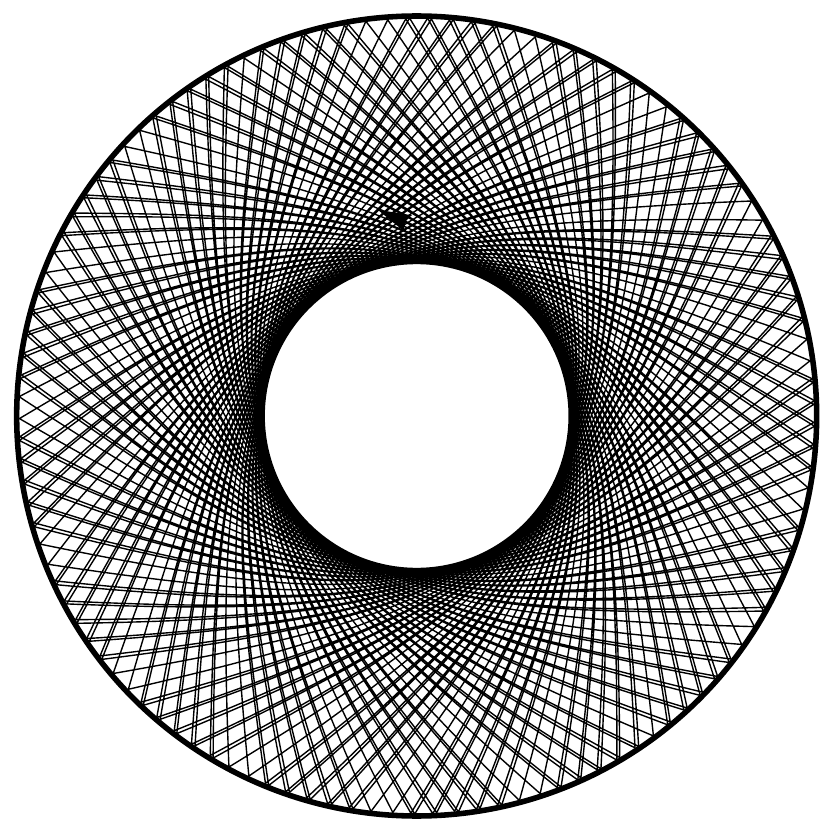}} \hspace{0.2cm}
   \subfloat[Top view of the state 157]{\includegraphics[width=0.25\textwidth]{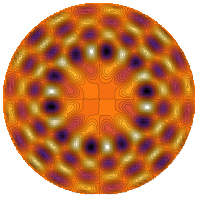}}\hspace{0.2cm}
   \subfloat[State 157]{\includegraphics[width=0.25\textwidth]{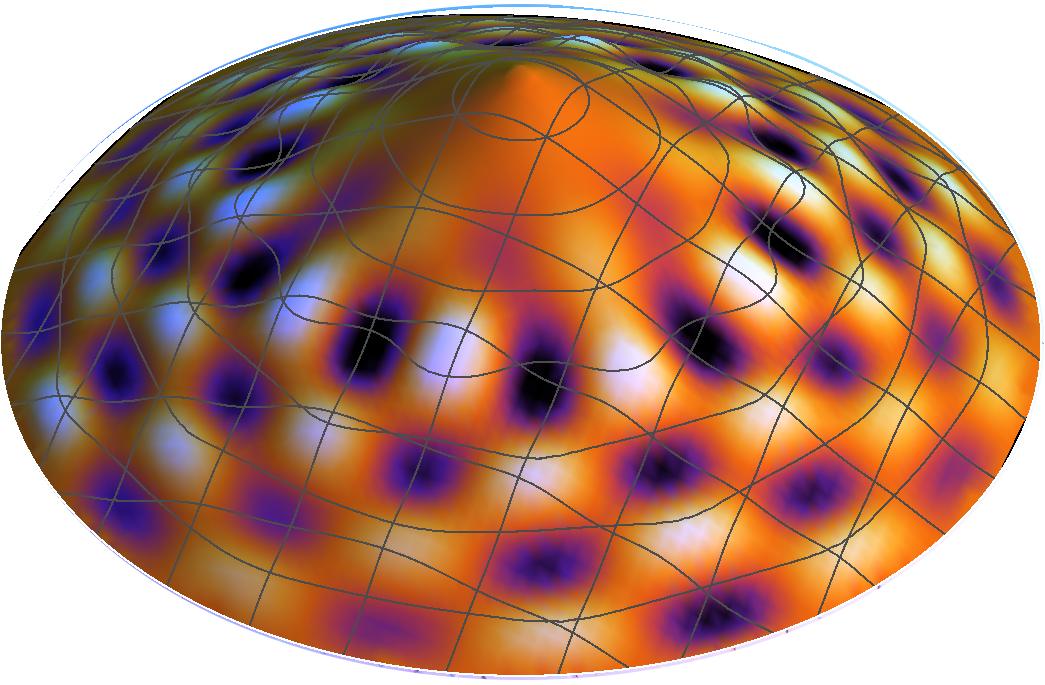}}\\
   \subfloat{\includegraphics[width=0.2\textwidth]{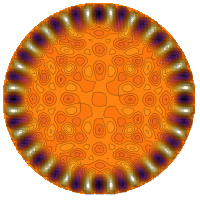}}
   \subfloat{\includegraphics[width=0.2\textwidth]{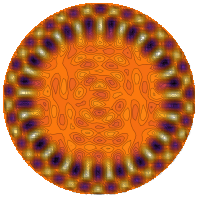}}
   \subfloat{\includegraphics[width=0.2\textwidth]{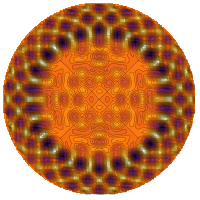}}
   \subfloat{\includegraphics[width=0.2\textwidth]{state157}}
   \subfloat{\includegraphics[width=0.2\textwidth]{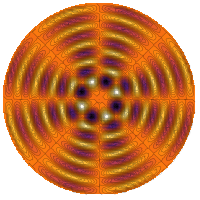}}
   \caption[States of the dunce hat billiard with a circular contour.]
           {\textbf{States of the dunce hat billiard with a circular
               contour.}  \textit{Upper panel}. Several wavefunctions
             of this integrable billiard exhibit the corresponding
             structure of classical dynamics. \textit{Lower panel from
               the left to the right}. Other states related to the
             caustics.}
 \label{DHBilliardStatesFig}
\end{figure}

This is probably the simplest case of study. The Gaussian curvature
$K$, given by (\ref{GaussianCurvatureEq}), is zero for
this billiard as happens with the typical billiards in the plane. This
makes a significant simplification in the Hamiltonian
\begin{equation}
\hat{H} = -\frac{\hbar^2}{2\mu}\left[\xi_o\left(\partial_{rr} + \frac{1}{r}\partial_r\right) + \frac{1}{r^2}\partial_{\phi\phi}\right] + \hat{U}_\Sigma 
\end{equation}
where the confining potential is 
\begin{equation}
U_\Sigma(r) = -\frac{\hbar^2}{8\mu}\left(\frac{f_o}{R}\right)^2\xi_o^2 = - \frac{L_{z_\Sigma}^2}{2\mu r^2} \hspace{0.5cm} \mbox{with} \hspace{0.5cm} L_{z_\Sigma}^2 := \frac{\hbar^2}{4}\left(\frac{f_o}{R}\right)^2\xi_o^2.
\end{equation}
We may expect that the effect of the confining potential should not be
relevant because it may be absorbed in
$-\frac{\hbar^2}{2\mu}\frac{1}{r^2}\partial_{\phi\phi} =
\frac{\hat{L}_z^2}{2\mu r^2}$ at least with a circular contour where
$\hat{L}_z = i\hbar\partial_\phi$ commutes with the Hamiltonian by
virtue of the radial symmetry of the billiard\footnote{This will not
  be the case for an square contour where only some discrete
  symmetries of the billiard may remain depending the position of the
  cone center in the box.}. The analytic solution for this system may
be found using the transformation defined by the equation
(\ref{transformationEq}). After the change of variable the Hamiltonian
takes the form
\begin{equation}
\hat{H} = -\frac{\hbar^2}{2\mu}\left[ \partial_{\tilde{r}\tilde{r}} + \frac{1}{\tilde{r}}\partial_{\tilde{r}} + \frac{1}{\tilde{r}^2}\partial_{\tilde{\phi}\tilde{\phi}}\right] - \frac{\tilde{L}_{z_\Sigma}^2}{2\mu \tilde{r}^2} 
\end{equation}
where $\partial_r = \frac{1}{\sqrt{\xi_o}}\partial_{\tilde{r}}$,
$\partial_\phi = \sqrt{\xi_o}\partial_{\tilde{\phi}}$ and
$\tilde{L}_{z_\Sigma} = \frac{1}{\sqrt{\xi_o}}L_{z_\Sigma}$. The
eigenvectors and eigenvalues may be obtained by separation of
variables. Substituting
$\psi({r},{\phi})=U({r})\Phi({\phi})$ in the
eigenvalue equation $\hat{H}\psi({r},{\phi}) = E
\psi({r},{\phi})$, one finds 
\begin{equation}
\Phi({\phi}) = C\exp\left(i \tilde{m}\tilde{\phi}\right). 
\end{equation}
The angular function is $\Phi(\phi) = C\exp\left(i
\tilde{m}\sqrt{\xi_o}\phi\right)$ and it must fulfill with the
condition $\Phi(\phi)=\Phi(\phi+2\pi)$, therefore
$\tilde{m}\sqrt{\xi_o}:=m \in \mathbb{Z}$. As result, the angular
solution is the same obtained in the planar circular billiard
\begin{equation}
\Phi(\phi) = C\exp\left(i m\phi\right) \hspace{0.5cm} \mbox{with} \hspace{0.5cm} m \in \mathbb{Z}.
\end{equation}
The radial part of the equation will lead as usual to the Bessel
differential equation, but with a non integer index
$\tilde{m}={m}/{\sqrt{\xi_o}}$,
\begin{equation}
\left(\tilde{r}^2\frac{d^2}{d\tilde{r}^2}+\tilde{r}\frac{d}{d\tilde{r}}\right)U({r}) + \left[\frac{2\mu E\tilde{r}^2}{\hbar^2} + \frac{\tilde{L}_{z_\Sigma}^2}{\hbar^2} - \tilde{m}^2\right]U({r}) =0.
\end{equation}
Defining
\begin{equation}
m_\Sigma(m)^2:= \tilde{m}^2-\frac{\tilde{L}_{z_\Sigma}^2}{\hbar^2} 
\,,
\label{mSigmaFirstDefinitionEq}
\end{equation}
the radial solution is 
\begin{equation}
U_{m,s}(r) = J_{m_{\Sigma(m)}}\left(\beta_{m_\Sigma(m),s}\frac{r}{R}\right),
\end{equation}  
where $\beta_{m_\Sigma(m),s}$ ($s=1,2,\cdots,\infty$) are the zeros of the Bessel function of
the first kind $J_{m_{\Sigma(m)}}$. The corresponding energy levels are
\begin{equation}
  E_{m,s} = \frac{\hbar^2}{2\mu
    R^2}\frac{\beta_{m_\Sigma(m),s}^2}{1+\left(\frac{fo}{R}\right)^2}
\,.
\label{DHBiliardEnergyLevelsEq}
\end{equation}
Some eigenstates are shown in Figure \ref{DHBilliardStatesFig} with a
comparison with some classical trajectories. 

Let us consider the \textit{energy staircase function}
$\mathcal{N}(E)$. This function counts the number of
states below the energy $E$ and it is defined by
\begin{equation}
\mathcal{N}(E) = \sum_{m,s} \theta\left(E-E_{m,s}\right)
\end{equation}\\
where $\theta(x)$ is the step function. Knowing the asymptotic behaviour
of the large zeros of the Bessel function, one can obtain the
asymptotic behaviour of $\mathcal{N}(E)$ for large $E$, following
similar steps as in the planar circular
billiard~\cite{EuclideanDiskReminder}. At the first order, we find that 
\begin{equation}
\mathcal{N}(E) = \frac{A_{cone}}{4\pi} \left(\frac{2\mu
  E}{\hbar^2}\right)
+O\left(\sqrt{E}\right)
\label{ourWeylFormula}
\end{equation}
where $A_{cone}=\pi R \sqrt{R^2+f_o^2}$ is the area of the cone without including its base. The equation (\ref{ourWeylFormula}) is in agreement with the well known \textit{Weyl's formula} usually employed for quantum billiards in the plane 
\begin{equation}
\mathcal{N}_{plane}(E) = \frac{A}{4\pi}\left(\frac{2\mu E}{\hbar^2}\right) - \frac{P}{4\pi}\sqrt{\frac{2\mu E}{\hbar^2}}+o\left(\sqrt{E}\right)
\label{waylFormulaEq}
\end{equation}
where $A$ and $P$ are the area and perimeter of the billiard. The
confining potential constitutes a remarkable difference between
quantum non-planar billiards and the quantum billiards in the
plane. In fact, the confining potential changes the spectrum and the
states. Nevertheless, the agreement between the Weyl's formula and
staircase function of the dunce hat billiard suggests that confining
potential does not change the $\mathcal{N}(E)$ at least at the first
order in the asymptotic limit.

\begin{figure}[h]
  \centering   
  \subfloat{\includegraphics[width=0.45\textwidth]{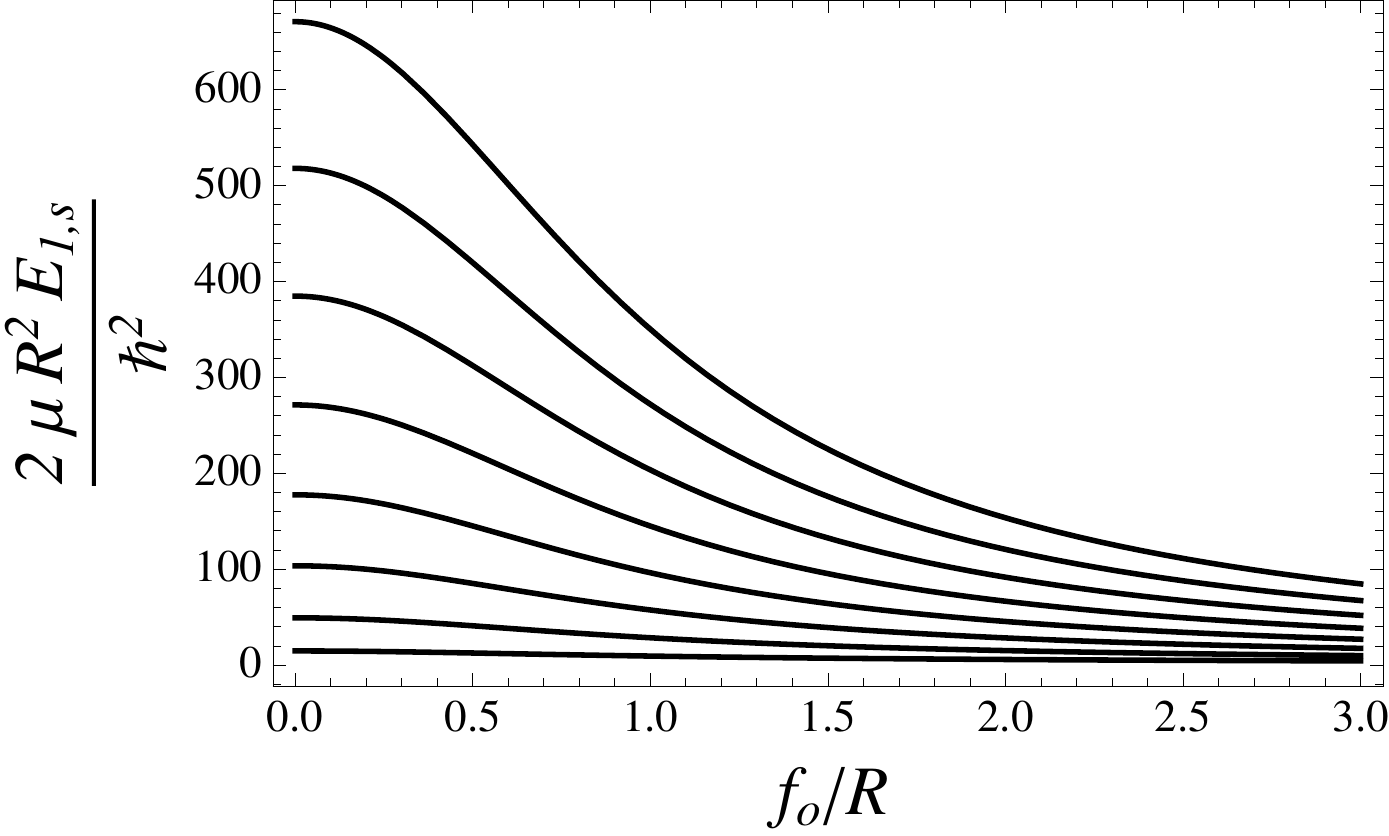}} \hspace{0.5cm} 
  \subfloat{\includegraphics[width=0.45\textwidth]{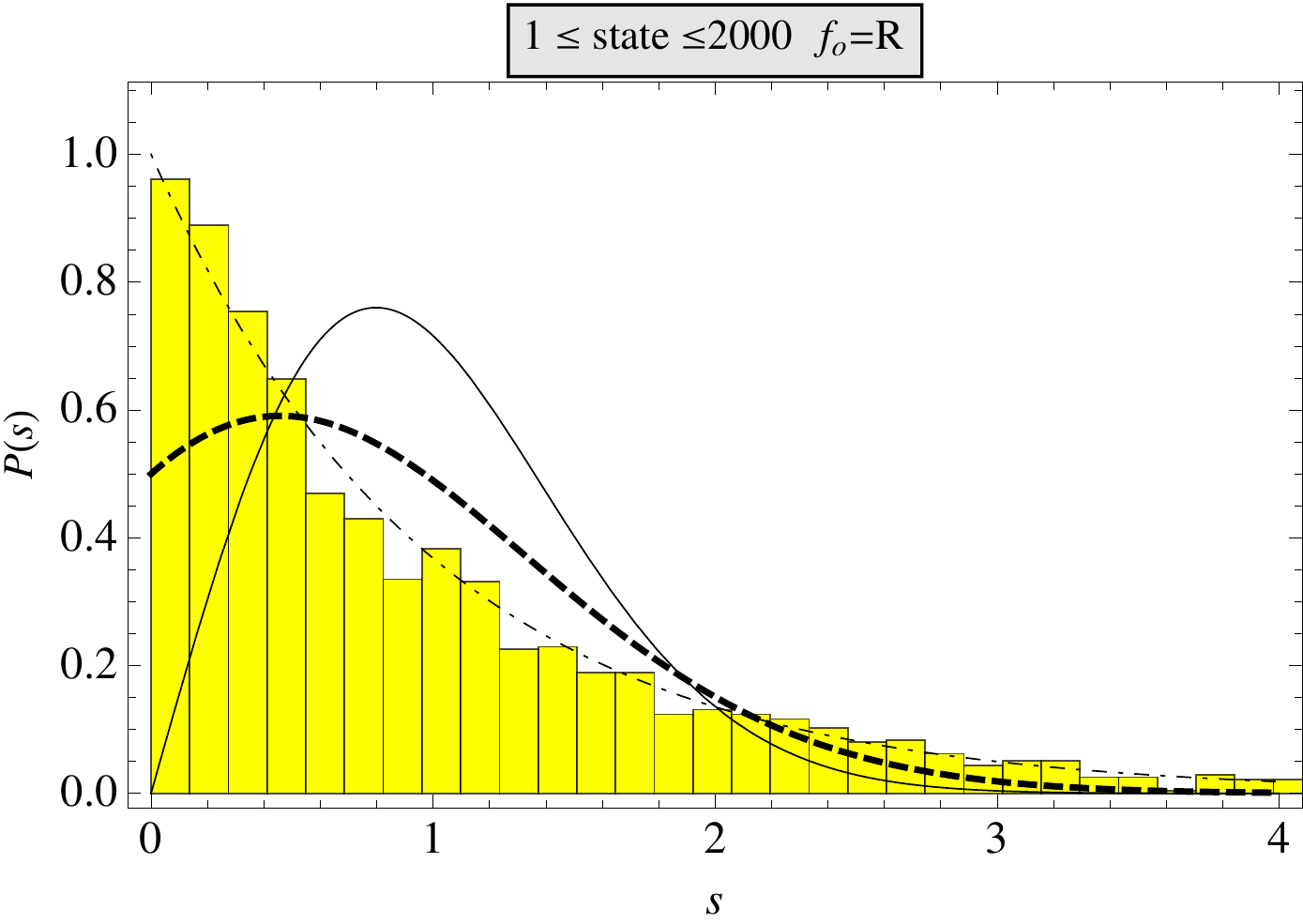}} 
  \caption[Dunce hat billiard spectrum with a circular contour]%
          {\textbf{Dunce hat billiard spectrum with a circular
              contour.} \textit{(left)} Changing of the billiard
            spectrum with the parameter $f_o/R$. \textit{(right)}
            Nearest neighbour spacing distribution of the dunce hat
            billiard energy levels. The dot-dashed, dashed and solid
            lines correspond to a Poisson, GOE2 and GOE
            distributions respectively. The nearest neighbour spacing
            distribution of the dunce hat billiard spectrum with a
            circular contour fits to the Poisson distribution.}
  \label{classicalDHBilliardFig} 
\end{figure}

Figure (\ref{classicalDHBilliardFig})-(left) shows how the energy
levels are somehow ``compressed'' when the parameter $f_o/R$ is
increased. Consequently, the number of energy levels below of a fixed
energy $E$ must increase with $f_o/R$. This is in agreement
with~(\ref{ourWeylFormula}).

The dunce hat billiard with a circular contour at the classical level
is an integrable system. Hence, we expect that nearest neighbour energy
level statistics will be a Poissonian in agreement with the
Bohigas-Giannoni-Schmit
  conjecture~\cite{sinaiBilliardAndBGSConjecture}. This can be
verified in Figure \ref{classicalDHBilliardFig}-(right).

\subsection{The quantum dunce hat billiard with a rectangular contour}

By changing the circular contour of the dunce hat billiard with
rectangular one, the rotational symmetry of the system is broken, and
the angular momentum $L_z$ is not conserved anymore. The system is no
longer integrable. A radial symmetric deflector as the cone in
combination with a square boundary induce complicated classical
dynamics. Therefore, at the quantum level, the spectrum and
eigenvectors are not simple in comparison with the ones obtained in
the circular boundary case. We expect that the quantum dunce hat
billiard in the square box will share some of the statistical
properties in its spectrum with other quantum billiards with a
classical chaotic counterpart.

\begin{figure}[h]
  \centering   
 \subfloat[state 1]{\includegraphics[width=0.20\textwidth]{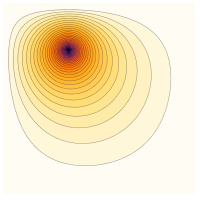}}
 \subfloat[state 2]{\includegraphics[width=0.20\textwidth]{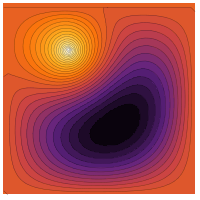}}  
 \subfloat[state 3]{\includegraphics[width=0.20\textwidth]{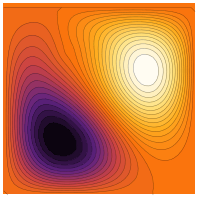}} 
 \subfloat[state 4]{\includegraphics[width=0.20\textwidth]{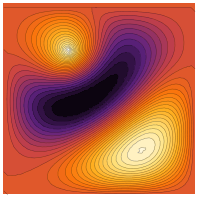}}\\
   \subfloat[state 5]{\includegraphics[width=0.20\textwidth]{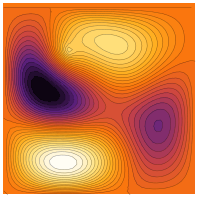}}
 \subfloat[state 6]{\includegraphics[width=0.20\textwidth]{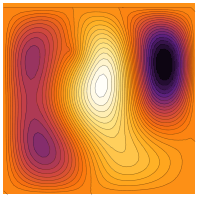}}  
 \subfloat[state 7]{\includegraphics[width=0.20\textwidth]{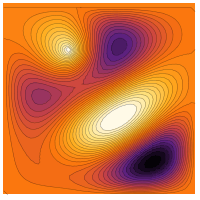}} 
 \subfloat[state 20]{\includegraphics[width=0.27\textwidth]{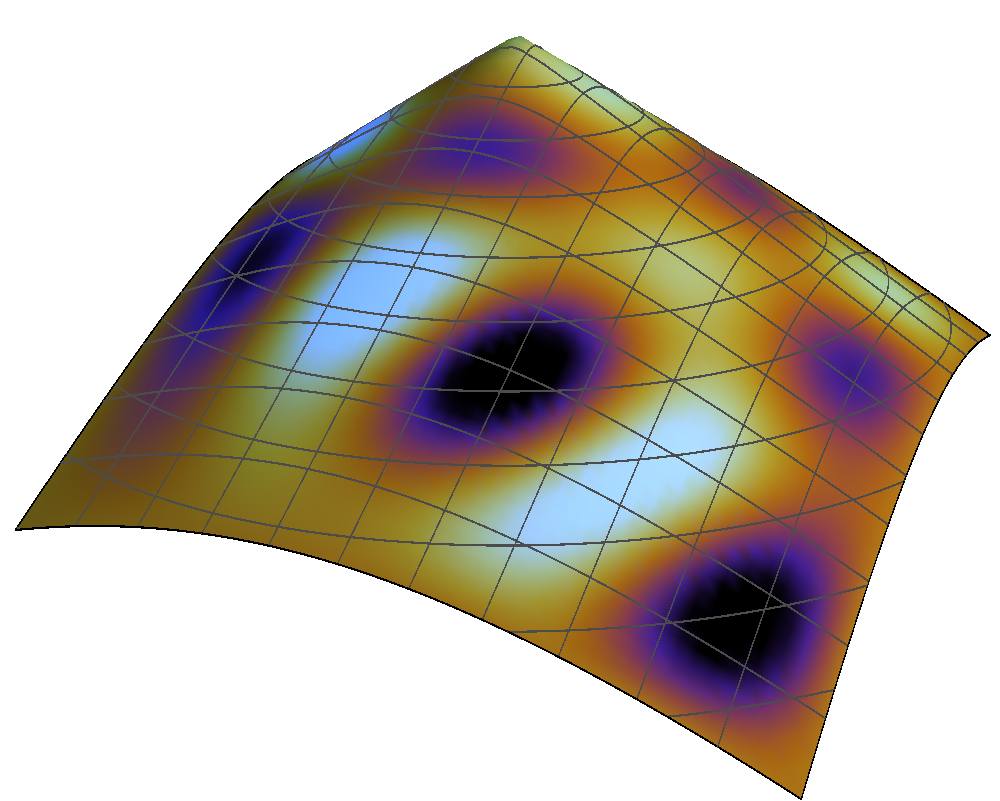}}

\caption[Eigenstates of the dunce hat billiard in a square box.]
        {\textbf{Eigenstates of the dunce hat billiard in a square box.} The figures
          from (a)-(g) are the first seven states of the non
          integrable dunce hat billiard (top view). The symmetries of the
          wavefunction were broken because the cone was not placed at
          the center of the box.}
\label{DHBilliardSquareStatesFig}
\end{figure}

\begin{figure}[h]
  \centering
 \subfloat{\includegraphics[width=0.45\textwidth]   
 {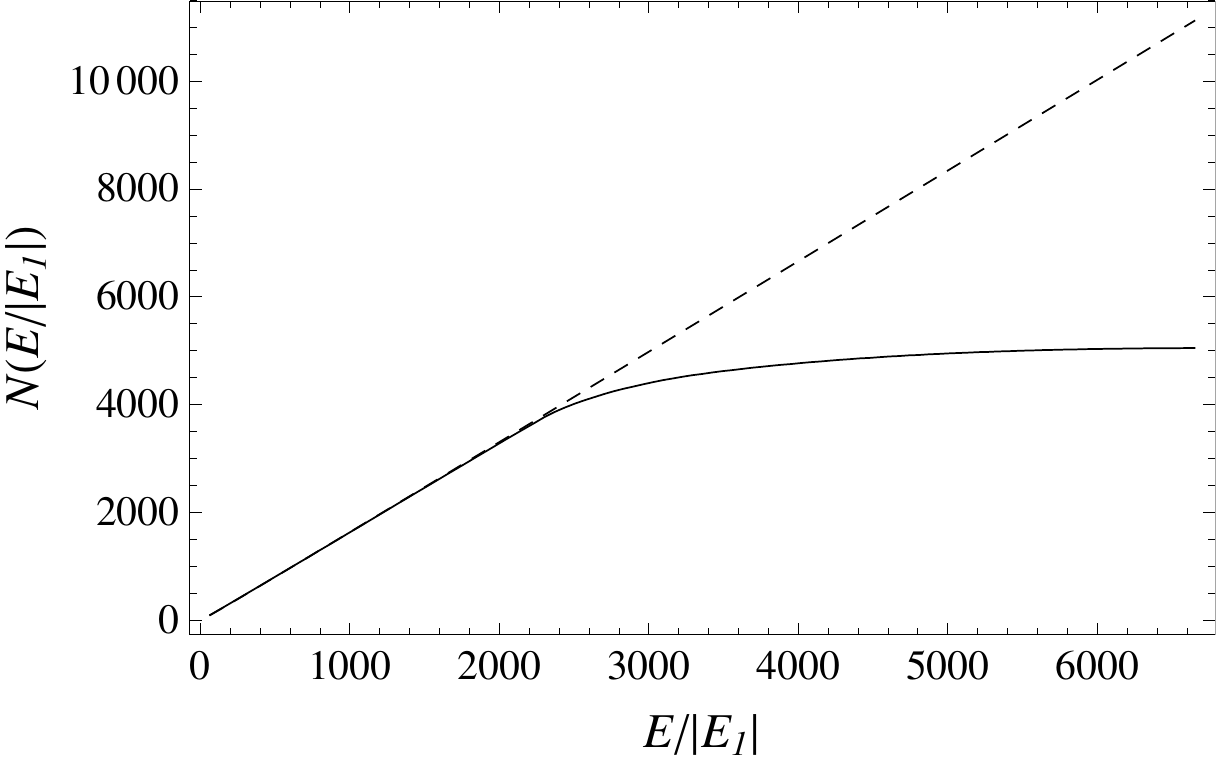}} \hspace{0.1cm}
\subfloat{\includegraphics[width=0.45\textwidth]{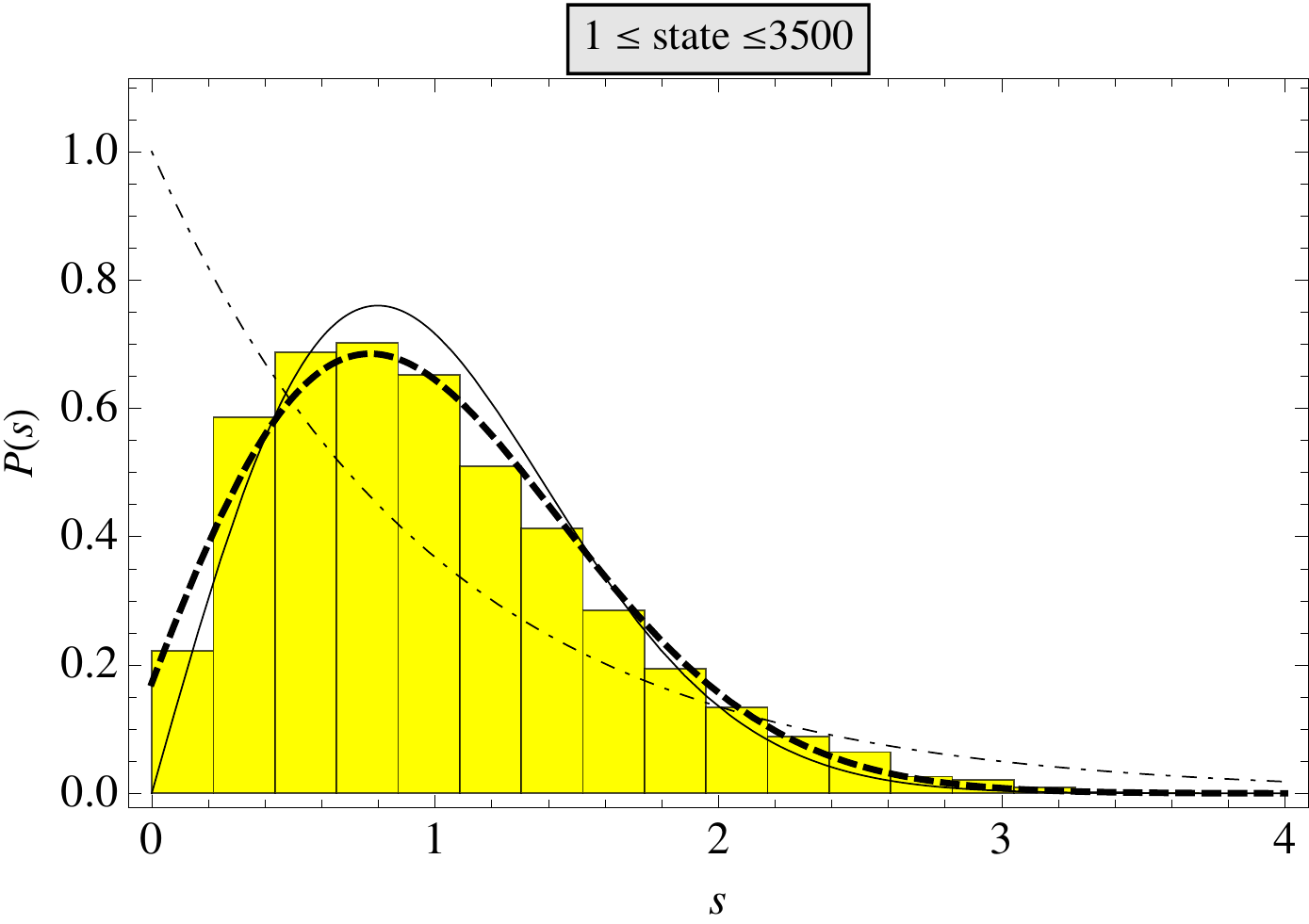}}  

\caption[Nearest neighbour spacing distribution for the dunce hat billiard in a
  square box.] 
 {\textbf{Nearest neighbour spacing distribution for the
    dunce hat billiard in a square box.} \textbf{(left)} The numerical staircase
  function (solid line) remains linear below 2000
  states. \textbf{(right)} Using this number of states the histogram
  of the level spacing was built. The total number of states computed
  with the finite difference method was 15625.  The dot-dashed, dashed
  and solid lines correspond to a Poissonian, GOE2 and GOE
  distributions respectively. The nearest neighbour spacing
  distribution of the dunce hat billiard spectrum with a rectangular contour fits
  the GOE distribution.}
\label{NNSDForDHBilliardSquareFig}
\end{figure}

In general, this billiard require a numerical treatment. We
implemented the \textit{Finite Difference Method} in order to solve
the eigenproblem for the Hamiltonian.  Some of the eigenstates are shown
in Figure \ref{DHBilliardSquareStatesFig} and the nearest neighbour
spacing distribution of the energy levels is shown in Figure
\ref{NNSDForDHBilliardSquareFig}. The cone has not been placed at the
center of the box in order to break the remaining discrete symmetries
and to avoid the symmetry classification of energy levels in the level
statistics computation.

We have demonstrated that the staircase function of the dunce hat
billiard with a circular box satisfies the Weyl's formula (see
equation (\ref{ourWeylFormula})). As a result, we may expect that
$\mathcal{N}(E)$ should be linear with $E$ for this billiard with an
arbitrary contour, for large $E$. We evidence this behaviour in the
numerical computation of $\mathcal{N}(E)$ (see Figure
\ref{NNSDForDHBilliardSquareFig}-(\textit{left})) at least for
$\mathcal{N}(E) \leq 2000$. After this value the numerical staircase
function was not linear because of the numerical error so we only use
the first $2000$ states in the level statistics computation.

The histogram of the nearest neighbour spacing distribution for the
energy levels below the state 2000 is shown in Figure
\ref{NNSDForDHBilliardSquareFig}-(\textit{right}). It fits the
Gaussian Orthogonal Ensemble (GOE) distribution~(\ref{eq:GOE}).
Therefore, the change of the inner geometry of the two-dimensional
square well from the plane to a cone produces the emerging of
classical chaos as well as a change in the distribution of the energy
levels of the billiard quantum counterpart.

\section{Quantum billiard with a Gaussian surface}
\label{sec:bigs}

We may use the finite difference method in order to solve the problem
for any surface with an arbitrary contour. For that purpose, the
Hamiltonian given by the equation (\ref{generalBilliardHamiltonianEq})
is expressed as a matrix on a lattice and then it is
diagonalized. However, in this section, an alternative way to solve
numerically the non-integrable problem will be presented. The idea is to
use the analytic eigenvectors of the planar billiard rectangular as a
basis to expand the wavefunctions of the non-planar rectangular
billiard. As an illustration, we will study the quantum rectangular
{billiard with a Gaussian surface}.

\subsection{Billiard with a rectangular contour}

The eigenvectors of the rectangular plane billiard will be used as a
basis $\mathfrak{B}:=\left\{ \mid
u\rangle:u=1,2,3,\cdots,\infty\right\}$ to expand the eigenfunctions
of the Hamiltonian (\ref{generalBilliardHamiltonianEq}), where
\begin{equation}
\langle\vec{r}\mid u\rangle = \frac{2}{\sqrt{A}}\sin\left(\kappa_u^{(x)}x\right)\sin\left(\kappa_u^{(y)}y\right)\hspace{0.1cm},
\end{equation}
$A=l_xl_y$ is the area of the box, and
\begin{equation}
\kappa_u^{(x)} = \frac{m(u)\pi}{l_x} \,,
\quad \kappa_u^{(y)} = \frac{n(u)\pi}{l_y} \hspace{0.5cm} \mbox{with} \hspace{0.5cm} \left(m(u),n(u) = 1,2,\cdots,\infty\right) \hspace{0.1cm}.
\end{equation}
The first kinetic term of the Hamiltonian $\xi(r)\hat{H}_o$ expressed in $\mathfrak{B}$ is
\begin{eqnarray}
\langle u\mid \xi(r)\hat{H}_o \mid v\rangle 
&=& \epsilon_v \int_{\mathfrak{D}} d^2\vec{r} \langle u \mid \vec{r} \rangle \xi(r) \langle \vec{r} \mid v \rangle \nonumber \\
&=&  \epsilon_v \sum_{n=0}^\infty(-1)^n \int_{\mathfrak{D}} d^2\vec{r} \langle u \mid \vec{r} \rangle (\partial_rf)^{2n}\langle \vec{r} \mid v \rangle \nonumber \\
&=& \epsilon_v \sum_{n=0}^\infty(-1)^n
\left(\frac{V_o}{2\pi\sigma^4}\right)^{2n} \int_{\mathfrak{D}}
d^2\vec{r} \langle u \mid \vec{r} \rangle r^{2n} \exp\left(-\frac{n
  r^2}{\sigma^2}\right) \langle \vec{r} \mid v \rangle \hspace{0.1cm}.
\nonumber\\
\label{largeEquationForjiHo}
\end{eqnarray}
where
\begin{equation}
\epsilon_v = \frac{\hbar^2}{2\mu}\left[\frac{m(v)^2\pi^2}{l_x^2}+\frac{n(v)^2\pi^2}{l_y^2}\right]
\end{equation}
are the eigenvalues of the plane
billiard. In~(\ref{largeEquationForjiHo}), we have used the expansion
\begin{equation}
\xi(r) = \frac{1}{1+\left(\partial_rf\right)^2} =
\sum_{n=0}^\infty(-1)^n (\partial_rf)^{2n}
\,,
 \label{geometricSerieEq}
\end{equation}
which requires $\mid\left(\partial_rf\right)^2\mid<1$, i.e., $\sigma >
\sigma_{min} = \left(\frac{V_o}{2\pi\sqrt{e}}\right)^{\frac{1}{3}} $
for the convergence of the geometric series.

Defining $\alpha_n:=\frac{n}{\sigma^2}$ and using the binomial formula
for $r^{2n}=(x+y)^{2n}$ in~(\ref{largeEquationForjiHo}), we obtain
\begin{eqnarray}
\langle u\mid \xi(r)\hat{H}_o \mid v\rangle 
&=& \epsilon_v\left[[ \delta_{u,v}  + \frac{4}{A}\sum_{n=1}^\infty(-1)^n
  \left(\frac{V_o}{2\pi\sigma^4}\right)^{2n}
  \times
  \right.
\nonumber\\
&&\left.  \sum_{k=0}^n {n \choose
    k}
  I_{2(n-k),n}^{[s^2,x]}(\alpha_n^{(1)};u,v)I_{2k,n}^{[s^2,y]}(\alpha_n^{(1)};u,v)
  \right]
\end{eqnarray}
where 
\begin{equation}
I^{[s^2,x_i]}_{q,n}(\alpha_n ;u,v) := \int_0^{l_{x_i}}x'^q \exp\left(-\alpha_n x'^2 \right)\sin\left(\kappa_u^{(x_i)} x'\right)\sin\left(\kappa_v^{(x_i)} x'\right)dx'\hspace{0.1cm}. \nonumber
\end{equation}
The super index $[s^2,x_i]$ means that the integrand has the product
$\sin\left(\kappa_u^{(x_i)} x'\right) \sin\left(\kappa_v^{(x_i)}
x'\right)$, and upper integration limit is $l_{x_i}$ with $x_1=x$ and
$x_2=y$.  All the other terms in the Hamiltonian may be expressed in a
similar way. After a lengthly, but simple algebra, we find 

\noindent \textit{Confining potential terms}
\begin{eqnarray}
\langle u\mid k_r^2\mid v \rangle &=& \frac{4}{A}\sum_{n=0}^\infty (-1)^{(n)}a_n^{(3)}\left(\frac{V_o}{2\pi\sigma^4}\right)^{2(n+1)} \left[\frac{1}{\sigma^4}\sum_{k=0}^{n+2} {n + 2\choose k} I_{2(n+2-k),n}^{[s^2,x]}I_{2k,n}^{[s^2,y]}-\right.\nonumber \\
&&-\left.\frac{2}{\sigma^2}\sum_{k=0}^{n+1} {n + 1\choose k} I_{2(n+1-k),n}^{[s^2,x]}I_{2k,n}^{[s^2,y]}+\sum_{k=0}^{n+2} {n\choose k} I_{2(n-k),n}^{[s^2,x]}I_{2k,n}^{[s^2,y]}\right]
\end{eqnarray}
and 
\begin{equation}
\langle u\mid \frac{-\hbar^2 k_\phi^2}{8\mu} \mid v \rangle = \frac{\hbar^2}{8\mu} \sum_{n=1}^\infty (-1)^{(n+1)}\left(\frac{V_o}{2\pi\sigma^4}\right)^{2(n+1)}\frac{4}{A}\sum_{k=0}^{n} {n \choose k} I_{2(n-k),n}^{[s^2,x]}I_{2k,n}^{[s^2,y]}
\end{equation}
where the functions $I$ must be evaluated at
$\left(\alpha_{(n+1)};u,v\right)$, and 
\begin{equation}
a_n^{(m)}=\frac{(m+n-1)!}{(m-1)!\,n!} \,.
\end{equation}

\noindent \textit{Radial kinetic term}
\begin{equation}
\langle u\mid k(r)\partial_r\mid v\rangle = \sum_{n=0}^{\infty} \Theta_n\left(\frac{1}{\sigma^2}\langle u\mid r^{2(n+1)}e^{-\alpha_n^{(2)}r^2}\vec{x}\cdot\partial_{\vec{x}}\mid v\rangle - \langle u\mid r^{2n}e^{-\alpha_n^{(2)}r^2}\vec{x}\cdot\partial_{\vec{x}}\mid v\rangle\right)
\end{equation}
with
\begin{equation}
\Theta_n:=(-1)^{3n+1} a_n^{(2)}\left(\frac{V_o}{2\pi\sigma^4}\right)^{2(n+1)}
\end{equation}
and
\begin{equation}
\langle u \mid r^{2m}e^{-\alpha_n r^2} \vec{x}\cdot\partial_{\vec{x}} \mid v \rangle = \frac{4}{A}\sum_{k=0}^{m} {m \choose k} \left[\kappa_v^{(x)}I_{2(m-k)+1,n}^{[sc,x]}I_{2k,n}^{[s^2,y]}+\kappa_v^{(y)}I_{2(m-k),n}^{[s^2,x]}I_{2k+1,n}^{[sc,y]}\right]
\end{equation}
where
\begin{equation}
I^{[sc,x_i]}_{q,n}(\alpha_n ;u,v) := \int_0^{l_{x_i}}x'^q \exp\left(-\alpha_n x'^2 \right)\sin\left(\kappa_u^{(x_i)} x'\right)\cos\left(\kappa_v^{(x_i)} x'\right)dx'\hspace{0.1cm}.
\end{equation}

\noindent \textit{Centrifugal term}
\begin{eqnarray}
\langle u\mid \frac{\zeta(r)\hat{L}_z^2}{r^2}\mid v \rangle &=& -\frac{4}{A}\left(\frac{\hbar^2}{2\mu}\right)\sum_{n=1}^\infty (-1)^{(n)}\left(\frac{V_o}{2\pi\sigma^4}\right)^{2n} \sum_{k=0}^{n-1} {n-1 \choose k} \cdot \nonumber\\&& \cdot\left[\left(\kappa_v^{(y)}\right)^2 I_{2(n-k),n}^{[s^2,x]}I_{2k,n}^{[sc,y]} + \left(\kappa_v^{(x)}\right)^2 I_{2(n-1-k),n}^{[s^2,x]}I_{2(k+1),n}^{[sc,y]}  + \right. \nonumber\\ & & + 2\kappa_v^{(x)}\kappa_v^{(y)} I_{2(n-k)-1,n}^{[sc,x]}I_{2k+1,n}^{[sc,y]} + \kappa_v^{(x)} I_{2(n-k)-1,n}^{[sc,x]}I_{2k,n}^{[s^2,y]} + \nonumber\\&& + \left.\kappa_v^{(y)} I_{2(n-1-k),n}^{[s^2,x]}I_{2k+1,n}^{[sc,y]}\right]
\end{eqnarray}
where the functions $I$ must be evaluated at
$\left(\alpha_1;u,v\right)$.

\noindent \textit{Electric potential term}
\begin{equation}
\langle u\mid q E_o f(r) \mid v \rangle = \frac{4\hbar^2}{A} \frac{q E_o V_o}{2\pi\sigma^2} I_{0,1}^{[s^2,x]}I_{0,1}^{[s^2,y]}
\end{equation}
where the functions $I$ must be evaluated at $(\alpha_1/2;u,v)$. 

The functions $I_{q,n}^{[s^2,x]}(\alpha;u,v)$ and
$I_{q,n}^{[sc,x]}(\alpha;u,v)$ can be computed by successive
differentiations with respect to $\alpha$ of the corresponding
function with $q=0$, which can be computed numerically. Using the
matrix elements previously computed, the Hamiltonian is diagonalized
numerically.

\begin{figure}[h]
  \centering   
  \subfloat[State 201]{\includegraphics[width=0.25\textwidth]{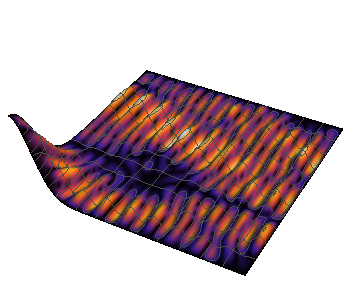}}
  \subfloat[Nearest neighbour spacing distribution]{\includegraphics[width=0.35\textwidth]{GoeDHSquared}}
  \subfloat[spectral staircase function]{\includegraphics[width=0.35\textwidth]{desimetrizedWaylProofDHSquared}}   
  \caption[Results for the rectangular billiard with a Gaussian surface.]   
  {\textbf{Results for the rectangular billiard with a Gaussian surface.}    (a) Absolute value of the
    wavefunction at the state 200. (b) Nearest neighbour spacing distribution of the billiard energy spectrum. It fits with the distribution $P_2 (S,\rho_1)$ setting $\rho_1=0.09$ (dashed line). (c) The numerical staircase function (solid line) remains linear below 3500 states.}
\label{rbgsFig}
\end{figure}

The numerical results for the rectangular billiard with the Gaussian
surface are shown in Figure \ref{rbgsFig}.  We have computed 5050
energy levels and the first 3500 of them were used to compute the
nearest neighbour spacing distribution. The Gaussian surface is located
in one of the rectangle corners. We have used a rectangle instead of a
square in order to avoid the energy level classification by each
symmetry. As in the case of a cone, the Gaussian surface introduces
chaos in the classical motion which is translated at the quantum scale
as a change of level statistics from a Poisson to a GOE
distribution. Although, the histogram does not fit with the GOE
distribution it does with the distribution function $P_2 (S,\rho_1)$
for mixture of chaos and regularity given by
\cite{berryRobnikPaper}
\begin{equation}
P_2 (S,\rho_1) = \rho_1^2\exp\left(-\rho_1 S\right) \mbox{erfc}\left(\frac{\sqrt{\pi}}{2}\bar{\rho}S\right) + \left(2\rho_1\bar{\rho}+\frac{1}{2}\pi\bar{\rho}^3 S\right)\exp\left(-\rho_1 S-\frac{1}{4}\pi\bar{\rho}^2  S^2\right) 
\end{equation}
with $\bar{\rho}:=1-\rho_1$. The function $P_2 (S,\rho_1)$ is a
Poisson distribution for $\rho_1 = 1$ and it characterizes the
regular behaviour of the classical counterpart. On the other hand, the
function $P_2 (S,\rho_1)$ is a GOE distribution~(\ref{eq:GOE}) for
$\rho_1 = 0$ when the classical counterpart is fully chaotic. In
general the Weyl's formula does not apply for this billiard because
the confining potential may change the spectral staircase
function. However, the effect of this potential is negligible for high
energy levels. This may be appreciated at the region where the
numerical spectral staircase function has a linear behaviour (see
Figure \ref{rbgsFig}-(c)).

\subsection{Quantum billiard immersed in a strong electric field}

If the applied electric field is strong, then it would be unlikely to
find the particle near the top of the Gaussian surface, at least for
small values of the kinetic energy. On the other hand, the function
$\zeta(r)=1-\xi(r)$ affects the particle motion around the top of the
surface where the slope of the Gaussian function is not
zero. Therefore, in this limit the contribution of the centrifugal
term is negligible because the particle hardly reach the region where
$\zeta(r)$ is important. The situation is similar for the $\kappa(r)$
term, so under the condition of large $|E_o|$ the kinetic energy will
be exclusively in $\xi\hat{H}_o$.  Then, the Hamiltonian takes the
form
\begin{equation}
\hat{H} = \xi(r)\hat{H}_o -\frac{\hbar^2}{8\mu}(k_r-k_\phi)^2 +
qE_of(r) + V_{box}\left(r_c(\phi)\right) \hspace{0.1cm}.
\end{equation}
The eigenvalues and eigenvectors can be found numerically either by using the
finite difference method, or the method explained in the previous section.

\begin{figure}[h]
  \centering   
  \subfloat[Eight discrete symmetries]{\includegraphics[width=0.3\textwidth]{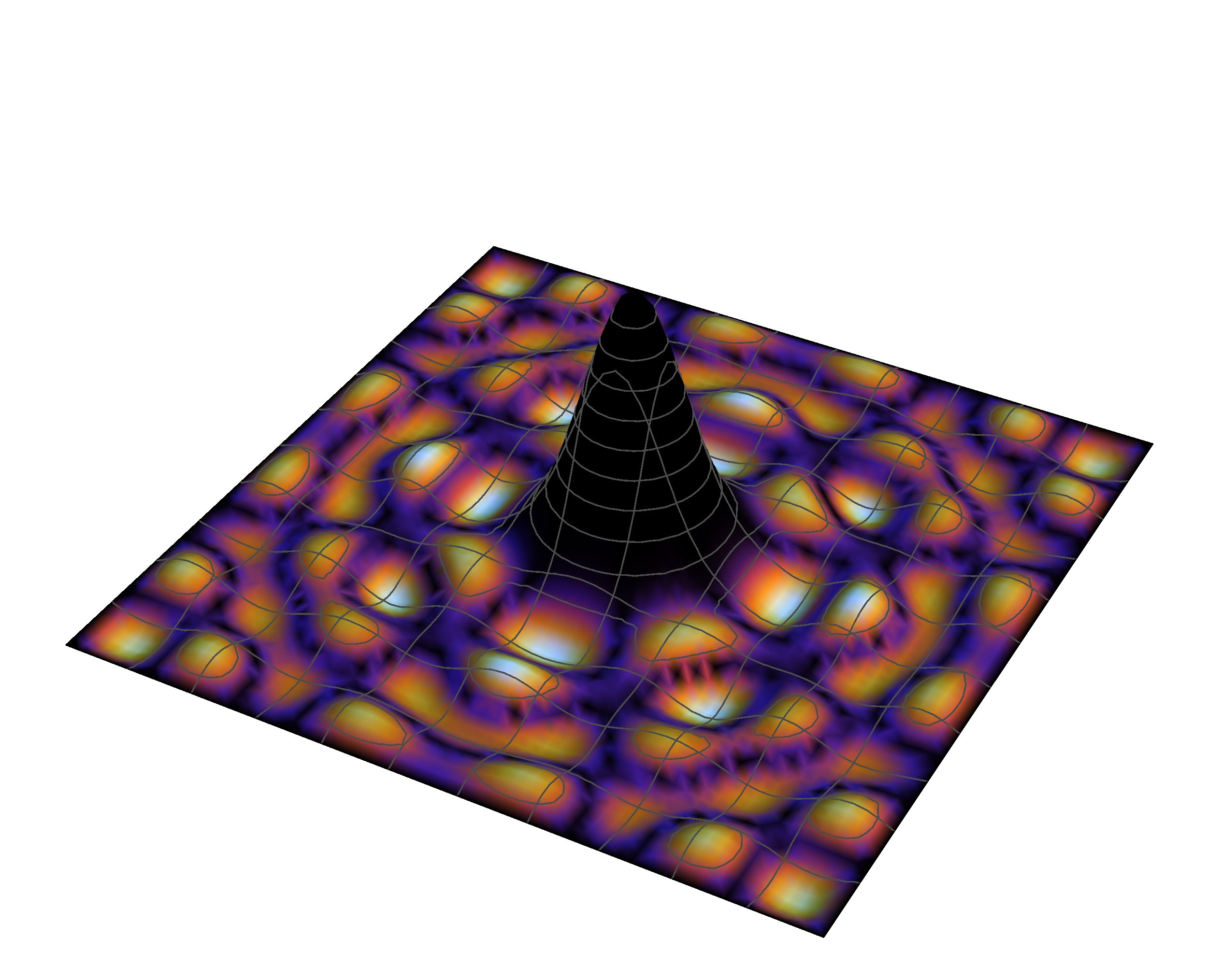}} 
  \subfloat[One discrete symmetry]{\includegraphics[width=0.3\textwidth]{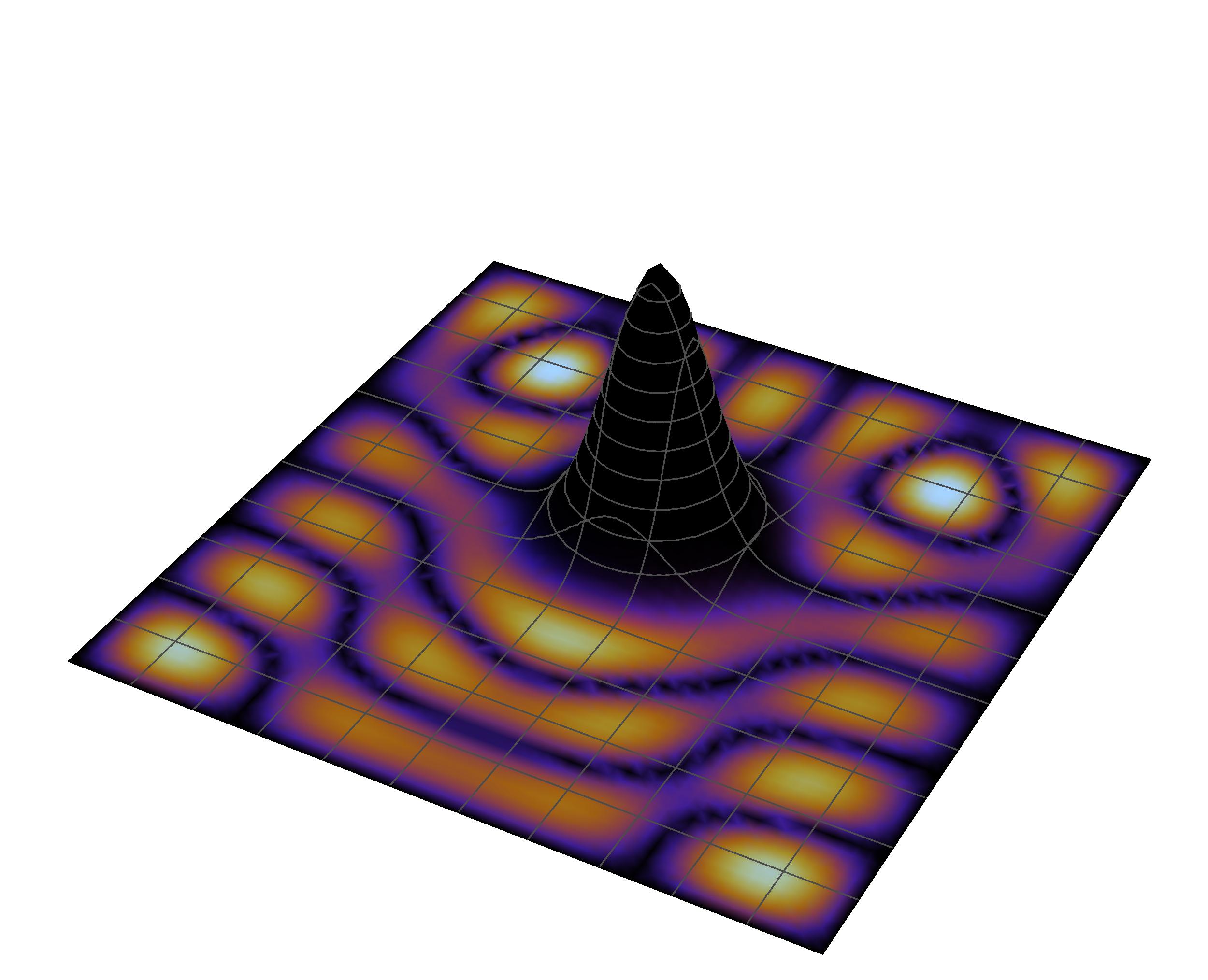}} 
  \subfloat[Billiard without symmetry axes]{\includegraphics[width=0.3\textwidth]{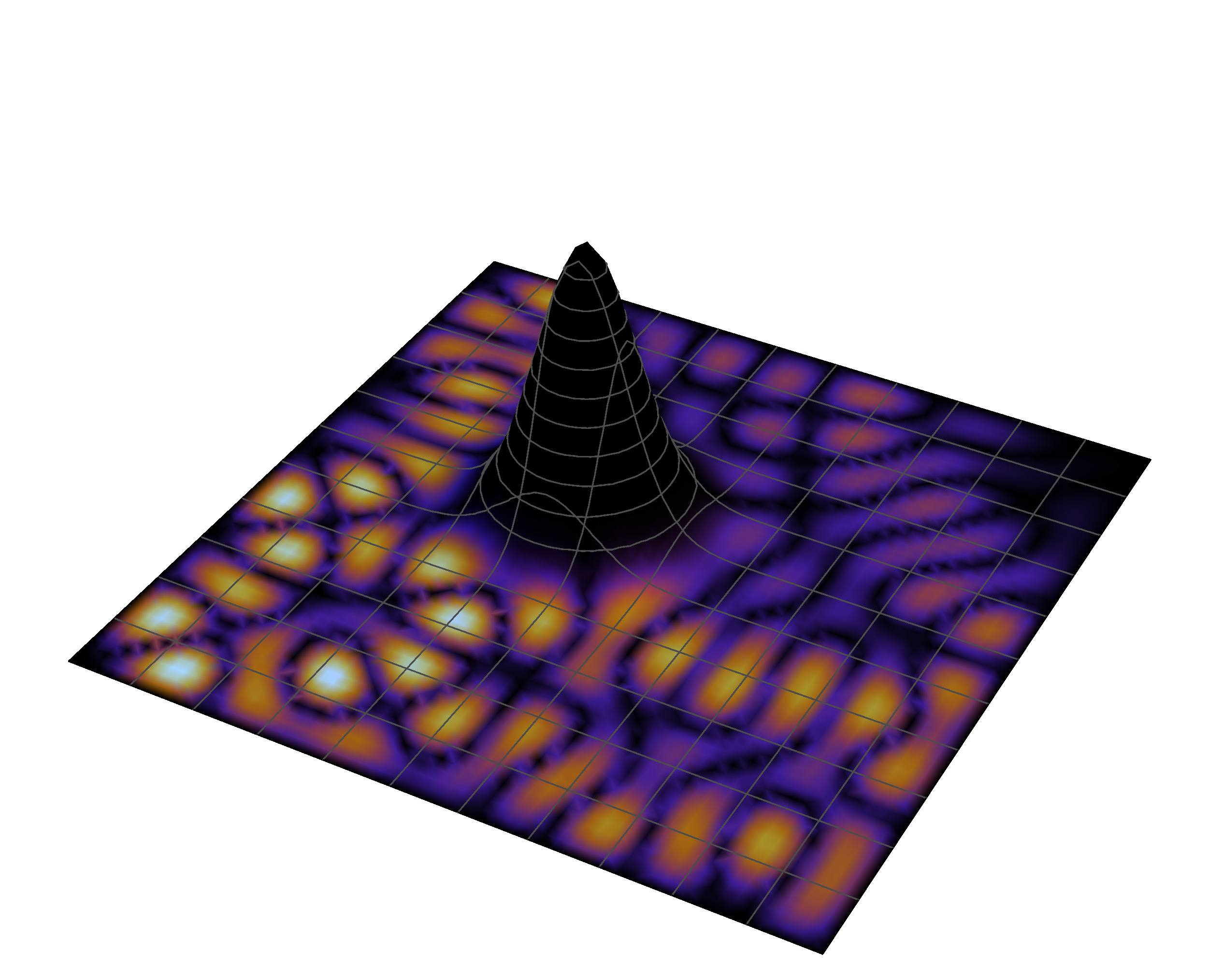}}\\
  \subfloat[Nearest neighbour spacing distribution holding one symmetry axes]{\includegraphics[width=0.35\textwidth]{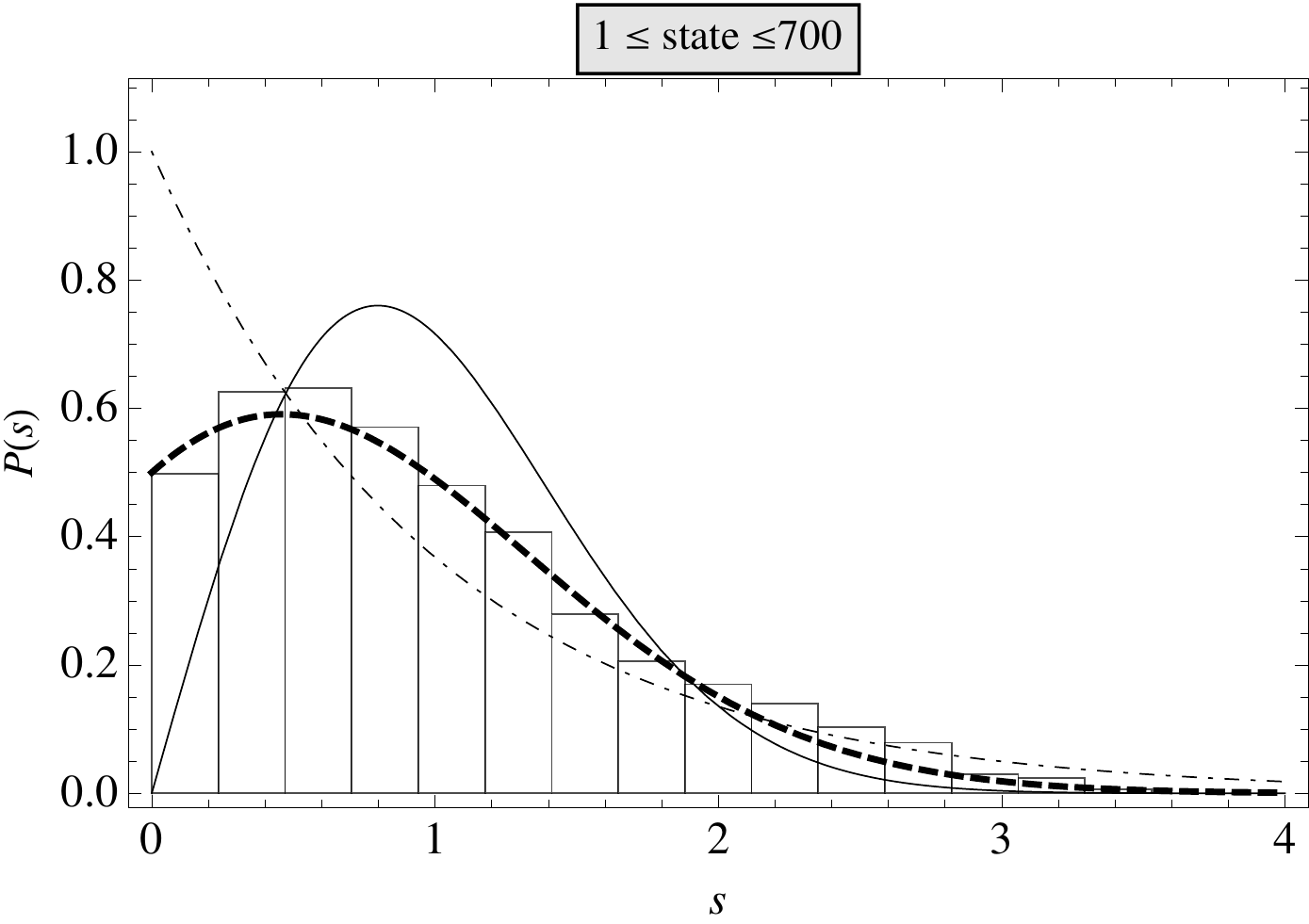}}\hspace{0.5cm}
  \subfloat[Nearest neighbour spacing distribution without symmetry axes]{\includegraphics[width=0.35\textwidth]{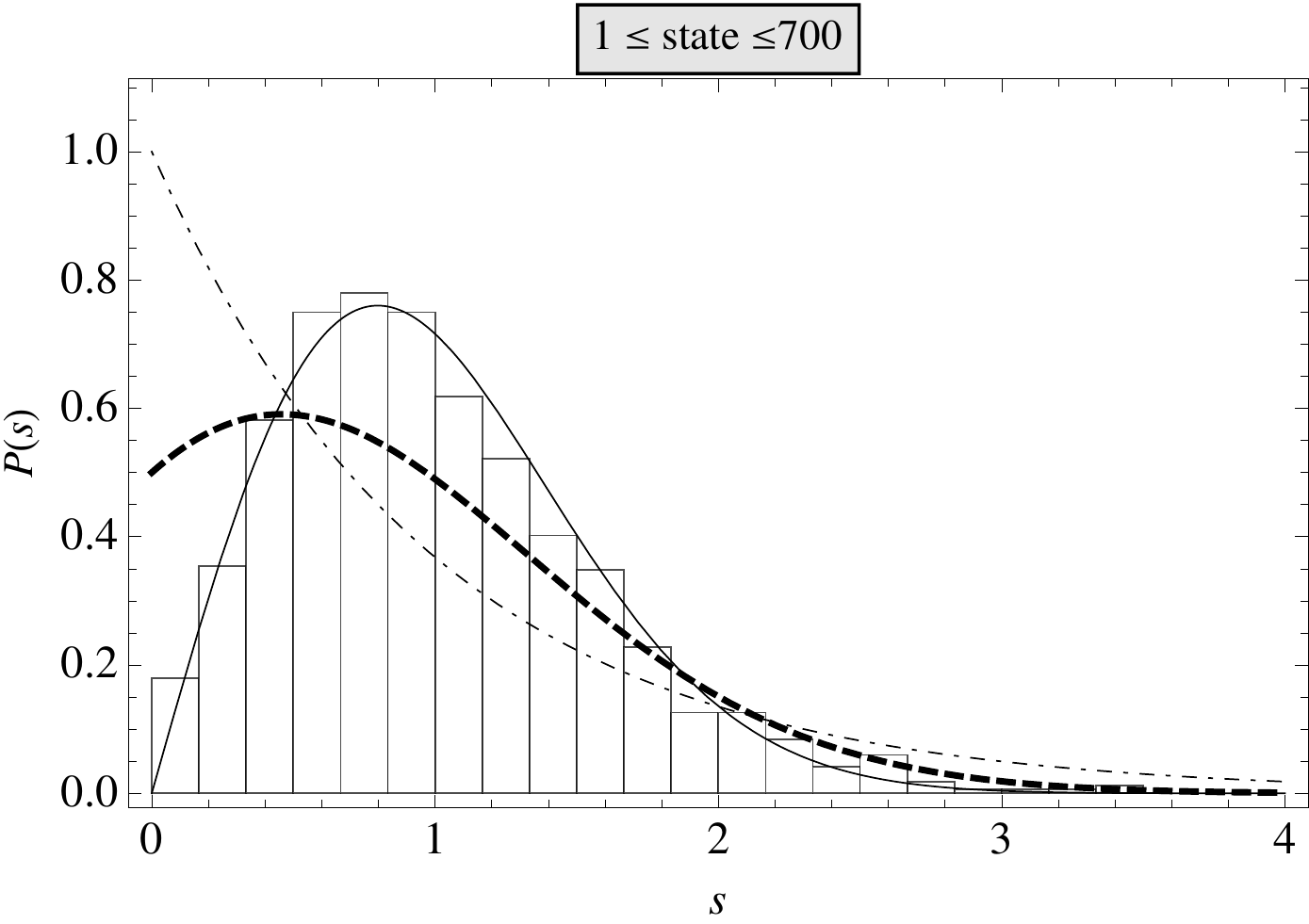}}   
  \caption[Eigenstates and level statistics of the quantum rectangular
    billiard with a Gaussian surface in a strong electric field]
  {\textbf{Eigenstates and level statistics of the quantum rectangular billiard with a Gaussian surface in a strong electric field.} }
  \label{rectangleStrongFieldFig}
\end{figure}

Two billiards will now be studied: the rectangular billiard with the
Gaussian surface and its counterpart with a circular contour, both
under the influence of a strong external electric field.  These
rectangular and circular billiards are analogous to the Sinai and
annular billiard with a soft inner disk respectively. 

For the circular billiard, if the center of the Gaussian surface is
located at the center of the circular boundary, the system has a
continuous rotational symmetry and the angular momentum $L_z$ is
conserved. This two-dimensional system, having two constants of motion
($H$ and $L_z$) is then integrable. The nearest neighbour spacing
distribution of the energy levels is a Poisson distribution, a
situation similar to the one shown in
Figure~\ref{classicalDHBilliardFig}-\textit{(right)} for the dunce hat
billiard with circular contour. If the rotational symmetry is broken
by not placing the center of the Gaussian on the center
of the circular boundary, then the system will be chaotic.

\begin{figure}[h]
  \centering   
  \subfloat[Continuous rotational symmetry]{\includegraphics[width=0.33\textwidth]{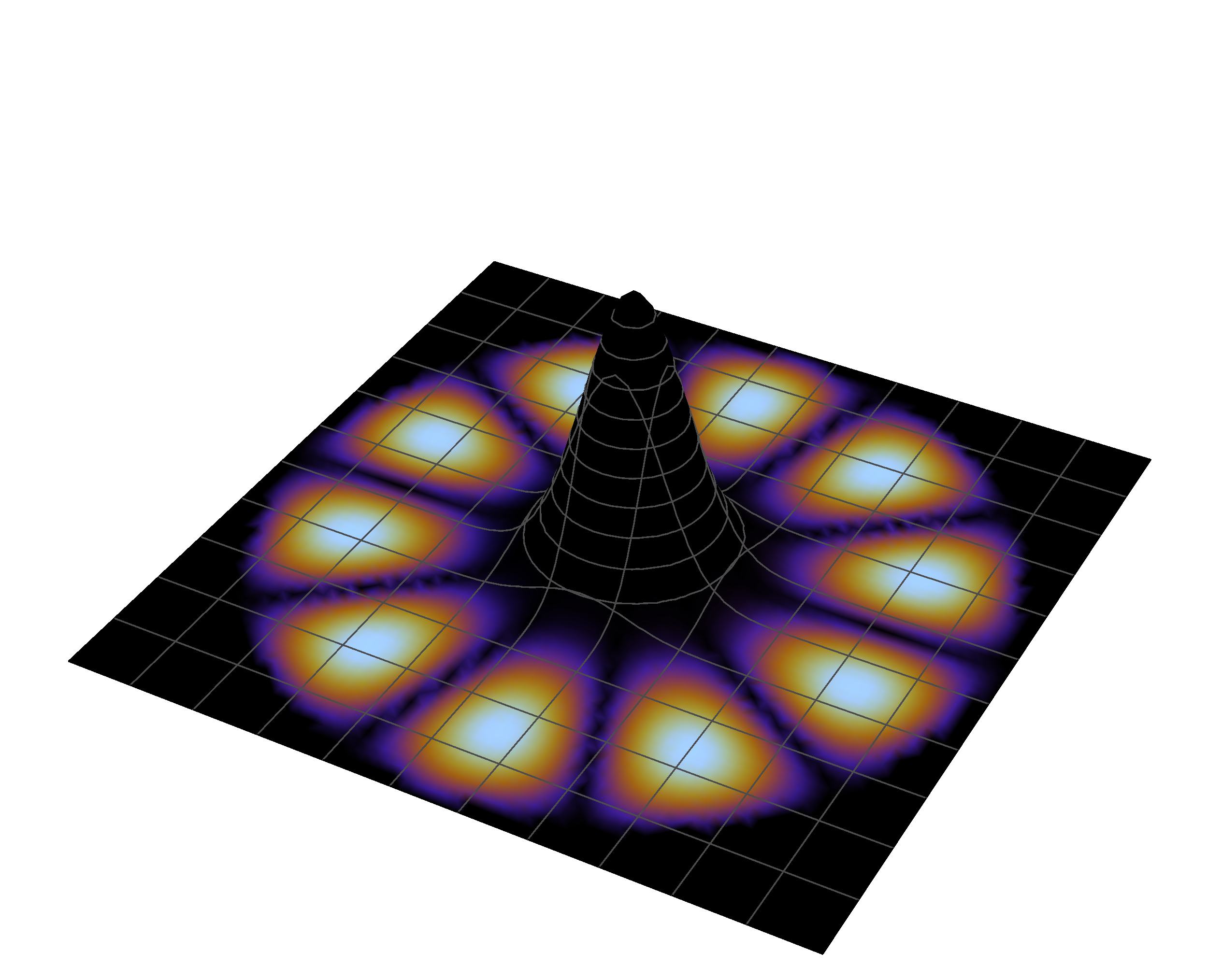}} 
  \subfloat[One discrete symmetry]{\includegraphics[width=0.33\textwidth]{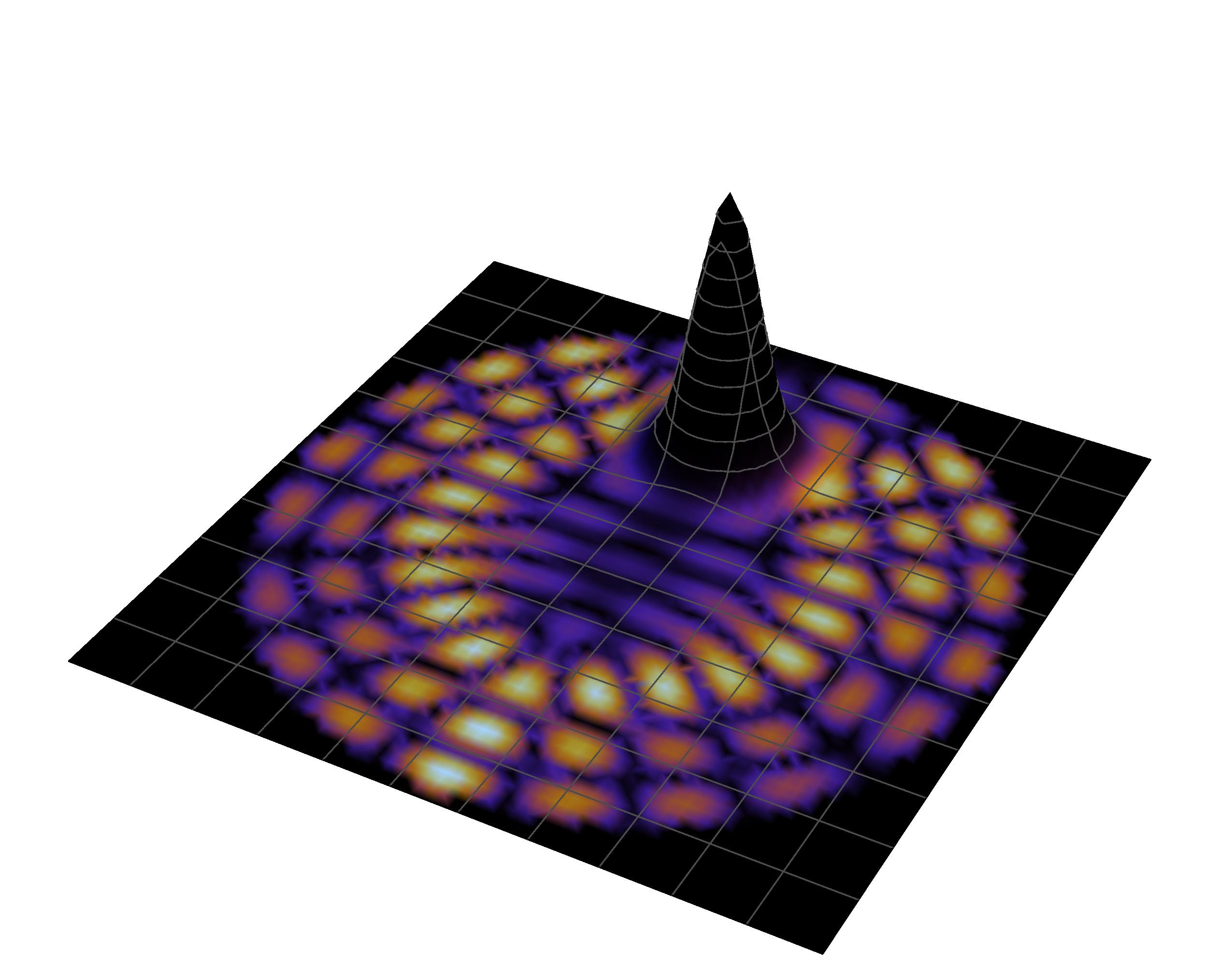}} 
  \subfloat[Nearest neighbour spacing distribution holding one symmetry axe]{\includegraphics[width=0.33\textwidth]{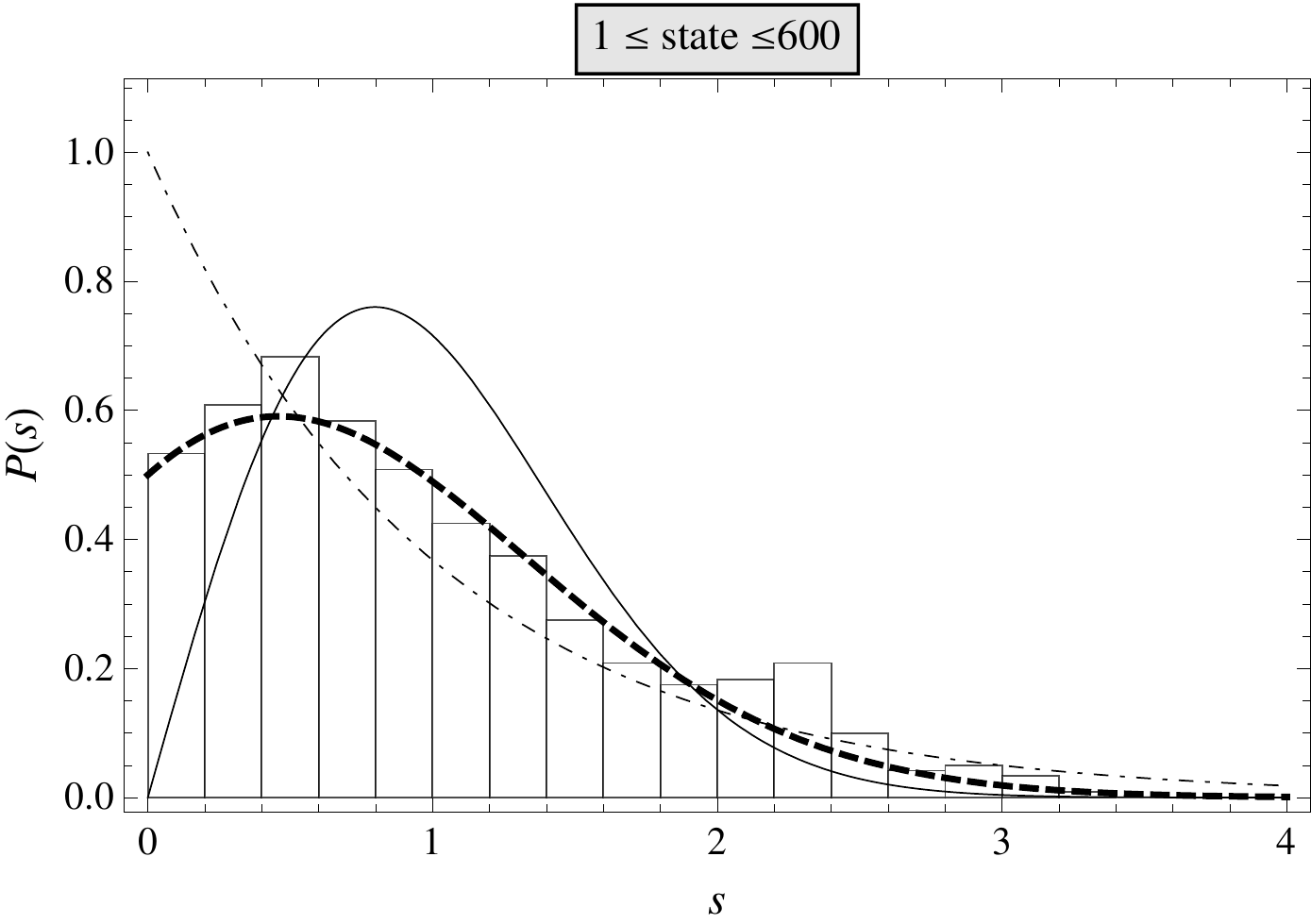}}    
  \caption[Eigenstates and level statistics of the quantum circular
    billiard with a Gaussian surface in a strong field]%
  {\textbf{Eigenstates and level statistics of the quantum circular billiard with a Gaussian surface with a strong field.} }
  \label{circleStrongFieldFig}
\end{figure}

The symmetry group of the rectangular billiard with the Gaussian
surface is the dihedral group $D_4$. This is the same symmetry group
of the Sinai billiard. Therefore, the rectangular billiard level
statistics requires the energy levels classification by each billiard
symmetry.  We may avoid this symmetry classification by placing the
Gaussian function away from the square center in order to break the
billiard symmetries. If a symmetry axe is hold as we show in Figure
\ref{rectangleStrongFieldFig}-b and Figure
\ref{circleStrongFieldFig}-b (this also applies for the circular
billiard), then the energy levels are divided into two sets according
to the parity of the wavefunction. The nearest neighbour spacing
distribution of the whole spectrum is the superposition of two
independent GOE distributions, known as the GOE2 distribution
\begin{equation}
P_{GOE2}(s) = \frac{1}{2}\exp{\left(-\frac{s^2\pi}{8}\right)} +
\frac{\pi s}{8}\exp{\left(-\frac{s^2\pi}{10}\right)}
\mathrm{erfc}\left(\frac{\sqrt{\pi} s}{4}\right)\hspace{0.1cm}.
\end{equation}
The histograms, for both the circular and rectangular billiard in this
situation, are shown in Figures \ref{rectangleStrongFieldFig}-d and
\ref{circleStrongFieldFig}-c. 

On the other hand, if we break all the geometrical symmetries then the
nearest neighbour spacing distribution of the rectangular billiard with
strong field is a GOE distribution~(\ref{eq:GOE}).  This is a feature
of classically chaotic systems with time reversal symmetry where the
energy levels are likely to repel to each other (see Figure
\ref{rectangleStrongFieldFig}-e). 

When one symmetry axe is kept, the GOE distribution can also be
obtained by taking the energy levels of the odd or even states
separately in the level statistics computation. Nevertheless, this
process requires the computation of more energy levels because the
parity classification enable us to use only approximately a half of
the numerically admissible energy levels computed for each nearest
neighbour spacing distribution.

\begin{figure}[h]
  \centering
\subfloat[closed unstable orbits]{\includegraphics[width=0.33\textwidth]{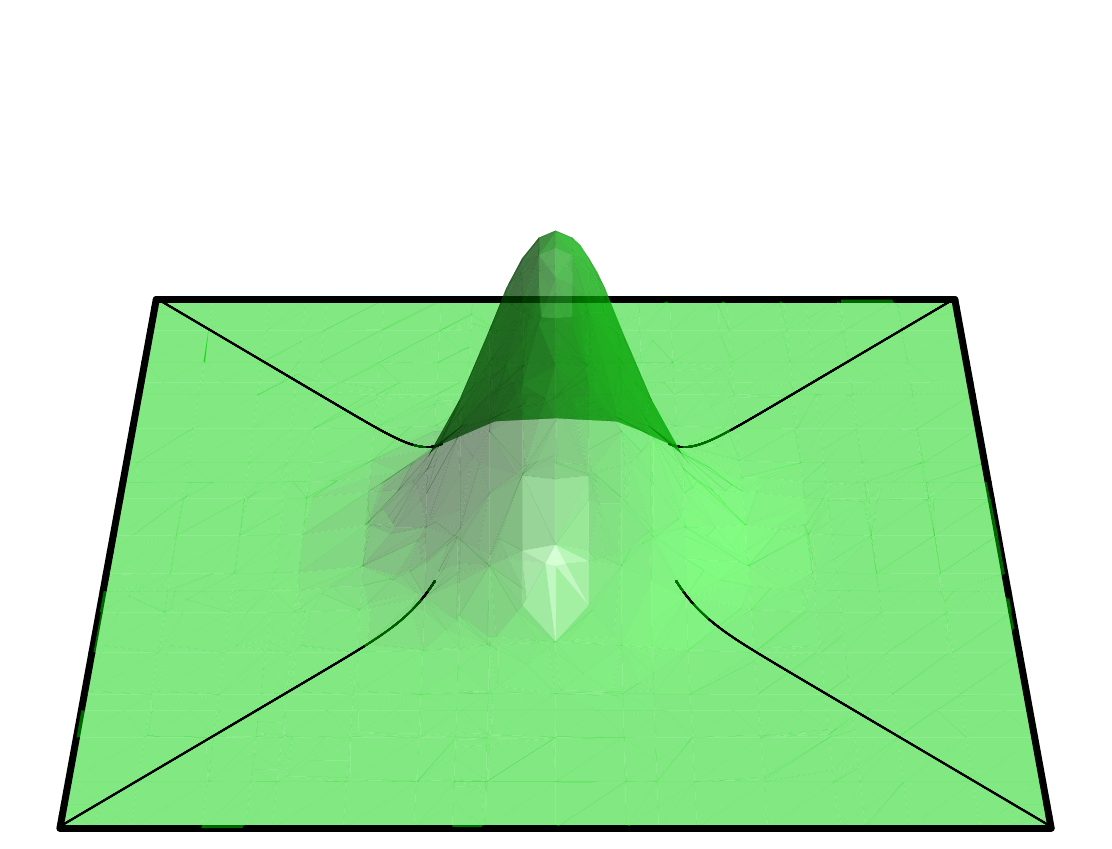}} 
  \subfloat[set of classical stable trajectories]{\includegraphics[width=0.33\textwidth]{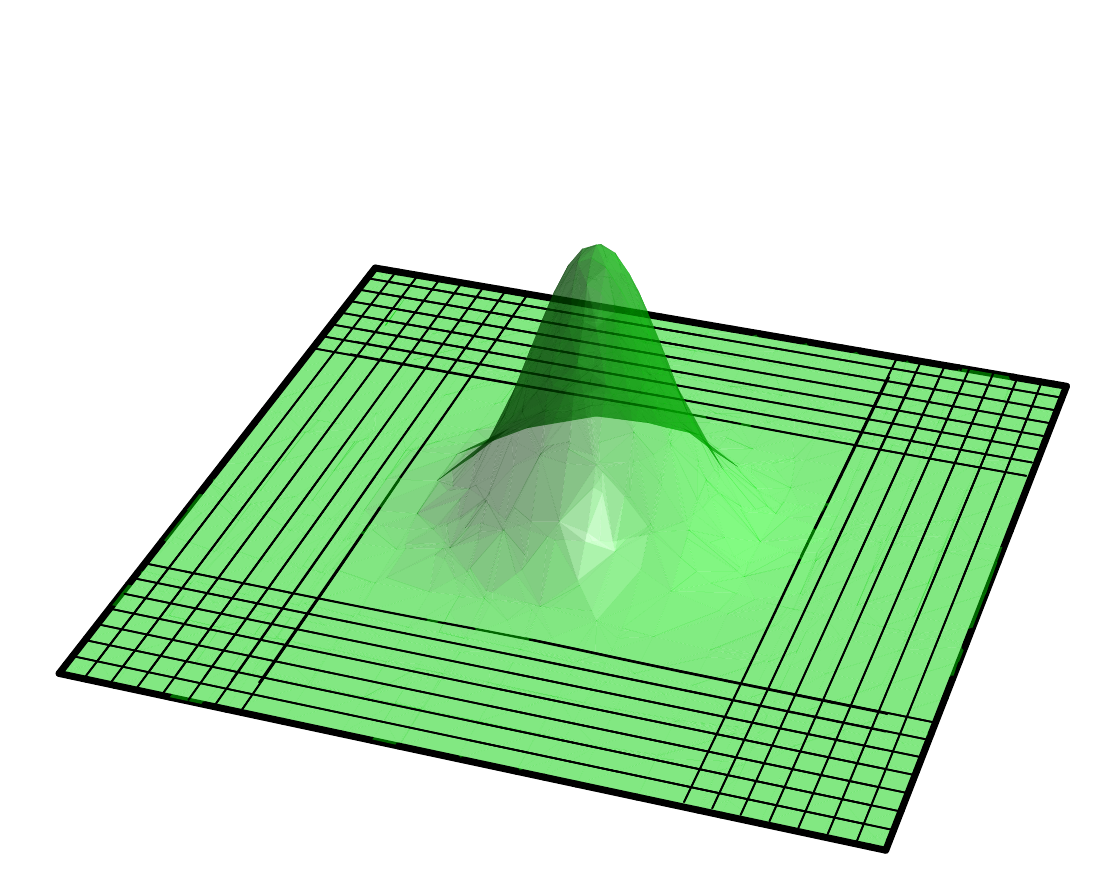}} 
  \subfloat[set of classical stable trajectories]{\includegraphics[width=0.33\textwidth]{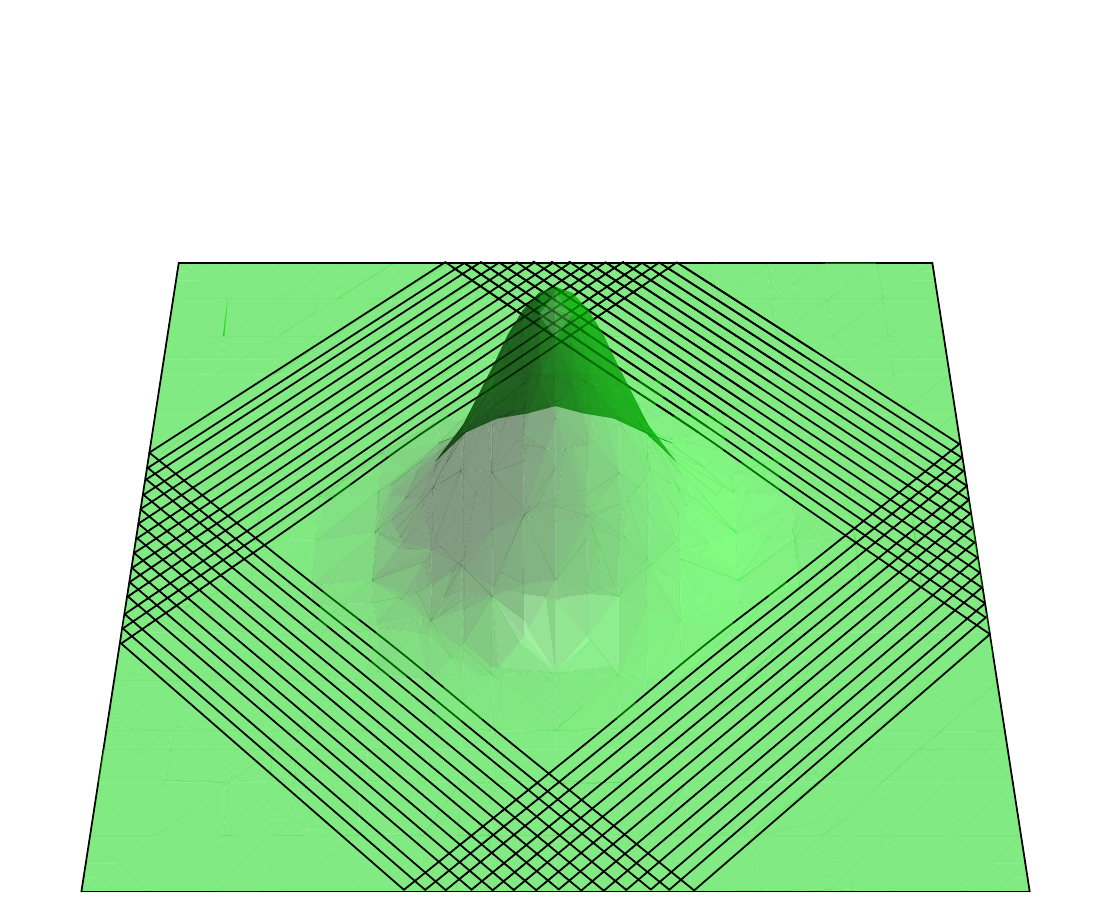}}\\     
  \subfloat[scarred state]{\includegraphics[width=0.33\textwidth]{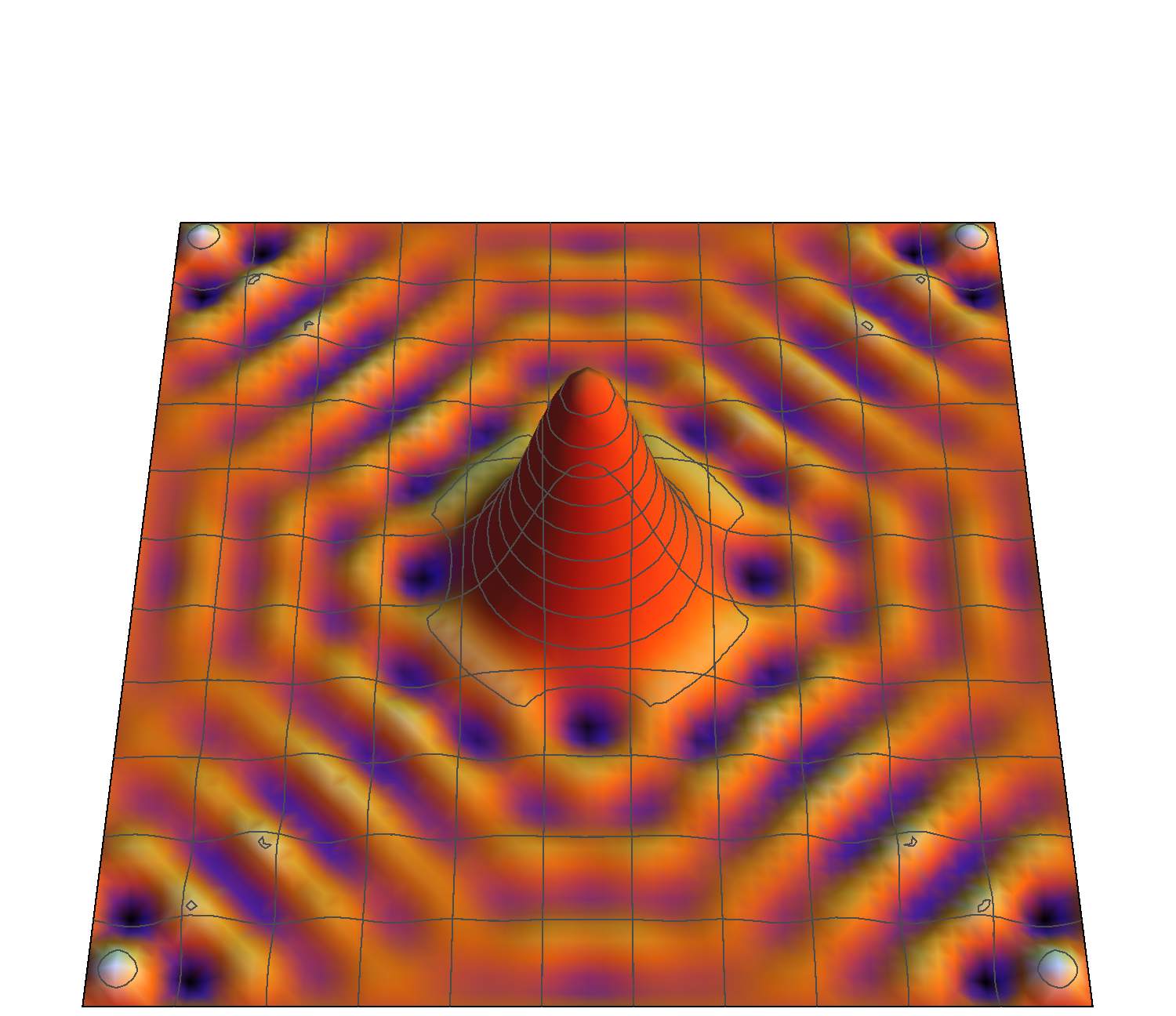}} 
  \subfloat[bouncing ball state]{\includegraphics[width=0.33\textwidth]{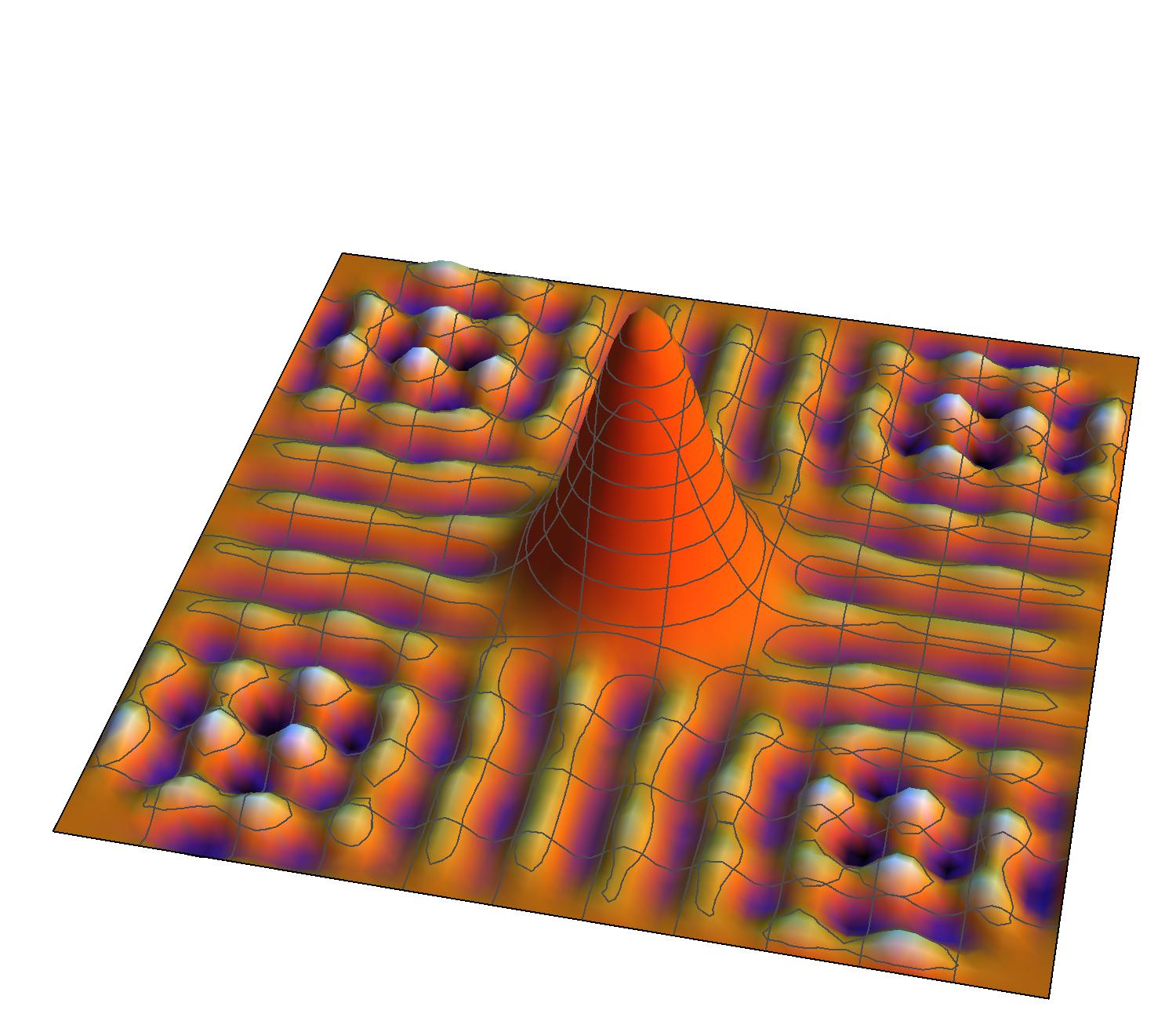}} 
  \subfloat[bouncing ball state]{\includegraphics[width=0.33\textwidth]{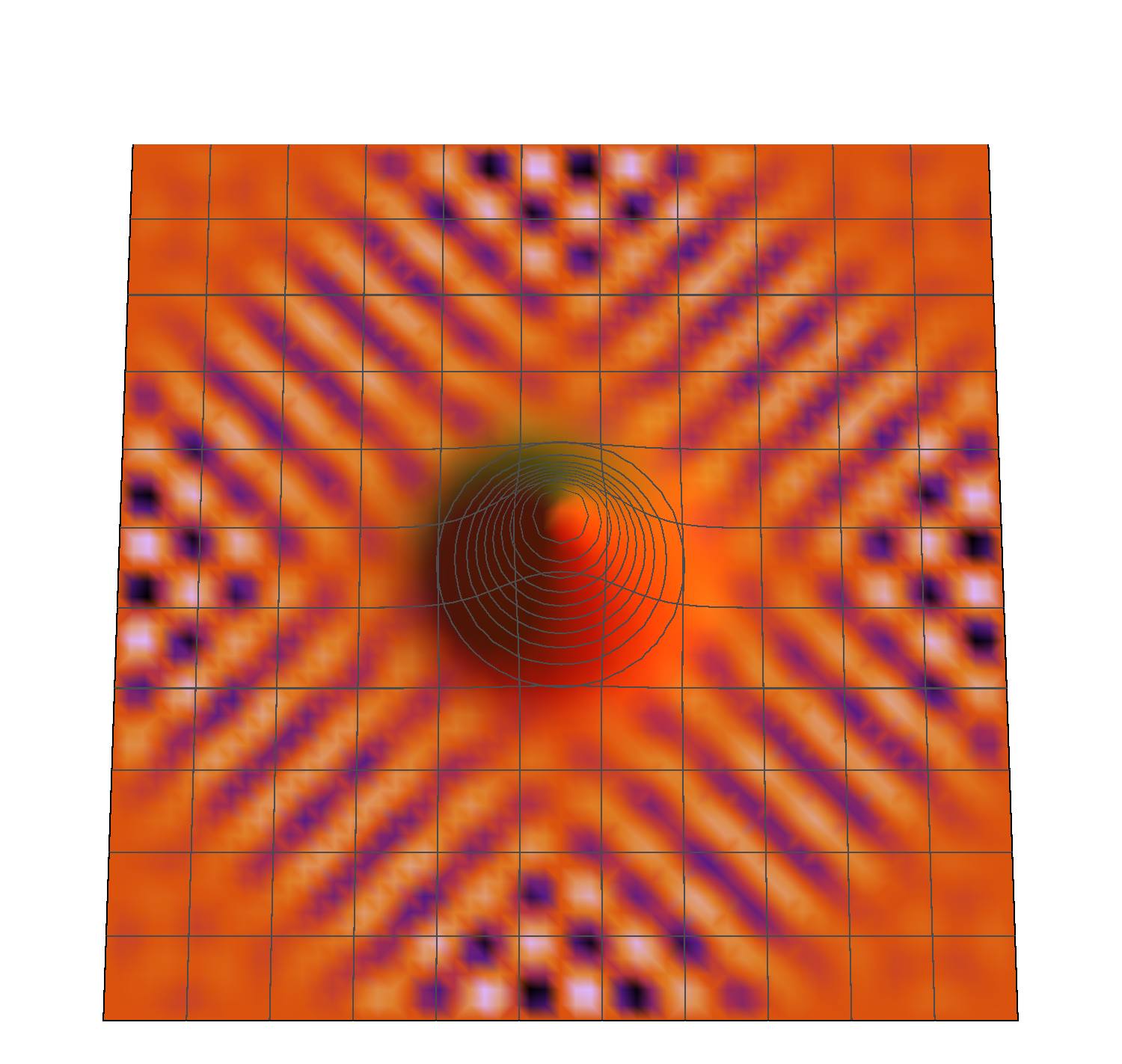}}    
  \caption[Wavefunction classical features of rectangular billiard
    with a Gaussian surface  in a
    strong electric field.] {\textbf{Wavefunction classical features
  of rectangular billiard with Gaussian surface in a strong electric
  field.} \textit{Upper panel}. Classical trajectories on the
    billiard. \textit{Lower panel}. The corresponding scar (left) and
    two bouncing ball states.}
  \label{scarsAndBouncingBallStatesOfRBGSSFfig}
\end{figure}

Scars and bouncing ball states are another interesting phenomena
observed in the standard quantum billiards. They are a manifestation
of the classical features on the wavefunction of states in the
semiclassical limit. The scarring of the wavefunction is a common
characteristic of quantum billiards with classical chaotic
counterpart. For the rectangular billiard with the Gaussian surface in
a strong electric field, a wave function presenting scars is shown in
Figure \ref{scarsAndBouncingBallStatesOfRBGSSFfig}-(d).  This state
corresponds to the closed unstable orbit presented in Figure
\ref{scarsAndBouncingBallStatesOfRBGSSFfig}-(a).  It is not a surprise
to find scars in the rectangular and circular billiards with the
Gaussian surface because of their similarity with the Sinai billiard
and the annular billiard which also exhibit scars. However, here the
scarring of the wavefunction of these billiards is a consequence of
the external field and the modification of the billiard interior
geometry. Taking into account that scarring does not appear in quantum
billiards with regular classic analogue, then this phenomenon is
another signature of chaos of the rectangular and circular billiards
with a Gaussian surface with strong field at the quantum scale.

\begin{figure}[h]
  \centering
\subfloat[]{\includegraphics[width=0.25\textwidth]{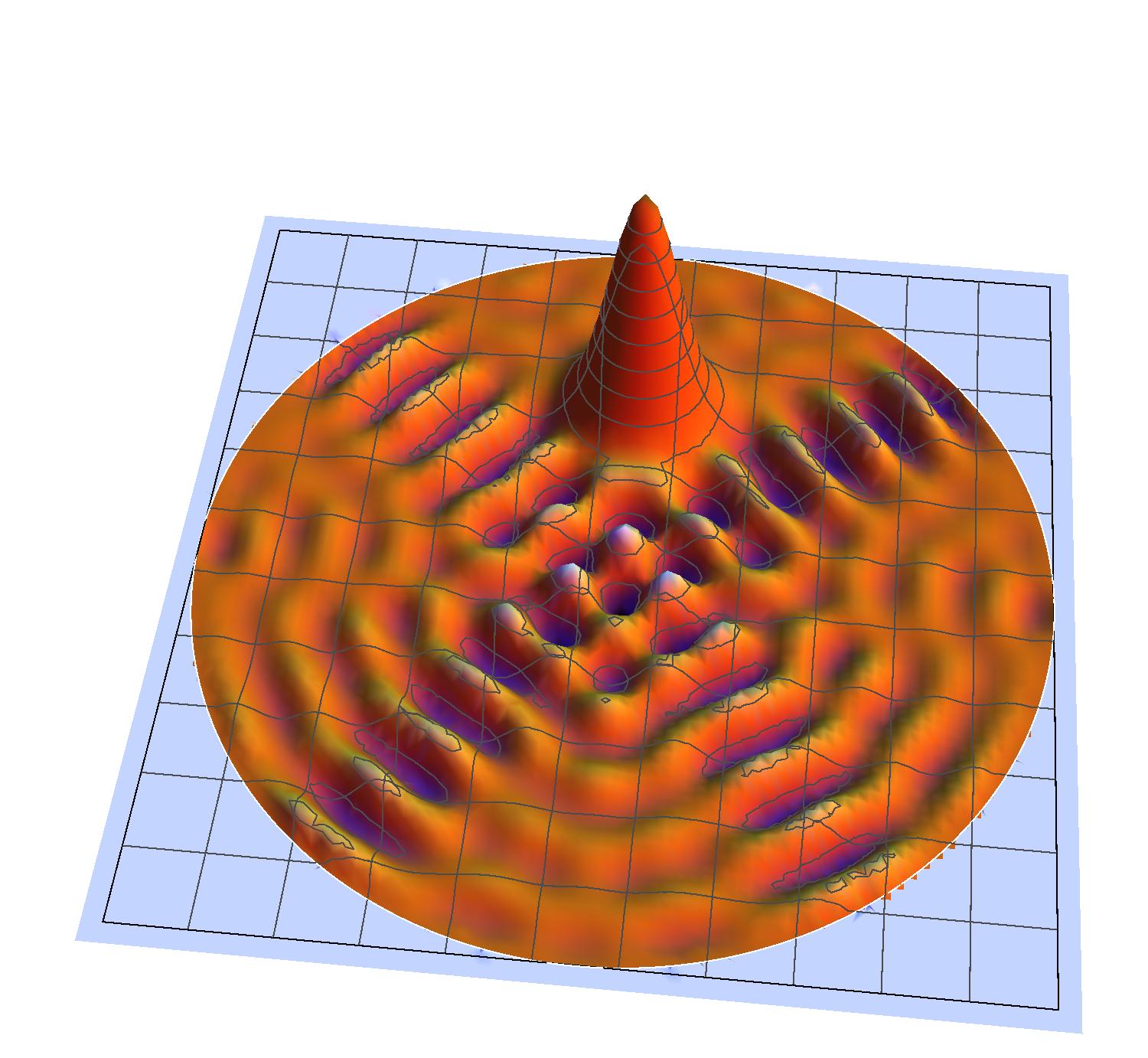}} 
\subfloat[]{\includegraphics[width=0.25\textwidth]{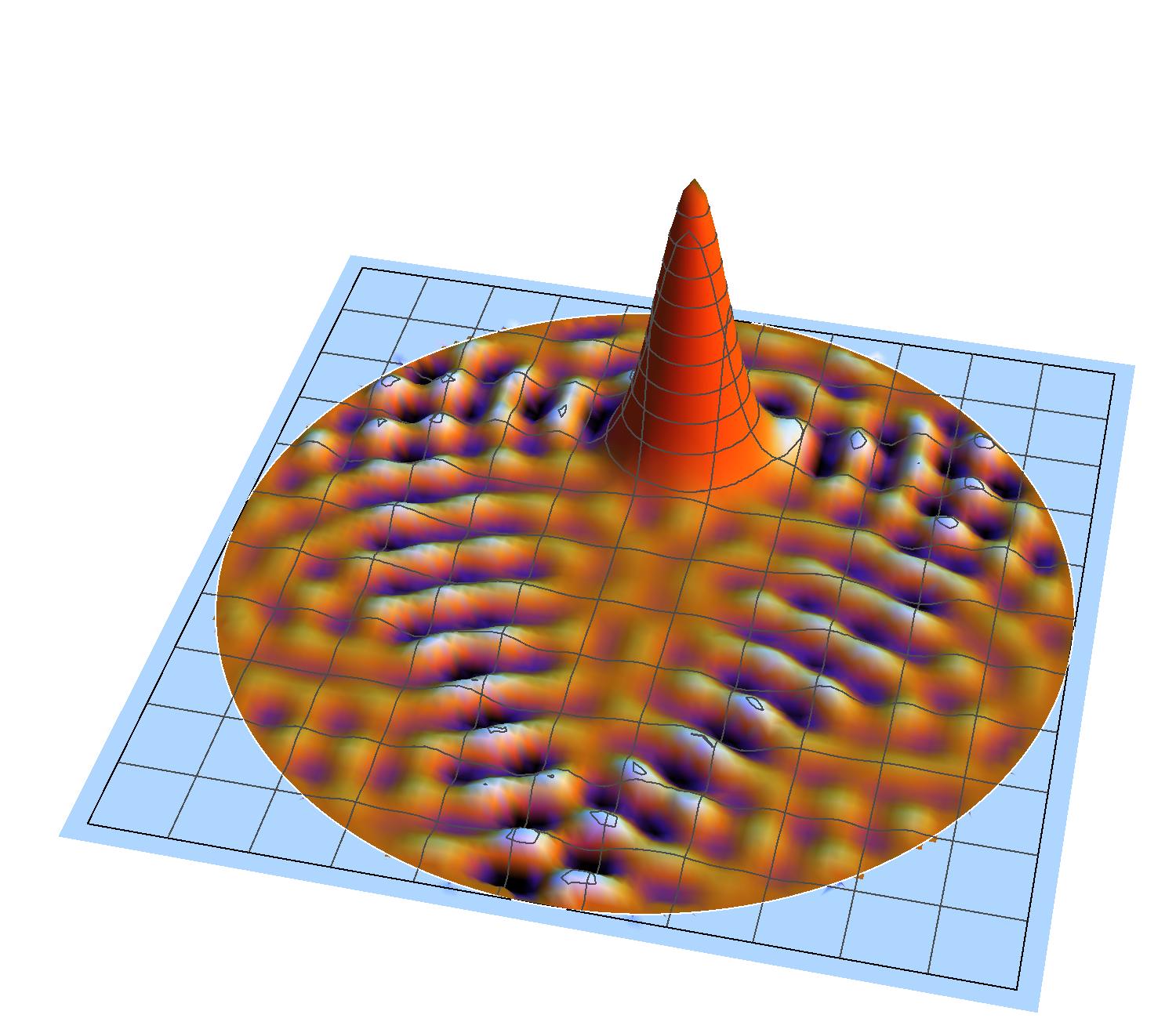}} 
\subfloat[]{\includegraphics[width=0.25\textwidth]{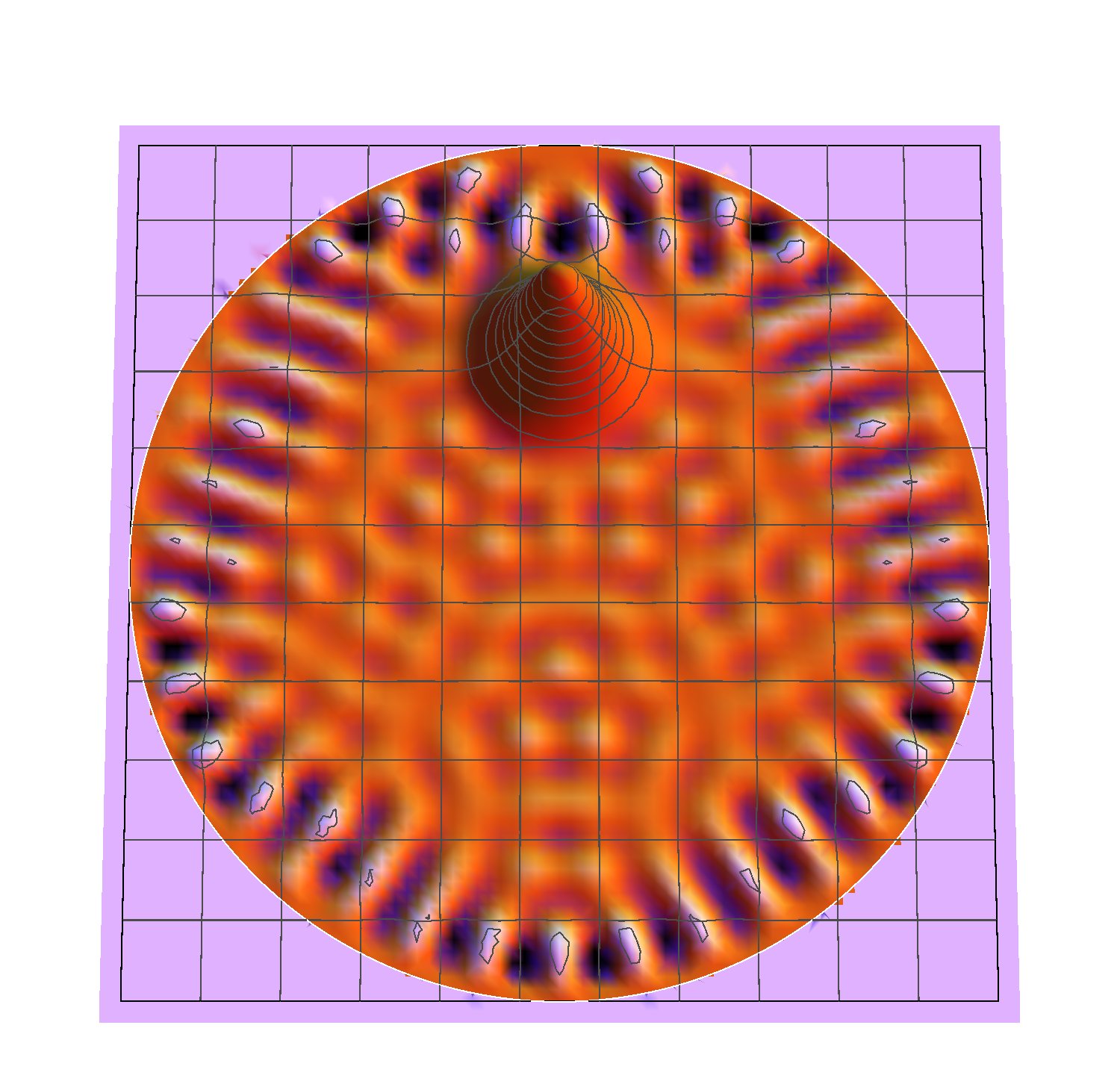}} 
\subfloat[]{\includegraphics[width=0.25\textwidth]{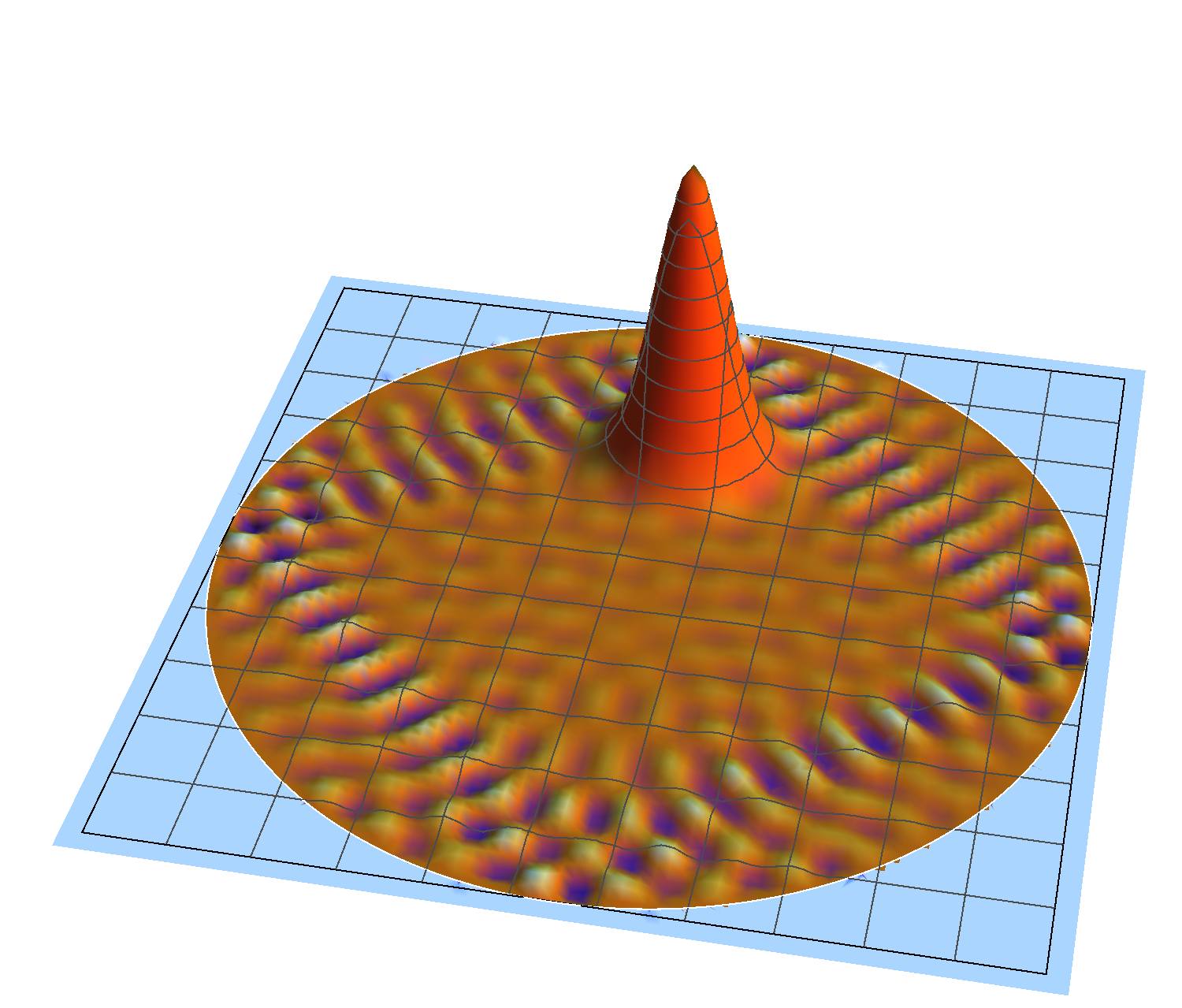}}\\ 
\subfloat[]{\includegraphics[width=0.25\textwidth]{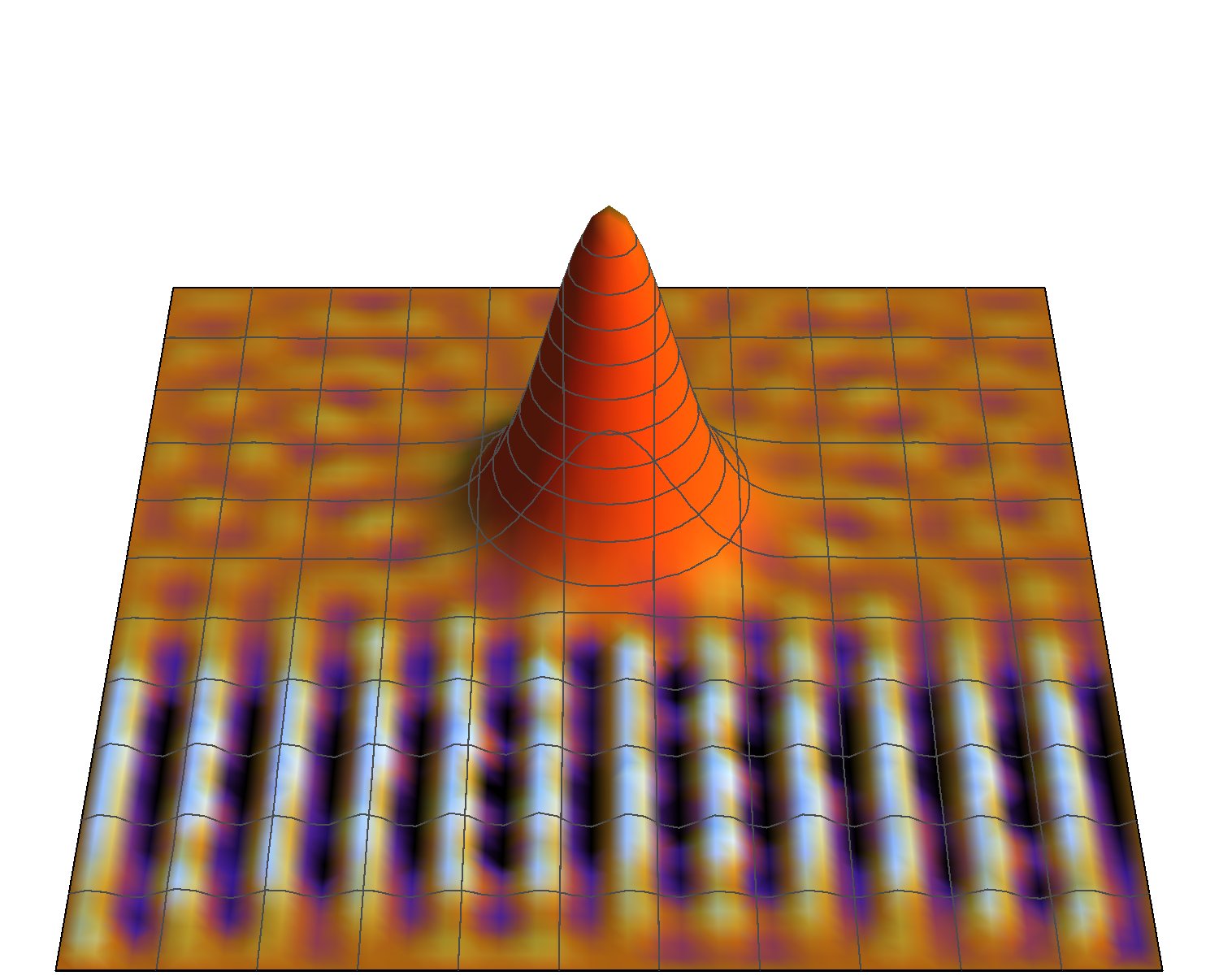}}  
\subfloat[]{\includegraphics[width=0.25\textwidth]{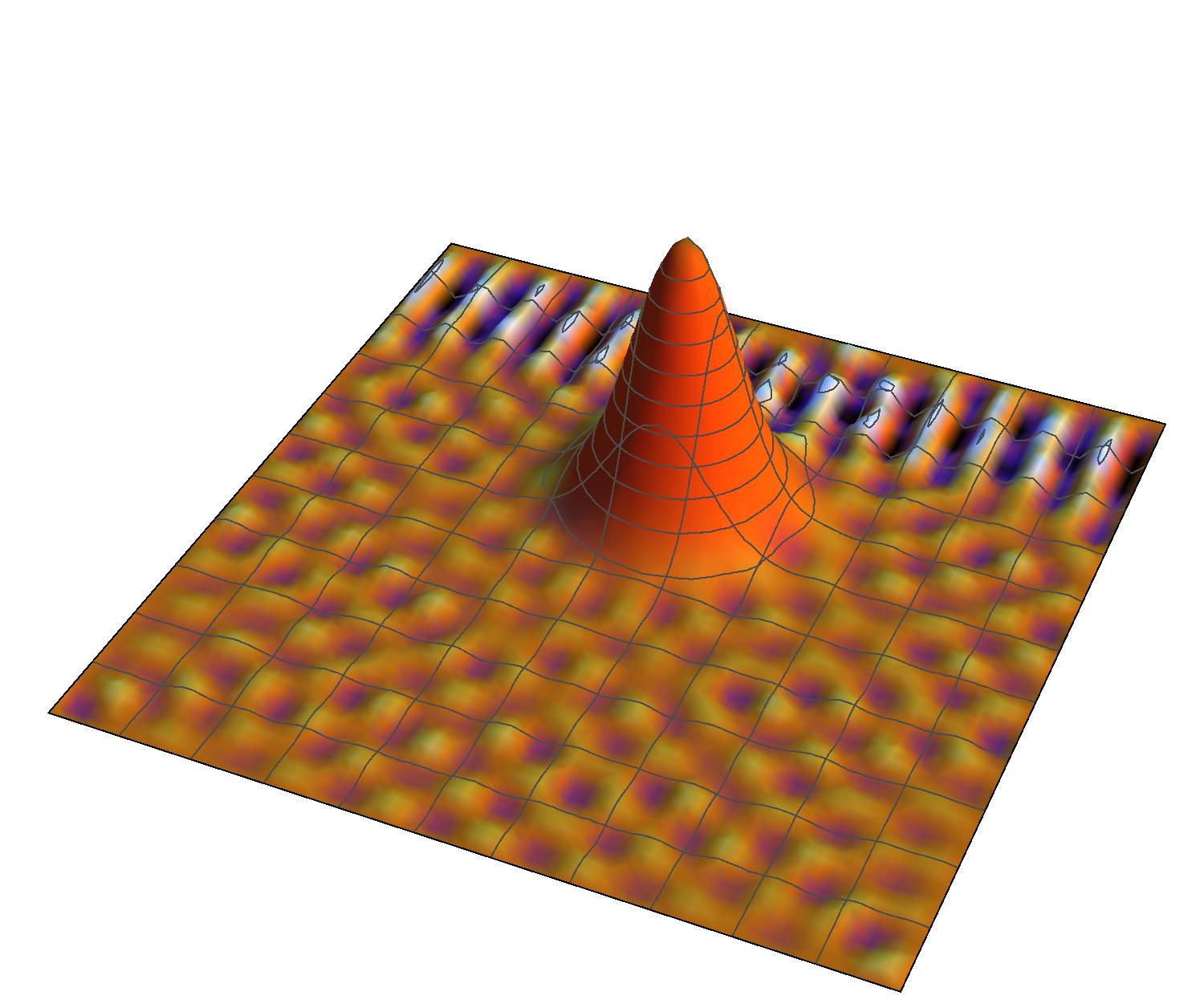}}  
\subfloat[]{\includegraphics[width=0.25\textwidth]{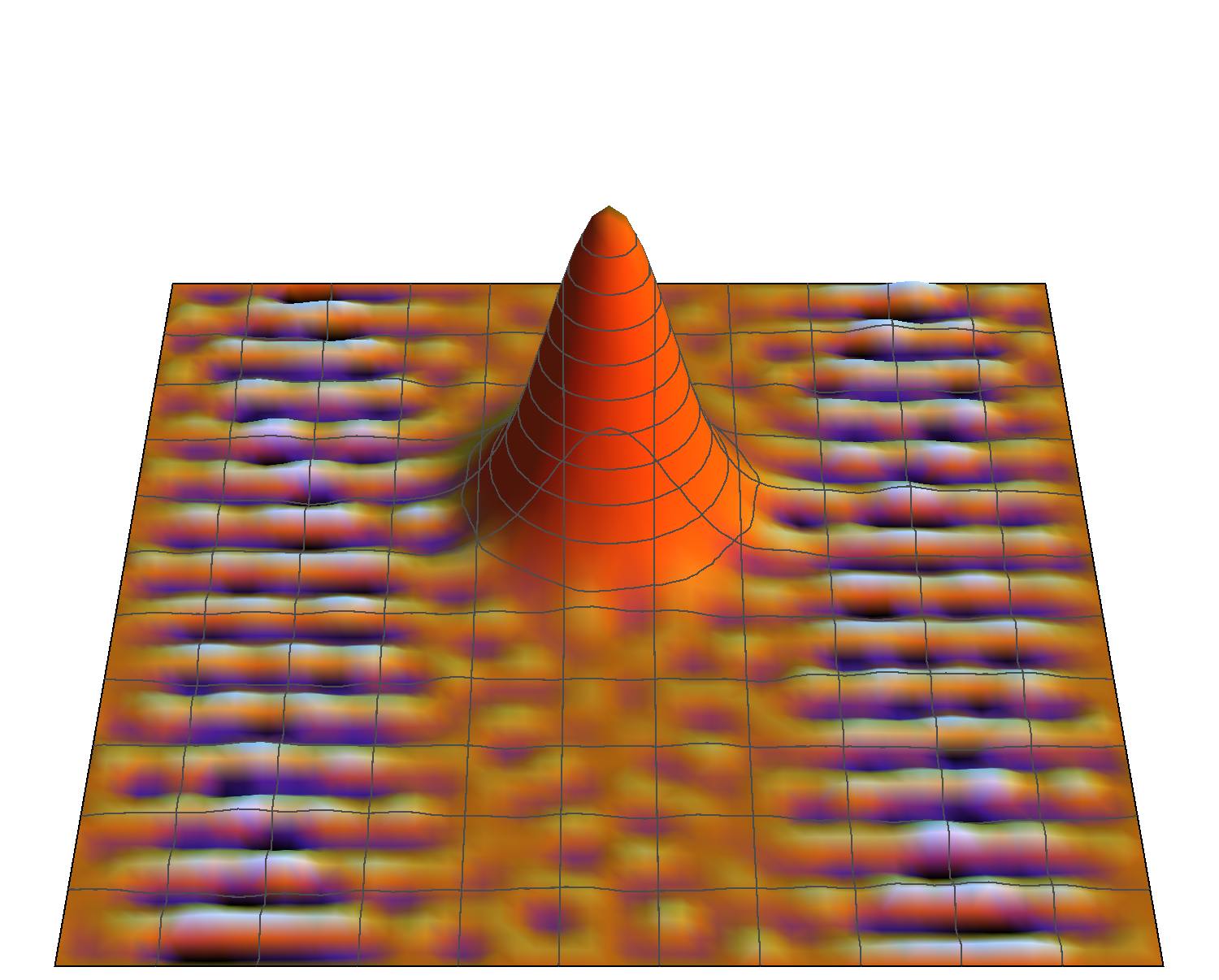}}  
\subfloat[]{\includegraphics[width=0.25\textwidth]{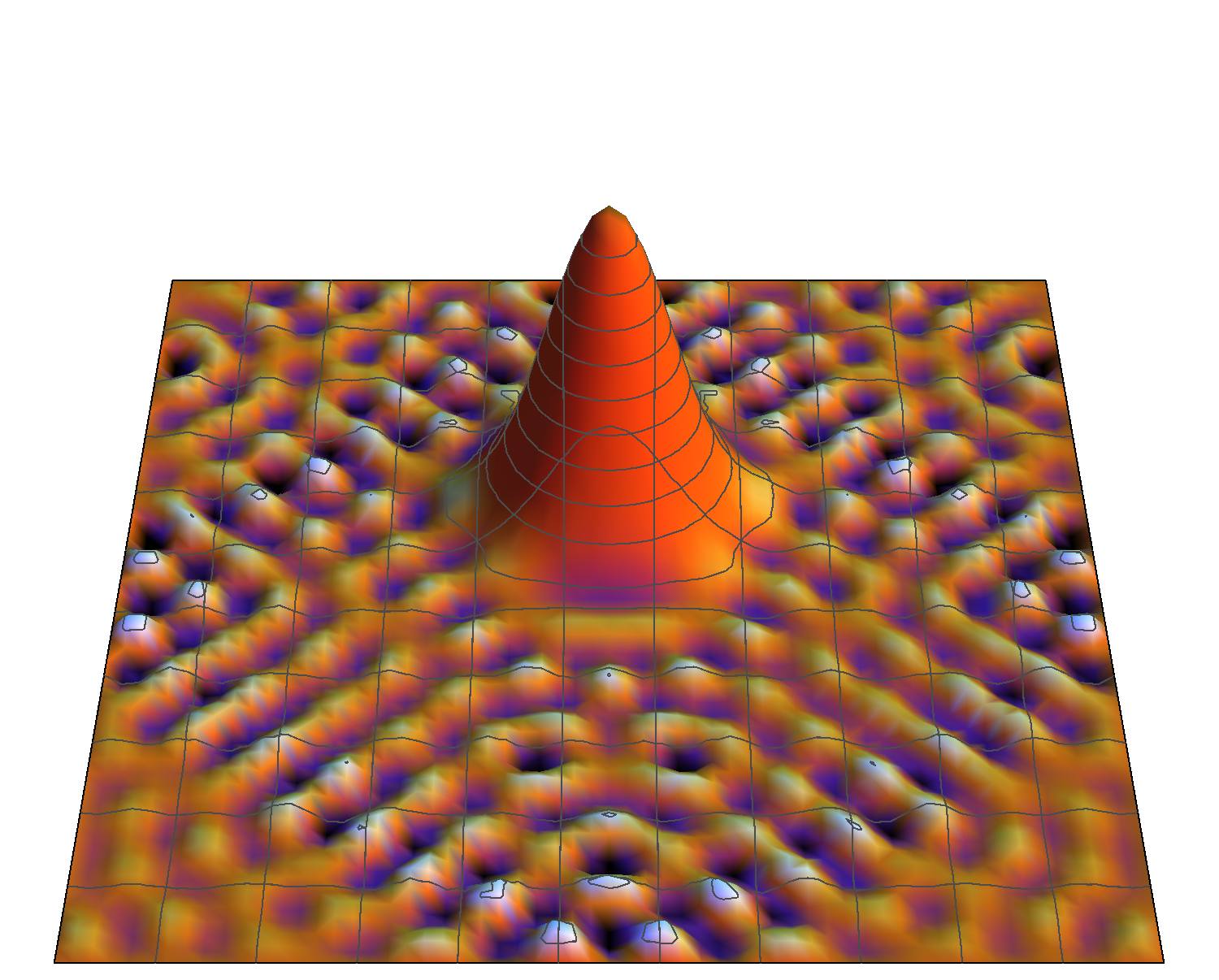}}\\
\subfloat[]{\includegraphics[width=0.35\textwidth]{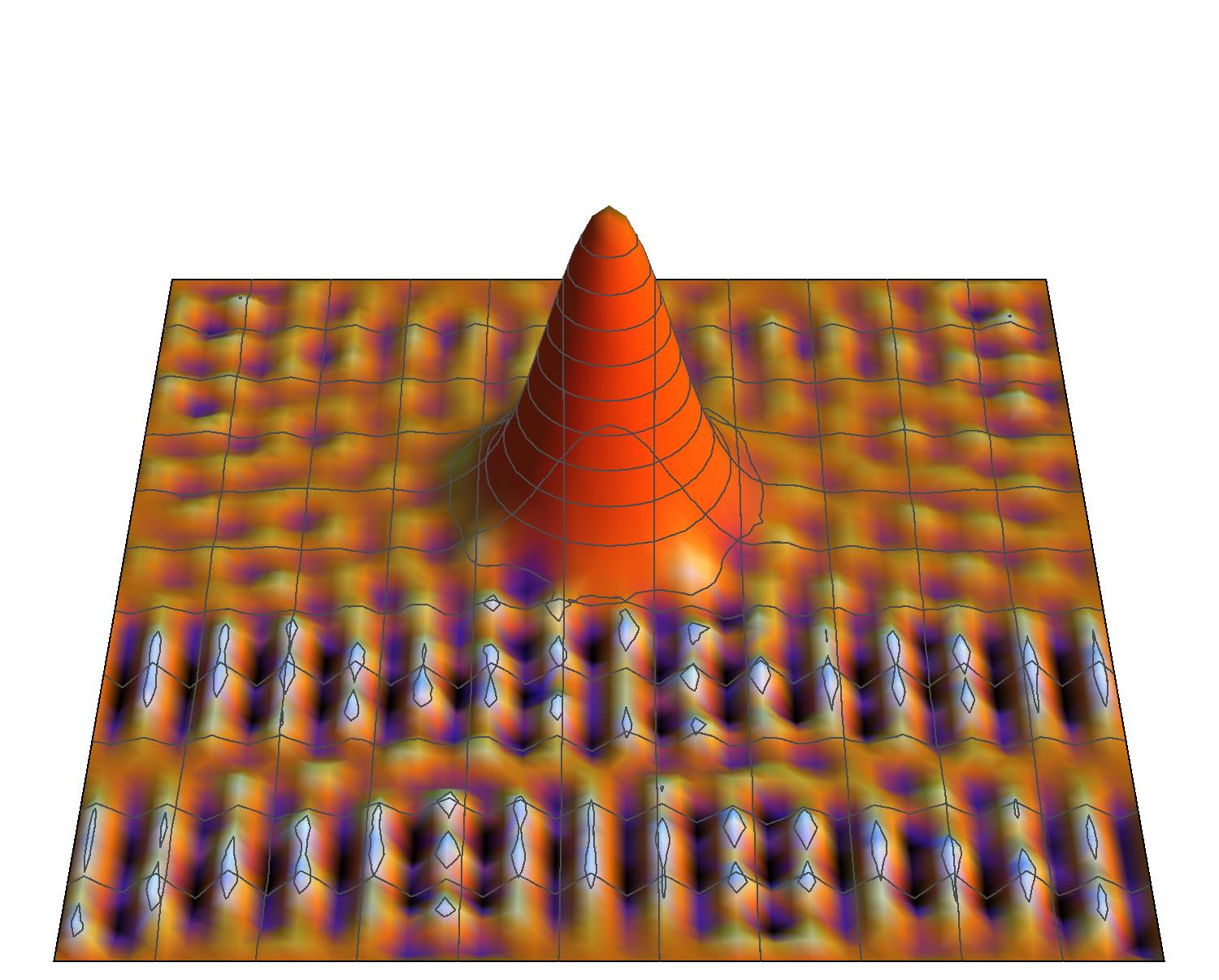}} \hspace{0.5cm} 
\subfloat[]{\includegraphics[width=0.35\textwidth]{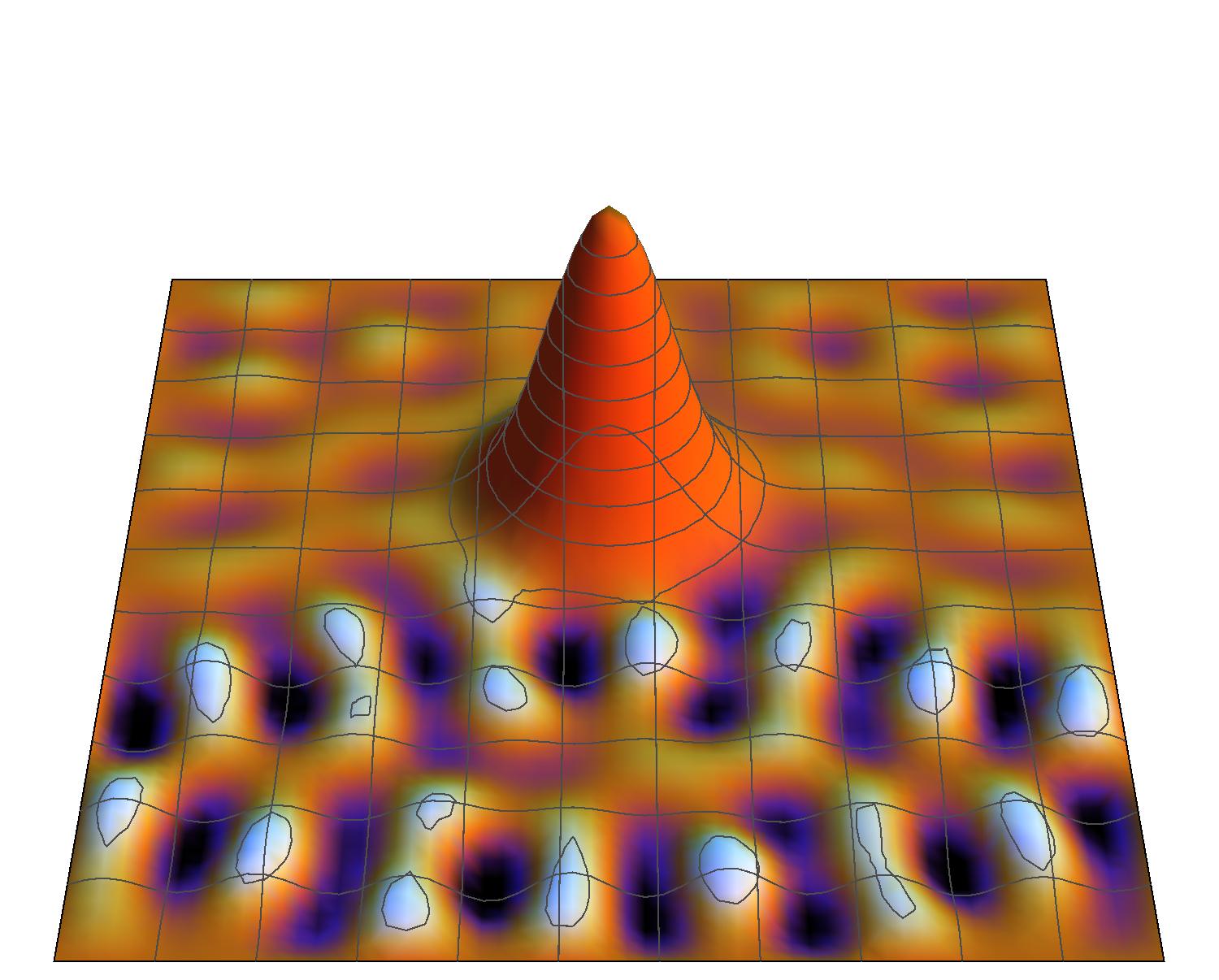}}   
  \caption[Scars and bouncing ball states of the rectangular and
    circular billiards with Gaussian surface immersed in a strong
    electric field.] {\textbf{Scars and bouncing ball states of the
  rectangular and circular billiards with Gaussian surface immersed in
  a strong electric field.} }
  \label{scarsAndBouncingBallStatesOfRBGSSFAndCBGSSFfig}
\end{figure}

Another evidence of the classical aspects on the
wavefunction are the bouncing ball states (see Figure
\ref{scarsAndBouncingBallStatesOfRBGSSFfig} and Figure
\ref{scarsAndBouncingBallStatesOfRBGSSFAndCBGSSFfig}). These states
differ from the scars because they represent a set of classical
stable trajectories on the wavefunction. Thus a bouncing ball state
represents a particle which has a well defined momentum but not a well
defined position.

\section{Concluding remarks}

The classical chaotic behaviour of a particle in a two dimensional
plane billiard depends of the billiard contour. However, the inner
geometry of the billiard is also a determinant factor. We show that if
we set a non planar surface into a billiard, then it may produce
classical chaos which may be identified at the quantum scale, even
with a regular contour (rectangular or circular). Some examples were
studied in this paper namely: the dunce hat billiard and the billiard
with a Gaussian surface with a rectangular or circular contours. We
found that the Bohigas-Giannoni-Schmit conjecture remained valid for
each of these billiards. Additionally, the billiards exhibited
scarring of the wavefunction.

For the circular billiard, if the center of the surface (cone or
Gaussian) is placed at the center of the boundary, the system is
integrable because it has two constant of motion, the Hamiltonian and
the $z$-component of the angular momentum. We explicitly found
analytically the spectrum and eigenvectors of the quantum dunce hat
billiard with a circular contour.  The effect of the surface was the
rescaling of the spectrum by a global factor with respect to the one
of a circular planar billiard. As a result, the nearest neighbour
spacing distribution is the Poisson distribution.

To confine the particle to a non-planar surface, a confining potential
should be added at the quantum level. This confining potential affects
the billiard energy levels. However, if the particle energy is high,
then the contribution of the confining potential term may be
dropped. As a result, although the Weyl's formula does not consider
the quantum confining potential, it may be used in the asymptotic
limit. This fact was analytically demonstrated with the quantum
circular dunce hat billiard and numerically observed in the dunce hat
and Gaussian billiards with a rectangular contour. 

The finite difference method and the expansion method were implemented
in order to diagonalize the Hamiltonian of the quantum rectangular
billiard with the Gaussian surface. The advantage of the second method
is that a comparison of the numerical staircase function of both
methods showed that the second one required the truncation of a
Hamiltonian to a smaller dimension in order to obtain a more energy
levels with an acceptable numerical error. For this aim, we use the
fact that the slope of the spectral staircase function is just
$A_{surface}/(4\pi)$ for a large energy where the confining potential
may be neglected.

This work was supported by Facultad de Ciencias de la Universidad de los Andes, and ECOS NORD/COLCIENCIAS-MEN-ICETEX.

\listoffigures

\end{document}